\documentclass[aps,prd,
nofootinbib,
twocolumn,
showpacs
]{revtex4-1}
\usepackage{amsmath}
\usepackage{amssymb}
\usepackage{epsfig}
\usepackage{graphicx}
\usepackage{multirow}
\usepackage{color}
\usepackage[normal]{subfigure}
\usepackage{rotating}

\newcommand{\capdef}{}
\newcommand{\mycaption}[2][\capdef]{\renewcommand{\capdef}{#2}
\caption[#1]{{\footnotesize #2}}}

\newcommand{\bwt}{\begin{widetext}}
\newcommand{\ewt}{\end{widetext}}
\newcommand{\be}{\begin{equation}}
\newcommand{\ee}{\end{equation}}
\newcommand{\bdm}{\begin{displaymath}}
\newcommand{\edm}{\end{displaymath}}
\newcommand{\bea}{\begin{eqnarray}}
\newcommand{\eea}{\end{eqnarray}}
\newcommand{\nn}{\nonumber}

\def\eq#1{{Eq.~(\ref{#1})}}
\def\eqs#1#2{{Eqs.~(\ref{#1})--(\ref{#2})}}
\def\fig#1{{Fig.~\ref{#1}}}

\def\sect#1{{Sect.~\ref{#1}}}

\def\app#1{{Appendix~\ref{#1}}}

\def\vev#1{\left\langle #1 \right\rangle}

\def\Im{\mbox{Im}\,}
\def\Re{\mbox{Re}\,}
\def\Tr{\mbox{Tr}\,}
\def\det{\mbox{det}\,}

\def\GSM{SU(3)_{c}\otimes SU(2)_{L}\otimes U(1)_{Y}}

\begin{document}


\title{Minimal Flipped $SO(10) \otimes U(1)$ Supersymmetric Higgs Model}
\date{November 8, 2010}
\pacs{12.10.-g, 12.60.Jv, 12.15.Ff}
\author{Stefano Bertolini}\email{bertolin@sissa.it}
\author{Luca Di Luzio}\email{diluzio@sissa.it}
\affiliation{INFN, Sezione di Trieste, and SISSA,
via Bonomea 265, 34136 Trieste, Italy}
\author{Michal Malinsk\'{y}}\email{malinsky@ific.uv.es}
\affiliation{AHEP Group, Instituto de F\'{\i}sica Corpuscular -- C.S.I.C./Universitat de Val\`encia, Edificio de Institutos de Paterna, Apartado 22085, E 46071 Val\`encia, Spain}
\begin{abstract}
We investigate the conditions on the Higgs sector that allow supersymmetric $SO(10)$ grand unified theories (GUT)
to break spontaneously to the standard electroweak model (SM) at the renormalizable level. If one considers Higgs representations of dimension up to the adjoint, a supersymmetric standard model vacuum
requires in most cases the presence of non-renormalizable (NR) operators.
The active role of Planck induced NR operators in the breaking of the gauge symmetry introduces a hierarchy in the mass spectrum at the GUT scale that may be an issue for gauge unification and proton decay.
We show that the minimal Higgs scenario that allows for a renormalizable breaking to the SM is obtained by considering flipped $SO(10) \otimes U(1)$ with one adjoint ($45_H$) and two pairs of $16_H\oplus \overline{16}_H$ Higgs
representations. We consider a non-anomalous matter content
and discuss the embedding of the model  in an $E_6$ grand unified scenario just above the flipped $SO(10)$ scale.
\end{abstract}
\maketitle

\section{Introduction}

It has been shown recently \cite{Bertolini:2009qj,Bertolini:2009es} that quantum effects solve the long-standing issue \cite{Yasue:1980fy} of the incompatibility between the dynamics of {the simplest} Higgs sectors in the renormalizable non-supersymmetric $SO(10)$ grand unified theory (GUT) and the gauge unification constraints. In particular, such a minimal grand unified scenarios not only support viable $SO(10)$ breaking patterns passing through intermediate {$SU(4)_{C}\otimes SU(2)_{L}\otimes U(1)_{R}$ or $SU(3)_{c}\otimes SU(2)_{L}\otimes SU(2)_{R}\otimes U(1)_{B-L}$} gauge symmetries (or their {$SU(3)_{c}\otimes SU(2)_{L}\otimes U(1)_{R}\otimes U(1)_{B-L}$} intersection), but {they also include all the ingredients necessary} for a potentially realistic description of the Standard Model (SM) flavor structure.

On the other hand, the simplest scenario featuring the Higgs scalars in $10_H\oplus 16_H\oplus 45_H$ of $SO(10)$ fails when addressing the neutrino  spectrum: in nonsupersymmetric models, the $B-L$ breaking scale $M_{B-L}$ turns out to be generally a few orders of magnitude below the GUT scale $M_{G}$. Thus, the scale of the right-handed (RH) neutrino masses $M_{N}\sim M_{B-L}^{2}/M_P$ emerging first at the $d=5$ level from an operator of the form $16_{F}^{2}(16^\ast_{H})^{2}/M_P$  (with $M_P$ typically identified with the Planck scale) undershoots by orders of magnitude the range of about $10^{12}$ to $10^{14}$ GeV naturally suggested by the seesaw mechanism.
{The same effective result is obtained in the nonsupersymmetric case within the radiative seesaw scheme~\cite{Witten:1979nr}.}

This issue can be somewhat alleviated by considering $126_H$ in place of $16_H$ in the Higgs sector, since in such a case the neutrino masses can be generated at the renormalizable level by the term $16_{F}^{2} 126^\ast_H$. This lifts the problematic $M_{B-L}/M_P$ suppression factor inherent to the $d=5$ effective mass and yields $M_{N}\sim M_{B-L}$, that might be, at least in principle, acceptable.  This scenario, though {conceptually simple}, c.f. \cite{Bertolini:2009es}, involves a detailed one-loop analysis of the scalar potential governing the dynamics of the $10_H\oplus 126_H\oplus 45_H$ Higgs sector that, to our knowledge, still remains to be done.

Invoking TeV-scale supersymmetry (SUSY), the qualitative picture changes dramatically. Indeed, the gauge running within the MSSM prefers $M_{B-L}$ in the proximity of $M_{G}$ and, hence, the Planck-suppressed $d=5$  RH neutrino mass operator {$16_{F}^{2}\overline{16}_{H}^{2}/M_{P}$, available whenever $16_H\oplus \overline{16}_H$ is present in the} Higgs sector, can naturally reproduce the desired range for $M_{N}$.
Let us recall that both $16_H$ as well as  $\overline{16}_H$ are required in order to retain SUSY below the GUT scale.

On the other hand, it is well known \cite{Buccella:1981ib,Babu:1994dc,Aulakh:2000sn}
that the relevant superpotential does not support,
at the renormalizable level, a supersymmetric breaking of the $SO(10)$
gauge group to the SM. This is due to the constraints on the vacuum manifold
imposed by the $F$- and $D$-flatness conditions which, apart from linking
the magnitudes of the $SU(5)$-singlet ${16}_H$ and ${\overline{16}_H}$
vacuum expectation values ({VEVs}), make the the adjoint {VEV $\vev{45_{H}}$}
aligned to {$\vev{16_H\overline{16}_{H}}$}.
As a consequence, an $SU(5)$ subgroup of the initial $SO(10)$ gauge symmetry remains unbroken. In this respect, a renormalizable Higgs sector with $126_H\oplus \overline{126}_H$ in place of $16_H\oplus \overline{16}_H$ suffers from the same ``$SU(5)$ lock'',
{because also in $\overline{126}_{H}$ the SM singlet direction is $SU(5)$-invariant.}

This issue {can} be addressed by giving up renormalizability. However, this option may be rather problematic since it introduces a delicate interplay between physics at two different scales, $M_{G}\ll M_{P}$, with the consequence of splitting the GUT-scale thresholds over several orders of magnitude around $M_{G}$. This may affect proton decay as well as the SUSY gauge unification, and {may} force the $B-L$ scale below the GUT scale. The latter is harmful for the setting with $16_H\oplus \overline{16}_H$ relying on a $d=5$ RH neutrino mass operator. The models with $126_H\oplus \overline{126}_H$ are also prone to trouble with gauge unification, due to the {number of large} Higgs multiplets spread around the GUT-scale.

Thus, in none of the cases above the simplest conceivable $SO(10)$ Higgs sector spanned over the lowest-dimensionality irreducible representations (up to the adjoint) seems to offer a natural scenario for realistic model building.
Since the option of a simple GUT-scale Higgs dynamics involving small representations governed by a simple renormalizable superpotential is particularly attractive, we aimed at studying the
conditions under which the seemingly ubiquitous {$SU(5)$ lock} can be {overcome}, while keeping only spinorial and adjoint $SO(10)$ representations.

Let us emphasize that the assumption that the gauge symmetry breaking is driven by the renormalizable part of the Higgs superpotential does not clash with the fact that, in models with $16_H\oplus \overline{16}_H$, the neutrino masses are generated at the non-renormalizable level, and other fermions may be sensitive to physics beyond the GUT scale.
As far as symmetry breaking is concerned, Planck induced {$d\geq 5$} effective interactions are irrelevant perturbations {in this picture}.

The simplest attempt to {breaking the $SU(5)$ lock by} doubling either $16_H\oplus \overline{16}_H$ or $45_H$ in order to relax the $F$-flatness constraints is easily shown not to work. {In the former case}, there is only one SM singlet field direction associated to each of the $16_H\oplus \overline{16}_H$ pairs. Thus, $F$-flatness makes the VEVs in $45_H$ align along this direction regardless of the number of {$16_{H}\oplus \overline{16}_{H}$'s contributing} to the relevant $F$-term, $\partial W/\partial 45_{H}$ (see for instance Eq. (6) in ref. \cite{Aulakh:2000sn}). Doubling the number of $45_H$'s does not help either.
Since there is no mixing among the 45's {besides} the mass term,
$F$-flatness aligns both $\vev{45_H}$ in the $SU(5)$ direction of  $16_H\oplus \overline{16}_H$.
For three (and more) adjoints a mixing term of the form $45_{1}45_{2}45_{3}$
is allowed, {but it turns out to be irrelevant to the minimization so that the alignment is maintained.}

From this brief excursus one might conclude that, {as far as the Higgs content is considered, the price for tractability and predictivity
is high on SUSY $SO(10)$ models,
as the desired group-theoretical simplicity
of the Higgs sector, with representations up to the
adjoint, appears not viable.}

In this paper, we point out that all these issues are alleviated if one considers a flipped variant of the SUSY $SO(10)$ unification.
In particular, we shall show that the flipped $SO(10)\otimes U(1)$ scenario~\cite{Kephart:1984xj,Rizos:1988jn,Barr:1988yj} offers an attractive option to break the gauge symmetry to the SM at the renormalizable level by means of a quite simple Higgs sector,
namely a couple of  $SO(10)$  spinors $16_{1,2}\oplus \overline{16}_{1,2}$ and one adjoint {$45_{H}$}.

Within the extended $SO(10)\otimes U(1)$ gauge algebra one finds in general {\em three} inequivalent embeddings of the SM hypercharge.
In addition to the two solutions with the hypercharge stretching over the $SU(5)$ or the $SU(5)\otimes U(1)$ subgroups of $SO(10)$ (respectively dubbed as the ``standard'' and ``flipped'' $SU(5)$ embeddings), there is a third, ``flipped'' $SO(10)$, solution inherent to the $SO(10)\otimes U(1)$ case, with a non-trivial projection of the SM hypercharge onto the $U(1)$ factor.

Whilst the difference between the standard and the flipped $SU(5)$ embedding is semantical from the $SO(10)$ point of view,  the flipped $SO(10)$ case is qualitatively different. In particular, the symmetry-breaking ``power'' of the $SO(10)$ spinor and adjoint representations is boosted with respect to the standard $SO(10)$ case,
increasing the number of SM singlet fields that may acquire non vanishing VEVs.
Technically, flipping allows for a {\em pair} of SM singlets in each of the $16_H$ and $\overline{16}_H$ ``Weyl'' spinors, together with {\em four} SM singlets within $45_H$.
This is at the root of the possibility of implementing the gauge symmetry breaking by means of a simple renormalizable Higgs sector.
Let us just remark that, if renormalizability is not required, the
breaking can be realized without the adjoint Higgs field, see for instance the flipped $SO(10)$ model with an additional anomalous $U(1)$ of Ref. \cite{Maekawa:2003wm}.

Nevertheless, flipping is not per-se sufficient to cure the {$SU(5)$ lock} of standard $SO(10)$ with $16_H\oplus \overline{16}_H\oplus 45_H$ in the Higgs sector. Indeed, the adjoint does not reduce the rank and the bi-spinor, in spite of the two qualitatively different SM singlets involved, can lower it only by a single unit, leaving a residual $SU(5)\otimes U(1)$ symmetry (the two SM singlet directions in the
$16_H$ still {retain} an $SU(5)$ algebra as {a} little group).
Only when two pairs of $16_H\oplus \overline{16}_H$ (interacting via $45_H$) are introduced the two pairs of SM singlet VEVs in the spinor multiplets may not generally be aligned and the little group is reduced to the SM.

Thus, the simplest renormalizable SUSY Higgs model that can provide the spontaneous breaking of the $SO(10)$ GUT symmetry to the SM by means of Higgs representations not larger than the adjoint, is the flipped $SO(10)\otimes U(1)$ scenario with {\em two copies} of the $16\oplus \overline{16}$ bi-spinor supplemented by the adjoint $45$. Notice further that in the flipped embedding the spinor representations include also weak doublets that may trigger the electroweak symmetry breaking
and allow for renormalizable Yukawa interactions with the chiral matter fields distributed in the flipped embedding over $16\oplus 10\oplus 1$.

Remarkably, the basics of the mechanism we advocate can be embedded in an underlying {\em non-renormalizable} $E_{6}$ Higgs model featuring a pair of $27_H\oplus \overline{27}_H$ and the adjoint $78_H$.

Technical similarities apart, there is, however, a crucial difference between the $SO(10)\otimes U(1)$ and $E_{6}$ scenarios, that is related  to the fact that the Lie-algebra of $E_{6}$ is larger than that of $SO(10)\otimes U(1)$.  It has been shown long ago \cite{Buccella:1987kc} that the renormalizable SUSY $E_{6}$ Higgs model spanned on a single copy of $27_H\oplus \overline{27}_H\oplus 78_H$ leaves an $SO(10)$ symmetry unbroken. Two pairs of $27_H \oplus \overline{27}_H$ are needed to reduce the rank by two units.
In spite of the fact that the two SM singlet directions in the $27_H$ are exactly those of the ``flipped'' $16_H$, the little group of the SM singlet directions $\vev{27_H\oplus \overline{27}_H}$ and $\vev{78_H}$
remains at the renormalizable level $SU(5)$, as we will explicitly show. 

Adding NR adjoint interactions allows for a disentanglement of the $\vev{78_H}$, such that the little group is reduced to the SM. {Since} a one-step $E_6$ breaking {is} phenomenologically problematic as mentioned earlier, we argue for a two-step breaking, via flipped $SO(10)\otimes U(1)$, with the $E_6$ scale near the Planck scale.

In summary, we make the case for an anomaly free flipped $SO(10)\otimes U(1)$ partial {unification} scenario. We provide a detailed discussion of the symmetry breaking pattern obtained within the minimal flipped $SO(10)$ SUSY Higgs model and consider its possible $E_{6}$ embedding. We finally present an {elementary} discussion of the flavour structure offered by these settings.

\section{The GUT-scale little hierarchy}
\label{littlehierarchy}

In supersymmetric $SO(10)$ models with just $45_H \oplus 16_H \oplus \overline{16}_H$ governing the GUT breaking, one way to obtain the misalignment between the adjoint and the spinors is that of invoking new physics at the Planck scale, parametrized in a model-independent way by a tower of effective operators suppressed by powers of $M_P$.

What we {call} the ``GUT-scale little hierarchy" is the hierarchy induced in the GUT spectrum by $M_{G}/M_{P}$ suppressed effective operators, which may split the GUT-scale thresholds over several orders of magnitude.
In turn this may be highly problematic for proton stability and the gauge unification in low energy SUSY scenarios (as discussed for instance in Ref. \cite{Chacko:1998jz}). It may also jeopardize the neutrino mass generation in the seesaw scheme. We briefly review the
relevant issues {here}.

\subsection{Proton decay and effective neutrino masses}

In Ref. \cite{Babu:1998wi} the emphasis is set on a class of {neutrino-mass-related} operators which turns out to be particularly dangerous for proton stability in scenarios with a nonrenormalizable GUT-breaking sector.
The relevant interactions can be schematically written as
\bea
W_Y
&\supset& \frac{1}{M_P} 16_F\ g\  16_F 16_H 16_H + \frac{1}{M_P} 16_F\ f\ 16_F \overline{16}_H \overline{16}_H \nn \\
&\supset& \frac{v_R}{M_P} \left( Q\ g\ L\ \overline{T} + Q\ f\ Q\ T \right)
\eea
where $g$ and $f$ are matrices in the family space, {$v_R \equiv |\vev{16_H}| = |\vev{\overline{16}_H}|$} and $T$ ($\overline{T}$) is the color triplet (anti-triplet)
contained in the $\overline{16}_H$ ($16_H$). Integrating out the color triplets, whose mass term is labelled {$M_T$}, one obtains the following effective superpotential involving fields belonging to $SU(2)_L$ doublets
\be
W_{eff}^{L} = \frac{v_R^2}{M_P^2 {M_{T}}} \left( u^T F d' \right) \left( u^T G V' \ell - d'^T G V' \nu' \right) \, ,
\label{Peffop}
\ee
where $u$ and $\ell$ denote the physical left-handed up quarks and charged lepton superfields in the basis in which neutral gaugino interactions are flavor diagonal. The $d'$ and $\nu'$ fields are related to the physical down quark and light neutrino fields $d$ and $\nu$ by $d'=V_{CKM} d$ and $\nu' = V_{PMNS} \nu$. In turn $V' = V^\dag_u V_\ell$, where $V_u$ and $V_\ell$ diagonalize the left-handed up quark and charged lepton mass matrices respectively. The $3 \times 3$ matrices $(G,F)$ {are given by} $(G,F) = V^T_u (g,f) V_u$.

By exploiting the correlations between the $g$ and $f$ matrices and the matter masses and mixings and by taking into account the uncertainties related to the low-energy SUSY spectrum, the GUT-thresholds and the hadronic matrix elements, the authors of Ref. \cite{Babu:1998wi} argue that the effective operators in \eq{Peffop} lead to a proton lifetime
\be
\Gamma^{-1} (\overline{\nu} K^+) \sim \left( 0.6 - 3 \right) \times 10^{33} \ \text{yrs} \, ,
\label{plife}
\ee
at the verge of the current
experimental lower bound of $0.67 \times 10^{33}$ years \cite{Amsler:2008zzb}.
In obtaining \eq{plife} the authors 
assume that the color triplet masses cluster about the GUT scale, $M_T \approx \vev{16_{H}}\sim\vev{45_{H}}\equiv M_G$. On the other hand, in scenarios where at the renormalizable level $SO(10)$ is broken to $SU(5)$ and the residual $SU(5)$ symmetry is broken to SM by means of non-renormalizable operators, the effective scale of the $SU(5)$ breaking physics is  typically suppressed by $\vev{16_{H}}/M_{P}$ or $\vev{45_{H}}/M_{P}$ with respect to $M_G$. As a consequence, the $SU(5)$-part of the colored triplet higgsino spectrum is effectively pulled down to the $M_{G}^{2}/M_{P}$ scale, in a clash with proton stability.

\subsection{GUT-scale thresholds and one-step unification}
The ``delayed'' residual $SU(5)$ breakdown has obvious implications for
the shape of the gauge coupling unification pattern.
Indeed, the {gauge bosons associated to the $SU(5)/{\rm SM}$ coset}, together with the relevant
part of the Higgs spectrum, tend to be uniformly shifted \cite{Babu:1994dc}
by a factor $M_G / M_P \sim 10^{-2}$ below the scale of the $SO(10)/SU(5)$
gauge spectrum, that sets the unification scale, $M_G$.
These thresholds may jeopardize the successful one-step gauge unification pattern favoured by the TeV-scale SUSY extension of the SM (MSSM).

\subsection{GUT-scale thresholds and neutrino masses}

With a non-trivial interplay among several GUT-scale thresholds \cite{Babu:1994dc} one may in principle end up with a viable gauge unification pattern. Namely, the threshold effects in different SM gauge sectors may be such that unification is preserved at a larger scale. In such a case the $M_{G}/M_{P}$ suppression is at least partially undone. This, in turn, is unwelcome for the neutrino mass scale because the VEVs entering the {$d=5$} effective operator responsible for the RH neutrino Majorana mass term {$16_{F}^{2}\overline{16}_{H}^{2}/M_{P}$}
are raised {accordingly} and thus $M_R \sim {M_G^{2}}/{M_P}$ tends to overshoot the upper limit $M_{R}\lesssim 10^{14}$ GeV implied by the light neutrino masses generated by the seesaw mechanism.

Thus, although the Planck-induced operators can provide a key to overcoming the {$SU(5)$ lock} of the minimal SUSY $SO(10)\to \GSM$ Higgs model with $16_{H}\oplus \overline{16}_{H}\oplus 45_{H}$, such an
effective scenario is prone to {failure} when addressing the measured proton stability and light neutrino phenomenology.

\section{Minimal flipped $SO(10)$ Higgs model}
\label{sect:minimalflippedSO10}

\renewcommand{\arraystretch}{1.3}
\begin{table*}[t]
\begin{tabular}{lllll}
\hline \hline
& \multicolumn{2}{l}{Standard $SO(10)$} & \multicolumn{2}{l}{Flipped $SO(10) \otimes U(1)$}
\\
\hline
Higgs superfields & R & NR & R & NR
\\
\hline
$16 \oplus \overline{16}$ & $SO(10)$ & $SU(5)$ & $SO(10) \otimes U(1)$ & $SU(5) \otimes U(1)$
\\
\null
$ 2 \times \left(16 \oplus \overline{16} \right)$ & $SO(10)$ & $SU(5)$ & $SO(10) \otimes U(1)$ & SM
\\
\null
$45 \oplus 16 \oplus \overline{16}$ & $SU(5)$ \cite{Buccella:1981ib} & SM \cite{Babu:1994dc} & $SU(5) \otimes U(1)$ & $\text{SM} \otimes U(1)$
\\
\null
$45 \oplus 2 \times \left( 16 \oplus \overline{16} \right) $ & $SU(5)$ & SM & {\bf SM} & SM
\\
\hline \hline
\end{tabular}
\mycaption{Comparative summary of supersymmetric vacua left invariant by the SM singlet VEVs in various combinations of spinorial and adjoint
Higgs representations of
standard $SO(10)$ and flipped $SO(10) \otimes U(1)$.
The results for a renormalizable (R) and a non-renormalizable (NR) Higgs superpotential are respectively listed.
}
\label{tab:standvsflipbreak}
\end{table*}

\renewcommand{\arraystretch}{1.3}
\begin{table}
\begin{tabular}{lll}
\hline \hline
Higgs superfields & R & NR
\\
\hline
$27 \oplus \overline{27}$ & $E_6$ & $SO(10)$
\\
\null
$2 \times \left( 27 \oplus \overline{27} \right)$ & $E_6$ & $SU(5)$
\\
\null
$78 \oplus 27 \oplus \overline{27}$ & $SO(10)$ \cite{Buccella:1987kc} & $\text{SM} \otimes U(1)$
\\
\null
$78 \oplus 2 \times \left( 27 \oplus \overline{27} \right)$ & $\mathbf{SU(5)}$ & SM
\\
\hline \hline
\end{tabular}
\mycaption{Same as in Table \ref{tab:standvsflipbreak} for the $E_6$ gauge group with fundamental and adjoint Higgs representations.
}
\label{tab:E6break}
\end{table}

As already anticipated in the previous sections, in a standard $SO(10)$ framework with a Higgs sector
built off {the} lowest-dimensional representations (up to the adjoint), it is rather difficult to achieve a phenomenologically viable symmetry breaking pattern even admitting multiple copies of each type of multiplets. Firstly, with a single $45_{H}$ at play, at the renormalizable-level the little group of all SM singlet VEVs is $SU(5)$ regardless of the number of $16_{H}\oplus \overline{16}_{H}$ pairs.
The {reason is} that one can not get anything more than an $SU(5)$ singlet out of a number of $SU(5)$ singlets. The same is true with a second $45_{H}$ added into the Higgs sector because there is no renormalizable mixing among the two $45_{H}$'s apart from the mass term that, without loss of generality, can be taken diagonal.
With a third adjoint Higgs representation at play a cubic $45_145_2 45_3$ interaction is allowed.
However, due to the total antisymmetry of the invariant and {to the fact that the adjoints commute on the SM vacuum,
the cubic term does not contribute to the F-term equations~\cite{Babu:1993we}.
This makes the simple flipped $SO(10)\otimes U(1)$ model proposed in this work a framework worth of consideration.}
For the sake of completeness, let us also recall that admitting Higgs representations
larger than the adjoint a renormalizable $SO(10)\to $ SM breaking can be devised with the Higgs sector of the form $54{_H} \oplus 45{_H} \oplus 16{_H} \oplus \overline{16}{_H}$ \cite{Buccella:2002zt}, or $54{_H}\oplus 45{_H} \oplus 126{_H} \oplus \overline{126}{_H}$ \cite{Aulakh:2000sn} for a renormalizable seesaw.

In Tables \ref{tab:standvsflipbreak} and \ref{tab:E6break} we collect
a list of the supersymmetric vacua that are obtained in {the basic $SO(10)$ Higgs}
models and {their} $E_6$ embeddings by considering a set of Higgs
representations of the dimension of the adjoint and smaller, with all SM singlet VEVs turned on. The cases of a renormalizable {(R) or non-renormalizable (NR)} Higgs potential are compared.
We quote reference papers where results relevant for the present study were obtained without any aim of exhausting the available literature.
The results without reference are either verified by us or follow by comparison with other cases and rank counting.
The {main results of this study} are shown in boldface.

We are going to show that by considering a non-standard hypercharge embedding in $SO(10) \otimes U(1)$ (flipped $SO(10)$) the breaking
to the SM is achievable at the renormalizable level with
$45{_H} \oplus 2 \times \left( 16{_H} \oplus \overline{16}{_H} \right)$ Higgs fields. Let us stress that what we require is that the GUT symmetry breaking is driven by the renormalizable part of the superpotential,
while Planck suppressed interactions may be relevant for the fermion
mass spectrum, in particular for the neutrino sector.


\subsection{Introducing the model}
\subsubsection{Hypercharge embeddings in $SO(10)\otimes U(1)$}
\label{Yembeds}

The so called flipped realization of the  $SO(10)$ gauge symmetry
requires an additional $U(1)_{X}$ gauge factor in order to provide an extra degree of freedom for the SM hypercharge identification.
For a fixed embedding of the $SU(3)_{c}\otimes SU(2)_{L}$ subgroup within $SO(10)$, the SM hypercharge can be generally spanned over the three remaining Cartans generating the abelian $U(1)^{3}$ subgroup of the $SO(10)\otimes U(1)_{X}/(SU(3)_{c}\otimes SU(2)_{L})$ coset.
There are two consistent implementations of the SM hypercharge
within the $SO(10)$ algebra (commonly denoted by standard and flipped $SU(5)$), while a third one becomes available due to the presence of $U(1)_{X}$.

In order to discuss the different embeddings
we find useful to consider two bases for the $U(1)^{3}$ subgroup.
Adopting the traditional left-right (LR) basis corresponding to the $SU(3)_{c}\otimes SU(2)_{L}\otimes SU(2)_{R}\otimes U(1)_{B-L}$ subalgebra of $SO(10)$, one can span the SM hypercharge on the generators of $U(1)_{R}\otimes U(1)_{B-L}\otimes U(1)_{X}$:
\be
\label{YPS}
Y=\alpha T_{R}^{(3)}+\beta(B-L)+\gamma X.
\ee
The normalization of the $T_{R}^{(3)}$ and $B-L$ charges is chosen so that the decompositions of the spinorial and vector representations of $SO(10)$ with respect to  $SU(3)_{c}\otimes SU(2)_{L}\otimes U(1)_{R}\otimes U(1)_{B-L}$ read
\bea
16 &= &(3,2;0,+\tfrac{1}{3})\oplus (\overline{3},1;+\tfrac{1}{2},-\tfrac{1}{3})\oplus (\overline{3},1;-\tfrac{1}{2},-\tfrac{1}{3}) \nn \\
& \oplus & (1,2;0,-1)\oplus (1,1;+\tfrac{1}{2},+1)\oplus (1,1;-\tfrac{1}{2},+1)\nn\;,\\
10 &= &(3,1;0,-\tfrac{2}{3})\oplus (\overline{3},1;0,+\tfrac{2}{3})\label{LRdecompositions}\\
&\oplus &(1,2;+\tfrac{1}{2},0)\oplus (1,2;-\tfrac{1}{2},0)\nn\;,
\eea
which account for the standard $B-L$ and $T_{R}^{(3)}$ assignments.

Alternatively, considering the $SU(5)\otimes U(1)_{Z}$ subalgebra of $SO(10)$, we identify the $U(1)_{Y'}\otimes U(1)_{Z}\otimes U(1)_{X}$ subgroup of $SO(10)\otimes U(1)_{X}$, and equivalently write:
\be
\label{YSU5}
Y=\tilde\alpha Y'+\tilde\beta Z+\tilde\gamma X\;,
\ee
where $Y'$ and $Z$ are normalized
so that the $SU(3)_{c}\otimes SU(2)_{L}\otimes U(1)_{Y'}\otimes U(1)_{Z}$ analogue of eqs.~(\ref{LRdecompositions}) reads:
\bea
16 &= &(3,2;+\tfrac{1}{6},+1)\oplus (\overline{3},1;+\tfrac{1}{3},-3)\oplus (\overline{3},1;-\tfrac{2}{3},+1) \nn \\
& \oplus & (1,2;-\tfrac{1}{2},-3)\oplus (1,1;+1,+1)\oplus (1,1;0,+5)\nn\;,\\
10 &= &(3,1;-\tfrac{1}{3},-2)\oplus (\overline{3},1;+\tfrac{1}{3},+2)\label{SU5decompositions}\\
&\oplus &(1,2;+\tfrac{1}{2},-2)\oplus (1,2;-\tfrac{1}{2},+2)\;.\nn
\eea
In both cases,  the $U(1)_X$ charge has been conveniently fixed to $X_{16}=+1$ for the spinorial representation (and thus $X_{10}=-2$ and also $X_{1}=+4$ for the $SO(10)$ vector and singlet, {respectively;
this is also the minimal way} to obtain an anomaly-free $U(1)_{X}$, that
allows $SO(10)\otimes U(1)_{X}$ to be naturally embedded into $E_{6}$).

It is a straightforward exercise to show that in order to accommodate the SM quark multiplets with  quantum numbers $Q=(3,2,{+}\frac{1}{6})$, $u^{c}=(\overline{3},1,-\frac{2}{3})$ and $d^{c}=(\overline{3},1,+\frac{1}{3})$ there are only three solutions.

On the $U(1)^3$ bases of \eq{YPS} {(and \eq{YSU5}, respectively)} one obtains,
\be\label{standardabc}
\alpha=1\;,\beta=\tfrac{1}{2}\;,\gamma=0\,, \ \ \ \left( \tilde\alpha=1\;,\tilde\beta=0\;,\tilde\gamma=0 \right)\,,
\ee
which is nothing but the ``standard'' embedding of the SM matter into $SO(10)$. Explicitly, $Y=T_{R}^{(3)}+\tfrac{1}{2}(B-L)$ in the LR basis (while $Y=Y'$ in the $SU(5)$ picture).

The second option is characterized by
\be\label{flippedSU5abc}
\alpha=-1\;,\beta=\tfrac{1}{2}\,,\gamma=0\;, \ \left(  \tilde\alpha=-\tfrac{1}{5}\;,\tilde\beta=\tfrac{1}{5}\;,\tilde\gamma=0 \right)\,,
\ee
which is usually denoted ``flipped $SU(5)$''~\cite{DeRujula:1980qc,Barr:1981qv}
embedding because the SM hypercharge is spanned non-trivially on the $SU(5)\otimes U(1)_{Z}$ subgroup\footnote{By definition, a flipped variant of a specific GUT model based on a simple gauge group $G$ is obtained by embedding the SM hypercharge nontrivially into the $G\otimes U(1)$ tensor product.} of $SO(10)$, $Y=\tfrac{1}{5}(Z-Y')$.
Remarkably, from the $SU(3)_{c}\otimes SU(2)_{L}\otimes SU(2)_{R}\otimes U(1)_{B-L}$ perspective this setting corresponds to a sign flip of the $SU(2)_{R}$ Cartan operator $T_{R}^{(3)}$, namely $Y=-T_{R}^{(3)}+\tfrac{1}{2}(B-L)$ which can be viewed as a $\pi$ rotation in the $SU(2)_{R}$ algebra.

A third solution corresponds to
\be\label{flippedSO10abc}
\!\alpha=0,\beta=-\tfrac{1}{4} ,\gamma=\tfrac{1}{4},  \left(  \tilde\alpha=-\tfrac{1}{5} ,\tilde\beta=-\tfrac{1}{20} ,\tilde\gamma=\tfrac{1}{4} \right),
\ee
denoted as ``flipped $SO(10)$''~\cite{Kephart:1984xj,Rizos:1988jn,Barr:1988yj} embedding of the SM hypercharge. Notice, in particular, the fundamental difference between the setting (\ref{flippedSO10abc}) with $\gamma =\tilde\gamma= \tfrac{1}{4}$ and the two previous cases (\ref{standardabc}) and (\ref{flippedSU5abc}) where $U(1)_{X}$ does not play any role.

Analogously to what is found for $Y$, once we consider the additional anomaly-free $U(1)_X$ gauge factor,
there are three SM-compatible ways of embedding the physical $\left( B-L \right)$ into $SO(10) \otimes U(1)_X$.
Using the $SU(5)$ compatible description they are respectively given by (see Ref. \cite{Harada:2003sb} for a complete set of relations)
\bea
\label{BLst}
\left( B-L \right)
&=& \tfrac{1}{5} \left( 4 Y' + Z \right)
\, , \\
\label{BLtw1}
\left( B-L \right)
&=& \tfrac{1}{20} \left( 16 Y' - Z + 5 X \right)
\, , \\
\label{BLtw2}
\left( B-L \right)
&=& - \tfrac{1}{20} \left( 8 Y' - 3 Z - 5 X \right)
\, .
\eea
where the first assignment is the standard $B-L$ embedding in \eq{YPS}.
Out of $3 \times 3$ possible pairs of $Y$ and $\left( B-L \right)$ charges only $6$ do correspond to
the quantum numbers of the SM matter \cite{Harada:2003sb}.
By focussing on the flipped $SO(10)$ hypercharge embedding
in \eq{flippedSO10abc}, the two SM-compatible $\left( B-L \right)$ assignments are those in \eqs{BLtw1}{BLtw2} (they are related by a sign flip in $T^{(3)}_R$).
In what follows we shall employ the $\left( B-L \right)$ assignment in \eq{BLtw2}.

\subsubsection{Spinor and adjoint SM singlets in flipped $SO(10)$}
The active role of the $U(1)_{X}$ generator in the SM hypercharge
(and $B-L$) identification within the flipped $SO(10)$ scenario has relevant consequences for model building. {In particular}, the SM decomposition of the $SO(10)$ representations change so that there are additional SM singlets both in $16_{H}\oplus \overline{16}_{H}$ as well as in $45_{H}$.

The pattern of SM singlet components in flipped $SO(10)$ has a simple and intuitive {interpretation} from the $SO(10)\otimes U(1)_{X}\subset E_{6}$ perspective, where $16_{+1}\oplus \overline{16}_{-1}$ (with the subscript indicating the $U(1)_{X}$ charge)
are contained in $27\oplus \overline{27}$ while $45_{0}$ is a part of the $E_{6}$ adjoint $78$. The point is that the flipped SM hypercharge assignment makes the various SM singlets within the complete $E_{6}$ representations ``migrate'' among their different $SO(10)$ sub-multiplets; namely, the two SM singlets in the $27$ of $E_{6}$ that in the standard embedding (\ref{standardabc}) reside in the $SO(10)$ singlet $1$ and spinorial $16$ components both happen to fall into just the single $16\subset 27$ {in the flipped $SO(10)$ case}.

Similarly, there are two additional SM singlet directions in $45_{0}$ in the flipped $SO(10)$ scenario, that, in the standard $SO(10)$ embedding, belong to the $16_{-3}\oplus \overline{16}_{+3}$ components of the $78$ of  $E_{6}$, thus accounting for a total of four adjoint SM singlets.

In Tables \ref{tab:10decomp}, \ref{tab:16decomp} and \ref{tab:45decomp}
we summarize the decomposition of the $10_{-2}$, $16_{+1}$ and $45_{0}$ representations of $SO(10)\otimes U(1)_{X}$ under the SM subgroup,
in both the standard and the flipped $SO(10)$ cases (and in both the LR
and $SU(5)$ descriptions). The pattern of the SM singlet components
is emphasized in boldface.

\renewcommand{\arraystretch}{1.3}
\begin{table}[ht]
\begin{tabular}{llll}
\hline \hline
\multicolumn{2}{c}{LR} & \multicolumn{2}{c}{$SU(5)$}
\\
\hline
$SO(10)$
& $SO(10)_f$
& $SO(10)$
& $SO(10)_f$
\\
\hline
$(3,1;-\tfrac{1}{3})_{6}$
& $(3,1;-\tfrac{1}{3})_{6}$
& $(3,1;-\tfrac{1}{3})_{5}$
& $(3,1;-\tfrac{1}{3})_{5}$
\\
\null
$(\overline{3},1;+\tfrac{1}{3})_{6}$
& $(\overline{3},1;-\tfrac{2}{3})_{6}$
& $(1,2;+\tfrac{1}{2})_{5}$
& $(1,2;-\tfrac{1}{2})_{5}$
\\
\null
$(1,2;+\tfrac{1}{2})_{1^+}$
& $(1,2;-\tfrac{1}{2})_{1^+}$
& $(\overline{3},1;+\tfrac{1}{3})_{\overline{5}}$
& $(\overline{3},1;-\tfrac{2}{3})_{\overline{5}}$
\\
\null
$(1,2;-\tfrac{1}{2})_{1^-}$
& $(1,2;-\tfrac{1}{2})_{1^-}$
& $(1,2;-\tfrac{1}{2})_{\overline{5}}$
& $(1,2;-\tfrac{1}{2})_{\overline{5}}$
\\
\hline \hline
\end{tabular}
\mycaption{Decomposition of the fundamental 10-dimensional representation under $SU(3)_c \otimes SU(2)_L \otimes U(1)_Y$,
for standard $SO(10)$ and flipped $SO(10) \otimes U(1)_X$ ($SO(10)_f$) respectively.
In the first two columns {(LR)} the subscripts keep track of the
$SU(4)_C$ origin of the multiplets (the extra symbols $\pm$ correspond to the eigenvalues of the $T^{(3)}_{R}$ Cartan generator)
while in the last two columns the $SU(5)$ content is shown.}
\label{tab:10decomp}
\end{table}

\renewcommand{\arraystretch}{1.3}
\begin{table}[ht]
\begin{tabular}{llll}
\hline \hline
\multicolumn{2}{c}{LR} & \multicolumn{2}{c}{$SU(5)$}
\\
\hline
$SO(10)$
& $SO(10)_f$
& $SO(10)$
& $SO(10)_f$
\\
\hline
$(3,2;+\tfrac{1}{6})_{4}$
& $(3,2;+\tfrac{1}{6})_{4}$
& $(\overline{3},1;+\tfrac{1}{3})_{\overline{5}}$
& $(\overline{3},1;+\tfrac{1}{3})_{\overline{5}}$
\\
\null
$(1,2; -\tfrac{1}{2})_{4}$
& $(1,2; +\tfrac{1}{2})_{4}$
& $(1,2; -\tfrac{1}{2})_{\overline{5}}$
& $(1,2; +\tfrac{1}{2})_{\overline{5}}$
\\
\null
$(\overline{3},1;+\tfrac{1}{3})_{\overline{4}^+}$
& $(\overline{3},1;+\tfrac{1}{3})_{\overline{4}^+}$
& $(3,2;+\tfrac{1}{6})_{10}$
& $(3,2;+\tfrac{1}{6})_{10}$
\\
\null
$(\overline{3},1;-\tfrac{2}{3})_{\overline{4}^-}$
& $(\overline{3},1;+\tfrac{1}{3})_{\overline{4}^-}$
& $(\overline{3},1;-\tfrac{2}{3})_{10}$
& $(\overline{3},1;+\tfrac{1}{3})_{10}$
\\
\null
$(1,1;+1)_{\overline{4}^+}$
& $\mathbf{(1,1;0)_{\overline{4}^+}}$
& $(1,1;+1)_{10}$
& $\mathbf{(1,1;0)_{10}}$
\\
\null
$\mathbf{(1,1;0)_{\overline{4}^-}}$
& $\mathbf{(1,1;0)_{\overline{4}^-}}$
& $\mathbf{(1,1;0)_{1}}$
& $\mathbf{(1,1;0)_{1}}$
\\
\hline \hline
\end{tabular}
\mycaption{The same as in Table \ref{tab:10decomp} for the spinor $16$-dimensional representation.
The SM singlets are emphasized in boldface and shall be denoted, in the the $SU(5)$ description, as $e \equiv (1,1;0)_{10}$ and $\nu \equiv (1,1;0)_{1}$.
The LR decomposition shows that $e$ and $\nu$ belong to an $SU(2)_R$ doublet.}
\label{tab:16decomp}
\end{table}

\renewcommand{\arraystretch}{1.3}
\begin{table}[ht]
\begin{tabular}{llll}
\hline \hline
\multicolumn{2}{c}{LR} & \multicolumn{2}{c}{$SU(5)$}
\\
\hline
$SO(10)$
& $SO(10)_f$
& $SO(10)$
& $SO(10)_f$
\\
\hline
$\mathbf{(1,1;0)_{1^0}}$
& $\mathbf{(1,1;0)_{1^0}}$
& $\mathbf{(1,1;0)_{1}}$
& $\mathbf{(1,1;0)_{1}}$
\\
\null
$\mathbf{(1,1;0)_{15}}$
& $\mathbf{(1,1;0)_{15}}$
& $\mathbf{(1,1;0)_{24}}$
& $\mathbf{(1,1;0)_{24}}$
\\
\null
$(8,1;0)_{15}$
& $(8,1;0)_{15}$
& $(8,1;0)_{24}$
& $(8,1;0)_{24}$
\\
\null
$(3,1;+\tfrac{2}{3})_{15}$
& $(3,1;-\tfrac{1}{3})_{15}$
& $(3,2;-\tfrac{5}{6})_{24}$
& $(3,2;+\tfrac{1}{6})_{24}$
\\
\null
$(\overline{3},1;-\tfrac{2}{3})_{15}$
& $(\overline{3},1;+\tfrac{1}{3})_{15}$
& $(\overline{3},2;+\tfrac{5}{6})_{24}$
& $(\overline{3},2;-\tfrac{1}{6})_{24}$
\\
\null
$(1,3;0)_{1}$
& $(1,3;0)_{1}$
& $(1,3;0)_{24}$
& $(1,3;0)_{24}$
\\
\null
$(3,2;+\tfrac{1}{6})_{6^+}$
& $(3,2;+\tfrac{1}{6})_{6^+}$
& $(3,2;+\tfrac{1}{6})_{10}$
& $(3,2;+\tfrac{1}{6})_{10}$
\\
\null
$(\overline{3},2;+\tfrac{5}{6})_{6^+}$
& $(\overline{3},2;-\tfrac{1}{6})_{6^+}$
& $(\overline{3},1;-\tfrac{2}{3})_{10}$
& $(\overline{3},1;+\tfrac{1}{3})_{10}$
\\
\null
$(1,1;+1)_{1^+}$
& $\mathbf{(1,1;0)_{1^+}}$
& $(1,1;+1)_{10}$
& $\mathbf{(1,1;0)_{10}}$
\\
\null
$(\overline{3},2;-\tfrac{1}{6})_{6^-}$
& $(\overline{3},2;-\tfrac{1}{6})_{6^-}$
& $(\overline{3},2;-\tfrac{1}{6})_{\overline{10}}$
& $(\overline{3},2;-\tfrac{1}{6})_{\overline{10}}$
\\
\null
$(3,2;-\tfrac{5}{6})_{6^-}$
& $(3,2;+\tfrac{1}{6})_{6^-}$
& $(3,1;+\tfrac{2}{3})_{\overline{10}}$
& $(3,1;-\tfrac{1}{3})_{\overline{10}}$
\\
\null
$(1,1;-1)_{1^-}$
& $\mathbf{(1,1;0)_{1^-}}$
& $(1,1;-1)_{\overline{10}}$
& $\mathbf{(1,1;0)_{\overline{10}}}$
\\
\hline \hline
\end{tabular}
\mycaption{{The same} as in Table \ref{tab:10decomp} for the $45$ representation.
The SM singlets are given in boldface and labeled throughout the text as $\omega_Y \equiv (1,1;0)_{15}$,
$\omega^+ \equiv (1,1;0)_{1^+}$, $\omega_R \equiv (1,1;0)_{1^0}$ and $\omega^- \equiv (1,1;0)_{1^-}$ {where again the LR notation has been used}.
The LR decomposition {also} shows that $\omega^+$, $\omega_R$ and $\omega^-$ belong to an $SU(2)_R$ triplet, while $\omega_Y$ is a $B-L$ singlet.}
\label{tab:45decomp}
\end{table}

\subsubsection{The supersymmetric flipped $SO(10)$ model}

The presence of additional SM singlets (some of them transforming {non-trivially} under $SU(5)$) in the lowest-dimensional representations of the flipped realisation of the $SO(10)$ gauge symmetry provides the
ground for obtaining a viable symmetry breaking with a significantly simplified renormalizable Higgs sector. Naively, one may guess that the pair of VEVs in $16_{H}$ (plus another conjugated pair in $\overline{16}_{H}$ to maintain the required $D$-flatness)
might be enough to break the GUT symmetry entirely, since
one component transforms as a $10$ of $SU(5)\subset SO(10)$, while the other one is identified with the $SU(5)$ singlet (c.f. Table~\ref{tab:16decomp}). Notice that even in the presence of an additional four-dimensional vacuum manifold of the adjoint Higgs multiplet, the little group is determined by the $16_H$ VEVs since, due to the simple form of the renormalizable superpotential $F$-flatness {makes the VEVs of $45_{H}$ align} with those of ${16}_{H}\overline{16}_{H}$, providing just {enough freedom for them to develop non-zero values}.

{Unfortunately, this is still not enough to support the desired symmetry breaking pattern.} The two VEV directions in $16_{H}$ are equivalent to one and a residual $SU(5) \otimes U(1)$ symmetry is always preserved by $\vev{16}_{H}$~\cite{Buccella:1980qb}.
Thus, even in the flipped $SO(10)\otimes U(1)$ setting the Higgs model spanned on ${16}_{H}\oplus\overline{16}_{H}\oplus 45_{H}$ suffers from
an $SU(5)\otimes U(1)$ lock {analogous} to the one of the standard SUSY $SO(10)$ models with the same Higgs sector.
This can be understood by taking into account the freedom in choosing the basis in the $SO(10)$ algebra so that the pair of VEVs within $16$ can be ``rotated'' onto a single component, which can be then viewed as the direction of the singlet in the decomposition of $16=\overline{5}\oplus 10\oplus 1$ with respect to an $SU(5)$ subgroup of the {original} $SO(10)$ gauge symmetry.

On the other hand, with a pair of interacting $16_{H}\oplus \overline{16}_{H}$'s the vacuum directions in the two $16_{H}$'s need not be aligned and the intersection of the two different invariant subalgebras ({e.g.,} standard and flipped $SU(5)$ for a specific VEV configuration) leaves as a little group the the $SU(3)_{c}\otimes SU(2)_{L}\otimes U(1)_{Y}$ of the SM.
$F$-flatness makes then the adjoint VEVs ($45_H$ is the needed carrier
of $16_H$ interaction at the renormalizable level) aligned to the SM vacuum.
Hence, as we will show in the next section, $2\times (16_{H}+ \overline{16}_{H})\oplus 45_{H}$ defines the minimal renormalizable Higgs setting for the SUSY flipped $SO(10)\otimes U(1)_{X}$ model.
For comparison, let us reiterate that in the standard renormalizable $SO(10)$ setting the SUSY vacuum is always $SU(5)$ regardless of how many copies of $16_{H}\oplus \overline{16}_{H}$ are employed together with at most a pair of adjoints.

\subsubsection{The matter sector}

Due to the flipped hypercharge assignment, the SM matter can no longer be fully embedded into the 16-dimensional $SO(10)$ spinor, as in the standard case. By inspecting Table~\ref{tab:16decomp} one can see that in the flipped setting the pair of the SM sub-multiplets of $16$ transforming as $u^{c}$ {and} $e^{c}$ is traded for an extra $d^{c}$-like state and an extra SM singlet.
The former pair is {instead} found in the $SO(10)$ vector and the singlet {(the lepton doublet as well appears in the vector multiplet)}. Thus, flipping {spreads} each of the SM matter generations across $16\oplus 10\oplus 1$ of $SO(10)$, which, by construction, {can be viewed as} the {complete} 27-dimensional fundamental representation of {$E_{6}\supset SO(10)\otimes U(1)_{X}$}. This brings in a set of additional degrees of freedom, in particular $(1,1,0)_{{16}}$, $(\overline{3},1,+\tfrac{1}{3})_{{16}}$,   $(1,2,+\tfrac{1}{2})_{{16}}$,  $(3,1,-\tfrac{1}{3})_{{10}}$ and $(1,2,-\tfrac{1}{2})_{{10}}$,  where the subscript indicates their ${SO(10)}$ origin. Notice, {however}, that {these} SM ``exotics'' can be grouped into superheavy vector-like pairs and thus {no extra states} appear in the low energy spectrum. Furthermore, the $U(1)_{X}$ anomalies associated with each of the $SO(10)\otimes U(1)_{X}$ matter multiplets cancel when summed over the entire reducible representation $16_{1}\oplus 10_{-2}\oplus 1_{4}$. An {elementary} discussion of the matter spectrum in this scenario is deferred to \sect{TRflavor}.

\subsection{Supersymmetric vacuum}
\label{vacuumFSO10}

The most general renormalizable Higgs superpotential, made of the representations {$45 \oplus 16_1 \oplus \overline{16}_1 \oplus 16_2  \oplus \overline{16}_2$} is given by
\be
\label{WHFSO10}
W_H = \frac{\mu}{2} \, \Tr 45^2 + \rho_{ij} 16_{i} \overline{16}_{j} + \tau_{ij} 16_{i} 45  \overline{16}_{j} \, ,
\ee
where $i,j = 1,2$ and the notation is explained in \app{flippedSO10notation}.
Without loss of generality we can take $\mu$ real by a global phase redefinition, while $\tau$ (or $\rho$) can be diagonalized by a bi-unitary transformation acting on the flavor indices of the $16$ and the $\overline{16}$. Let us choose, for instance, $\tau_{ij} = \tau_i \delta_{ij}$, with $\tau_i$ real.
We label the SM-singlets contained in the $16$'s in the following way:
$e \equiv (1,1;0)_{10}$ (only for flipped $SO(10)$) and $\nu \equiv (1,1;0)_{1}$
(for {all} embeddings).

By plugging in the SM-singlet VEVs $\omega_R$, $\omega_Y$, $\omega^+$, $\omega^-$, $e_{1,2}$, $\overline{e}_{1,2}$, $\nu_{1,2}$ and $\overline{\nu}_{1,2}$
(c.f. \app{flippedSO10notation}), the superpotential on the vacuum reads
\bea
\label{vevWHFSO10}
\vev{W_H} &=& \mu \left(2 \omega _R^2+3 \omega _Y^2+4 \omega ^- \omega ^+\right) \nn \\
    &+& \rho _{11} \left(e_1 \overline{e}_1+\nu _1 \overline{\nu }_1\right)+\rho _{21} \left(e_2
   \overline{e}_1+\nu _2 \overline{\nu }_1\right) \nn \\
    &+& \rho _{12} \left(e_1 \overline{e}_2+\nu _1 \overline{\nu }_2\right)+\rho _{22} \left(e_2 \overline{e}_2+\nu _2 \overline{\nu }_2\right) \nn \\
   &+& \tau _1 \left[ - \omega ^- e_1 \overline{\nu }_1- \omega ^+ \nu _1 \overline{e}_1 - \frac{\omega _R}{\sqrt{2}} \left(e_1 \overline{e}_1-\nu _1 \overline{\nu }_1\right) \right. \nn \\
   &+& \left.  \frac{3}{2} \frac{\omega _Y}{\sqrt{2}} \left(e_1 \overline{e}_1
   +\nu _1 \overline{\nu }_1\right) \right] \nn \\
   &+& \tau _2 \left[- \omega ^- e_2 \overline{\nu }_2- \omega ^+ \nu _2  \overline{e}_2 - \frac{\omega _R}{\sqrt{2}} \left(e_2
   \overline{e}_2-\nu _2 \overline{\nu }_2\right) \right. \nn \\
   &+& \left. \frac{3}{2} \frac{\omega _Y}{\sqrt{2}} \left( e_2 \overline{e}_2+ \nu _2 \overline{\nu }_2\right)\right] \, .
\eea
{In order to retain SUSY down to the TeV scale}
we must require
that the GUT gauge symmetry breaking preserves supersymmetry.
In \app{DFtermFSO10} we work out the relevant $D$- and $F$-term equations.
We find that the existence of a nontrivial vacuum requires $\rho$ (and $\tau$ for consistency) to be hermitian matrices.
This is a consequence of the fact that $D$-term flatness for the flipped $SO(10)$ embedding implies
$\vev{16_i} = \vev{\overline{16}_i}^*$ ({see \eq{complexDtermsFSO10} and the discussion next to it}).
With this restriction the vacuum manifold is given by
{
\bea
\label{vacmanifoldFSO10}
8 \mu\, \omega^+ &=& \tau _1 r_1^2 \sin{2\alpha_1} e^{i (\phi_{e_1} - \phi_{\nu_1})} \nn \\
&+& \tau _2 r_2^2 \sin{2\alpha_2} e^{i (\phi_{e_2} - \phi_{\nu_2})} \, , \nn \\
8 \mu\, \omega^- &=& \tau _1  r_1^2 \sin{2\alpha_1} e^{-i (\phi_{e_1} - \phi_{\nu_1})} \nn \\
&+& \tau _2 r_2^2 \sin{2\alpha_2} e^{-i (\phi_{e_2} - \phi_{\nu_2})} \, , \nn \\
4 \sqrt{2} \mu\, \omega_R &=& \tau _1 r_1^2 \cos{2\alpha_1} + \tau _2 r_2^2 \cos{2\alpha_2} \, , \nn \\
4 \sqrt{2} \mu\, \omega_Y &=& -\tau _1 r_1^2 - \tau _2 r_2^2 \, , \nn \\
e_{1,2} &=& r_{1,2} \cos \alpha_{1,2} \ e^{i \phi_{e_{1,2}}} \, , \nn \\
\nu_{1,2} &=& r_{1,2} \sin \alpha_{1,2}\ e^{i \phi_{\nu_{1,2}}} \, , \nn \\
\overline{e}_{1,2} &=& r_{1,2} \cos \alpha_{1,2}\ e^{- i \phi_{e_{1,2}}} \, , \nn \\
\overline{\nu}_{1,2} &=& r_{1,2} \sin \alpha_{1,2}\ e^{- i \phi_{\nu_{1,2}}} \, ,
\eea
}
where $r_{1,2}$ and $\alpha^{\pm} \equiv \alpha_1 \pm \alpha_2$ are fixed in terms of the superpotential parameters,
\bea
\label{r1sqFSO10}
&& r_{1}^{2} = -\frac{2 \mu \left( \rho _{22} \tau _1-5 \rho _{11} \tau _2\right)}{3 \tau _1^2 \tau _2} \, , \\
\label{r2sqFSO10}
&& r_{2}^{2} = -\frac{2 \mu \left(\rho _{11} \tau _2-5 \rho _{22} \tau _1\right)}{3 \tau _1 \tau _2^2} \, , \\
\label{cosalpha1m2FSO10}
&& \cos{\alpha^{-}} = \xi \ \frac{\sin \Phi_\nu - \sin \Phi_e}{\sin \left(\Phi_\nu - \Phi_e \right)} \, , \\
\label{cosalpha1p2FSO10}
&& \cos{\alpha^{+}} = \xi \ \frac{\sin \Phi_\nu + \sin \Phi_e}{\sin \left(\Phi_\nu - \Phi_e \right)}  \, ,
\eea
with
\be
\label{xiFSO10}
\xi = \frac{6 |\rho _{12}|}{\sqrt{-\frac{5 \rho _{11}^2 \tau _2}{\tau _1}-\frac{5 \rho _{22}^2 \tau _1}{\tau _2}+26 \rho _{22} \rho _{11}}} \, .
\ee
The phase factors $\Phi_\nu$ and $\Phi_e$ are defined as
\be
\label{PhienuFSO10}
\Phi_\nu \equiv \phi_{\nu_1}-\phi_{\nu_2}+\phi_{\rho _{12}} \, , \quad \Phi_e \equiv \phi_{e_1}-\phi_{e_2}+\phi_{\rho _{12}} \, ,
\ee
in terms of the relevant phases $\phi_{\nu_{1,2}}$, $\phi_{e_{1,2}}$ and $\phi_{\rho_{12}}$.
\eqs{cosalpha1m2FSO10}{cosalpha1p2FSO10} imply that for $\Phi_{\nu} = \Phi_{e} = \Phi$, \eq{cosalpha1m2FSO10} reduces
to $\cos{\alpha^{-}} \rightarrow \xi \cos{\Phi}$ while $\alpha^{+}$ is undetermined (thus parametrizing an orbit of isomorphic vacua).

In order to determine the little group of the vacuum manifold
we explicitly compute the corresponding gauge boson spectrum {in \app{gaugespectrum}}.
We find that, for $\alpha^- \neq 0$ {and/or} $\Phi_\nu \neq \Phi_e$, the vacuum in
\eq{vacmanifoldFSO10} does preserve the SM algebra.

As already mentioned in the introduction this result is a consequence
of the misalignement of the spinor VEVs, that is made possible
at the renormalizable level by the interaction with the $45_H$.
If we choose to align the $16_1 \oplus \overline{16}_1$ and $16_2 \oplus \overline{16}_2$ VEVs ($\alpha^- = 0$ and $\Phi_{\nu} = \Phi_{e}$)
or equivalently, to decouple one of the Higgs spinors from the vacuum ($r_2 = 0$ for instance)
the little group is $SU(5)\otimes U(1)$.

This result can be easily understood by observing that in the case {with} just one pair of $16_{H} \oplus \overline{16}_{H}$
(or {with} two pairs of $16_{H} \oplus \overline{16}_{H}$ {aligned}) the two SM-singlet directions, $e_H$ and $\nu_H$, are connected by
an $SU(2)_R$ transformation. This freedom can be used to rotate one of the VEVs to zero, so that the little group is {standard or flipped $SU(5) \otimes U(1)$, depending on which of the two VEVs is zero}.

In this respect, the Higgs adjoint plays the role of a renormalizable agent that prevents the two pairs of spinor vacua from aligning with each other along the $SU(5) \otimes U(1)$ direction.
Actually, by decoupling the adjoint Higgs, $F$-flatness makes the (aligned) $16_i \oplus \overline{16}_i$ vacuum trivial, as one verifies by inspecting the $F$-terms in \eq{FtermsFSO10} of \app{DFtermFSO10} for $\vev{45_H}=0$ and $\det{\rho}\neq 0$.

The same result with just two pairs of $16_{H} \oplus \overline{16}_{H}$ Higgs multiplets is obtained by adding NR spinor interactions,
at the {cost of}
introducing a potentially critical GUT-scale threshold hierarchy.
In the flipped $SO(10)$ setup here proposed the GUT symmetry breaking is driven by the renormalizable part of the Higgs superpotential,
thus allowing naturally for a one-step matching with the minimal supersymmetric extension of the SM (MSSM).

Before addressing the possible embedding of the model in a
unified $E_6$ scenario, we comment {in brief} on the naturalness
of the doublet-triplet mass splitting in flipped embeddings.

\subsection{Doublet-Triplet splitting in flipped models}
\label{DTsplittingflipped}

Flipped embeddings offers a rather economical way to implement the Doublet-Triplet (DT) splitting through the so called Missing Partner (MP) mechanism \cite{Antoniadis:1987dx,Barr:2010tv}. In order to show the relevat features let us consider first {the}
flipped $SU(5) \otimes U(1)_Z$.

In order to implement the MP mechanism in {the} flipped $SU(5) \otimes U(1)_Z$ the Higgs superpotential is required to have the couplings
\be
\label{MPflippedSU5}
W_H \supset 10_{+1} 10_{+1} 5_{-2} + \overline{10}_{-1} \overline{10}_{-1} \overline{5}_{+2} \, ,
\ee
where the subscripts correspond to the $U(1)_Z$ quantum numbers,
but not {the} $5_{-2} \overline{5}_{+2}$ mass term.
From \eq{MPflippedSU5}
we extract the relevant terms that lead to a mass for the Higgs triplets
\bea
\label{MPflippedSU5vev}
W_H &\supset& \vev{(1,1;0)_{10}} (\overline{3},1;+\tfrac{1}{3})_{10} (3,1;-\tfrac{1}{3})_{5} \nn \\
&+& \vev{(1,1;0)_{\overline{10}}} (3,1;-\tfrac{1}{3})_{\overline{10}} (\overline{3},1;+\tfrac{1}{3})_{\overline{5}} \, .
\eea
On the other hand, the Higgs doublets, contained in the $5_{-2} \oplus \overline{5}_{+2}$ remain massless since they have no partner in the
$10_{+1} \oplus \overline{10}_{-1}$ to couple with.

The MP mechanism cannot be implemented in standard $SO(10)$.
The relevant interactions, analogue of \eq{MPflippedSU5}, are contained into the $SO(10)$ invariant term
\be
W_H \supset 16\, 16\, 10 + \overline{16}\, \overline{16}\, 10 \, ,
\ee
which, however, gives a mass to the doublets as well, via the superpotential terms
\bea
W_H &\supset& \vev{(1,1;0)_{1_{16}}} (1,2; -\tfrac{1}{2})_{\overline{5}_{16}} (1,2;+\tfrac{1}{2})_{5_{10}} \nn \\
&+& \vev{(1,1;0)_{1_{\overline{16}}}} (1,2; +\tfrac{1}{2})_{5_{\overline{16}}} (1,2;-\tfrac{1}{2})_{\overline{5}_{10}} \, .
\eea

Flipped $SO(10) \otimes U(1)_X$, on the other hand,  offers again the possibility of implementing the MP mechanism.
The prize to pay is the necessity of avoiding a large number of terms, both bilinear and trilinear, in the Higgs superpotential.
In particular, the analogue of \eq{MPflippedSU5} is given by the NR term \cite{Maekawa:2003wm}
\be
\label{MPflippedSO10}
W_H \supset \frac{1}{M_P} \overline{16}_1 16_2 16_2 \overline{16}_1 + \frac{1}{M_P} 16_1 \overline{16}_2 \overline{16}_2 16_1 \, .
\ee
By requiring that $16_1$ ($\overline{16_1}$) takes a VEV in the $1_{16}$ ($1_{\overline{16}}$) direction while $16_2$ ($\overline{16_2}$) in the $10_{16}$ ($\overline{10}_{\overline{16}}$) component, one gets
\bea
W_H &\supset& \frac{1}{M_P} \vev{1_{\overline{16}_1}} \vev{{10}_{16_2}} 10_{16_2} 5_{\overline{16}_1} \nn \\
&+& \frac{1}{M_P} \vev{1_{16_1}} \vev{{\overline{10}}_{\overline{16}_2}} \overline{10}_{\overline{16}_2} \overline{5}_{16_1} \, ,
\eea
which closely resembles \eq{MPflippedSU5}, leading to massive triplets and massless doublets.
In order to have minimally one pair of electroweak doublets, one must further require that the $2 \times 2$
mass matrix of the $16$'s has rank equal to one.
Due to the active role of NR operators,
the Higgs triplets turn out to be two orders of magnitude below the flipped $SO(10) \otimes U(1)_X$ scale,
reintroducing the issues discussed as in \sect{littlehierarchy}.

An alternative possibility for naturally implementing the DT splitting in $SO(10)$ is the Dimopoulos-Wilczek (DW) (or the missing VEV) mechanism \cite{Dimopoulos:1981xm}.
In order to explain the key features it is convenient to decompose the relevant $SO(10)$ representations in terms of the $SU(4)_C \otimes SU(2)_L \otimes SU(2)_R$ group
\bea
\label{SO10decomp}
&& 45 \equiv (1,1,3) \oplus (15,1,1) \oplus \ldots \nn \\
&& 16 \equiv (4,2,1) \oplus (\overline{4},1,2) \, , \nn \\
&& \overline{16} \equiv (\overline{4},2,1) \oplus (4,1,2) \, , \nn \\
&& 10 \equiv (6,1,1) \oplus (1,2,2) \, ,
\eea
where $\omega_R \equiv \vev{(1,1,3)}$ and $\omega_Y \equiv \vev{(15,1,1)}$.
In the standard $SO(10)$ case (see \cite{Babu:1994dq,Barr:1997hq} and \cite{Babu:2010ej} for a recent discussion)
one assumes that the $SU(2)_L$ doublets are contained in two vector
multiplets ($10_1$ and $10_2$).
From the decompositions in \eq{SO10decomp} it's easy to see that the interaction $10_1 45\, 10_2$
(where the antisymmetry of $45$ requires the presence of two $10$'s)
leaves the $SU(2)_L$ doublets massless provided that $\omega_R = 0$.
For the naturalness of the setting
other superpotential terms must not appear, as a direct mass term for one of the $10$'s
and the interaction term $16\, 45 \, \overline{16}$. The latter aligns the SUSY vacuum in the $SU(5)$ direction ($\omega_R = \omega_Y$),
thus destabilizing the DW solution.

On the other hand, the absence of the $16\, 45 \, \overline{16}$
interaction enlarges the global symmetries of the
scalar potential with the consequent appearance of a set of light pseudo-Goldstone bosons in the
spectrum.
To avoid that the adjoint and the spinor sector must be coupled in an indirect way by adding extra fields and symmetries (see for instance \cite{Babu:1994dq,Barr:1997hq,Babu:2010ej}).

Our flipped $SO(10) \otimes U(1)_X$ setting offers the rather economical possibility of embedding the electroweak doublets directly into the spinors without the need of $10_H$
(see \sect{TRflavor}).
As a matter of fact, there exists a variant of the DW mechanism
where the $SU(2)_L$ doublets, contained in the $16{_H} \oplus \overline{16}{_H}$, are kept massless
by the condition $\omega_Y = 0$ (see e.g. \cite{Dvali:1996wh}).
However, in order to satisfy in a natural way the $F$-flatness for the configuration $\omega_Y = 0$,
again a contrived superpotential is required, when compared to that in \eq{WHFSO10}.
In conclusion, we cannot implement in our simple setup any of the natural mechanisms so far proposed {(see also \cite{Maekawa:2003ka})}
and we have to resort to the standard minimal fine-tuning.

\section{Minimal $E_6$ embedding}
\label{minimalE6Higgs}

The natural and minimal unified embedding of the flipped $SO(10)\otimes U(1)$ model is $E_6$ with one $78_H$ and two pairs of $27_H \oplus \overline{27}_H$
in the Higgs sector. The three matter families are contained in three $27_F$ chiral superfields.
The decomposition of the $27$ and $78$ representations under the SM quantum numbers is detailed
in Tables \ref{tab:27decompSU(5)}, \ref{tab:27decompPS}, \ref{tab:78decompSU(5)} and \ref{tab:78decompPS}, according to
the different hypercharge embeddings.

\renewcommand{\arraystretch}{1.3}
\begin{table}[ht]
\begin{tabular}{lll}
\hline \hline
$SU(5)$ \ \
& $SU(5)_f$\ \
& $SO(10)_f$
\\
\hline
$(\overline{3},1;+\tfrac{1}{3})_{\overline{5}_{16}}$
& $(\overline{3},1;-\tfrac{2}{3})_{\overline{5}_{16}}$
& $(\overline{3},1;+\tfrac{1}{3})_{\overline{5}_{16}}$
\\
\null
$(1,2; -\tfrac{1}{2})_{\overline{5}_{16}}$
& $(1,2; -\tfrac{1}{2})_{\overline{5}_{16}}$
& $(1,2; +\tfrac{1}{2})_{\overline{5}_{16}}$
\\
\null
$(3,2;+\tfrac{1}{6})_{10_{16}}$
& $(3,2;+\tfrac{1}{6})_{10_{16}}$
& $(3,2;+\tfrac{1}{6})_{10_{16}}$
\\
\null
$(\overline{3},1;-\tfrac{2}{3})_{10_{16}}$
& $(\overline{3},1;+\tfrac{1}{3})_{10_{16}}$
& $(\overline{3},1;+\tfrac{1}{3})_{10_{16}}$
\\
\null
$(1,1;+1)_{10_{16}}$
& $(1,1;0)_{10_{16}}$
& $(1,1;0)_{10_{16}}$
\\
\null
$(1,1;0)_{1_{16}}$
& $(1,1;+1)_{1_{16}}$
& $(1,1;0)_{1_{16}}$
\\
\hline
$(3,1;-\tfrac{1}{3})_{5_{10}}$
& $(3,1;-\tfrac{1}{3})_{5_{10}}$
& $(3,1;-\tfrac{1}{3})_{5_{10}}$
\\
\null
$(1,2;+\tfrac{1}{2})_{5_{10}}$
& $(1,2;-\tfrac{1}{2})_{5_{10}}$
& $(1,2;-\tfrac{1}{2})_{5_{10}}$
\\
\null
$(\overline{3},1;+\tfrac{1}{3})_{\overline{5}_{10}}$
& $(\overline{3},1;+\tfrac{1}{3})_{\overline{5}_{10}}$
& $(\overline{3},1;-\tfrac{2}{3})_{\overline{5}_{10}}$
\\
\null
$(1,2;-\tfrac{1}{2})_{\overline{5}_{10}}$
& $(1,2;+\tfrac{1}{2})_{\overline{5}_{10}}$
& $(1,2;-\tfrac{1}{2})_{\overline{5}_{10}}$
\\
\hline
$(1,1;0)_{1_{1}}$
& $(1,1;0)_{1_{1}}$
& $(1,1;+1)_{1_{1}}$
\\
\hline \hline
\end{tabular}
\mycaption{Decomposition of the fundamental representation $27$ of $E_6$ under $SU(3)_c \otimes SU(2)_L \otimes U(1)_Y$, according to the three SM-compatible different embeddings of the hypercharge {($f$ stands for flipped)}. The numerical subscripts keep track of the $SU(5)$ and $SO(10)$ origin.}
\label{tab:27decompSU(5)}
\end{table}

\renewcommand{\arraystretch}{1.3}
\begin{table}[ht]
\begin{tabular}{lll}
\hline \hline
$SU(5)$ \ \
& $SU(5)_f$\ \
& $SO(10)_f$
\\
\hline
$(3,2;+\tfrac{1}{6})_{4_{16}}$
& $(3,2;+\tfrac{1}{6})_{4_{16}}$
& $(3,2;+\tfrac{1}{6})_{4_{16}}$
\\
\null
$(1,2;-\tfrac{1}{2})_{4_{16}}$
& $(1,2;-\tfrac{1}{2})_{4_{16}}$
& $(1,2;+\tfrac{1}{2})_{4_{16}}$
\\
\null
$(\overline{3},1;+\tfrac{1}{3})_{\overline{4}^{+}_{16}}$
& $(\overline{3},1;-\tfrac{2}{3})_{\overline{4}^{+}_{16}}$
& $(\overline{3},1;+\tfrac{1}{3})_{\overline{4}^{+}_{16}}$
\\
\null
$(\overline{3},1;-\tfrac{2}{3})_{\overline{4}^{-}_{16}}$
& $(\overline{3},1;+\tfrac{1}{3})_{\overline{4}^{-}_{16}}$
& $(\overline{3},1;+\tfrac{1}{3})_{\overline{4}^{-}_{16}}$
\\
\null
$(1,1;+1)_{\overline{4}^{+}_{16}}$
& $(1,1;0)_{\overline{4}^{+}_{16}}$
& $(1,1;0)_{\overline{4}^{+}_{16}}$
\\
\null
$(1,1;0)_{\overline{4}^{-}_{16}}$
& $(1,1;+1)_{\overline{4}^{-}_{16}}$
& $(1,1;0)_{\overline{4}^{-}_{16}}$
\\
\hline
$(3,1;-\tfrac{1}{3})_{6_{10}}$
& $(3,1;-\tfrac{1}{3})_{6_{10}}$
& $(3,1;-\tfrac{1}{3})_{6_{10}}$
\\
\null
$(\overline{3},1;+\tfrac{1}{3})_{6_{10}}$
& $(\overline{3},1;+\tfrac{1}{3})_{6_{10}}$
& $(\overline{3},1;-\tfrac{2}{3})_{6_{10}}$
\\
\null
$(1,2;+\tfrac{1}{2})_{1_{10}^{+}}$
& $(1,2;-\tfrac{1}{2})_{1_{10}^{+}}$
& $(1,2;-\tfrac{1}{2})_{1_{10}^{+}}$
\\
\null
$(1,2;-\tfrac{1}{2})_{1_{10}^{-}}$
& $(1,2;+\tfrac{1}{2})_{1_{10}^{-}}$
& $(1,2;-\tfrac{1}{2})_{1_{10}^{-}}$
\\
\hline
$(1,1;0)_{1_{1}}$
& $(1,1;0)_{1_{1}}$
& $(1,1;+1)_{1_{1}}$
\\
\hline \hline
\end{tabular}
\mycaption{{The same} as in {Table} \ref{tab:27decompSU(5)},
where the subscripts keep track of the $SU(4)_C$
and $SO(10)$ origin.
The symbols $\pm$ refer to the eigenvalues of $T^{(3)}_{R}$. }
\label{tab:27decompPS}
\end{table}

\renewcommand{\arraystretch}{1.3}
\begin{table}[ht]
\begin{tabular}{lll}
\hline \hline
$SU(5)$ \ \
& $SU(5)_f$\ \
& $SO(10)_f$
\\
\hline
$(1,1;0)_{1_{1}}$
& $(1,1;0)_{1_{1}}$
& $(1,1;0)_{1_{1}}$
\\
\hline
$(1,1;0)_{1_{45}}$
& $(1,1;0)_{1_{45}}$
& $(1,1;0)_{1_{45}}$
\\
\null
$(8,1;0)_{24_{45}}$
& $(8,1;0)_{24_{45}}$
& $(8,1;0)_{24_{45}}$
\\
\null
$(3,2;-\tfrac{5}{6})_{24_{45}}$
& $(3,2;+\tfrac{1}{6})_{24_{45}}$
& $(3,2;+\tfrac{1}{6})_{24_{45}}$
\\
\null
$(\overline{3},2;+\tfrac{5}{6})_{24_{45}}$
& $(\overline{3},2;-\tfrac{1}{6})_{24_{45}}$
& $(\overline{3},2;-\tfrac{1}{6})_{24_{45}}$
\\
\null
$(1,3;0)_{24_{45}}$
& $(1,3;0)_{24_{45}}$
& $(1,3;0)_{24_{45}}$
\\
\null
$(1,1;0)_{24_{45}}$
& $(1,1;0)_{24_{45}}$
& $(1,1;0)_{24_{45}}$
\\
\null
$(3,2;+\tfrac{1}{6})_{10_{45}}$
& $(3,2;-\tfrac{5}{6})_{10_{45}}$
& $(3,2;+\tfrac{1}{6})_{10_{45}}$
\\
\null
$(\overline{3},1;-\tfrac{2}{3})_{10_{45}}$
& $(\overline{3},1;-\tfrac{2}{3})_{10_{45}}$
& $(\overline{3},1;+\tfrac{1}{3})_{10_{45}}$
\\
\null
$(1,1;+1)_{10_{45}}$
& $(1,1;-1)_{10_{45}}$
& $(1,1;0)_{10_{45}}$
\\
\null
$(\overline{3},2;-\tfrac{1}{6})_{\overline{10}_{45}}$
& $(\overline{3},2;+\tfrac{5}{6})_{\overline{10}_{45}}$
& $(\overline{3},2;-\tfrac{1}{6})_{\overline{10}_{45}}$
\\
\null
$(3,1;+\tfrac{2}{3})_{\overline{10}_{45}}$
& $(3,1;+\tfrac{2}{3})_{\overline{10}_{45}}$
& $(3,1;-\tfrac{1}{3})_{\overline{10}_{45}}$
\\
\null
$(1,1;-1)_{\overline{10}_{45}}$
& $(1,1;+1)_{\overline{10}_{45}}$
& $(1,1;0)_{\overline{10}_{45}}$
\\
\hline
$(\overline{3},1;+\tfrac{1}{3})_{\overline{5}_{16}}$
& $(\overline{3},1;-\tfrac{2}{3})_{\overline{5}_{16}}$
& $(\overline{3},1;-\tfrac{2}{3})_{\overline{5}_{16}}$
\\
\null
$(1,2; -\tfrac{1}{2})_{\overline{5}_{16}}$
& $(1,2; -\tfrac{1}{2})_{\overline{5}_{16}}$
& $(1,2; -\tfrac{1}{2})_{\overline{5}_{16}}$
\\
\null
$(3,2;+\tfrac{1}{6})_{10_{16}}$
& $(3,2;+\tfrac{1}{6})_{10_{16}}$
& $(3,2;-\tfrac{5}{6})_{10_{16}}$
\\
\null
$(\overline{3},1;-\tfrac{2}{3})_{10_{16}}$
& $(\overline{3},1;+\tfrac{1}{3})_{10_{16}}$
& $(\overline{3},1;-\tfrac{2}{3})_{10_{16}}$
\\
\null
$(1,1;+1)_{10_{16}}$
& $(1,1;0)_{10_{16}}$
& $(1,1;-1)_{10_{16}}$
\\
\null
$(1,1;0)_{1_{16}}$
& $(1,1;+1)_{1_{16}}$
& $(1,1;-1)_{1_{16}}$
\\
\hline
$(3,1;-\tfrac{1}{3})_{5_{\overline{16}}}$
& $(3,1;+\tfrac{2}{3})_{5_{\overline{16}}}$
& $(3,1;+\tfrac{2}{3})_{5_{\overline{16}}}$
\\
\null
$(1,2; +\tfrac{1}{2})_{5_{\overline{16}}}$
& $(1,2; +\tfrac{1}{2})_{5_{\overline{16}}}$
& $(1,2; +\tfrac{1}{2})_{5_{\overline{16}}}$
\\
\null
$(\overline{3},2;-\tfrac{1}{6})_{\overline{10}_{\overline{16}}}$
& $(\overline{3},2;-\tfrac{1}{6})_{\overline{10}_{\overline{16}}}$
& $(\overline{3},2;+\tfrac{5}{6})_{\overline{10}_{\overline{16}}}$
\\
\null
$(3,1;+\tfrac{2}{3})_{\overline{10}_{\overline{16}}}$
& $(3,1;-\tfrac{1}{3})_{\overline{10}_{\overline{16}}}$
& $(3,1;+\tfrac{2}{3})_{\overline{10}_{\overline{16}}}$
\\
\null
$(1,1;-1)_{\overline{10}_{\overline{16}}}$
& $(1,1;0)_{\overline{10}_{\overline{16}}}$
& $(1,1;+1)_{\overline{10}_{\overline{16}}}$
\\
\null
$(1,1;0)_{1_{\overline{16}}}$
& $(1,1;-1)_{1_{\overline{16}}}$
& $(1,1;+1)_{1_{\overline{16}}}$
\\
\hline \hline
\end{tabular}
\mycaption{{The same} as in Table \ref{tab:27decompSU(5)} for the $78$ representation.}
\label{tab:78decompSU(5)}
\end{table}

\renewcommand{\arraystretch}{1.3}
\begin{table}[ht]
\begin{tabular}{lll}
\hline \hline
$SU(5)$ \ \
& $SU(5)_f$\ \
& $SO(10)_f$
\\
\hline
$(1,1;0)_{1_{1}}$
& $(1,1;0)_{1_{1}}$
& $(1,1;0)_{1_{1}}$
\\
\hline
$(1,1;0)_{15_{45}}$
& $(1,1;0)_{15_{45}}$
& $(1,1;0)_{15_{45}}$
\\
\null
$(8,1;0)_{15_{45}}$
& $(8,1;0)_{15_{45}}$
& $(8,1;0)_{15_{45}}$
\\
\null
$(3,1;+\tfrac{2}{3})_{15_{45}}$
& $(3,1;+\tfrac{2}{3})_{15_{45}}$
& $(3,1;-\tfrac{1}{3})_{15_{45}}$
\\
\null
$(\overline{3},1;-\tfrac{2}{3})_{15_{45}}$
& $(\overline{3},1;-\tfrac{2}{3})_{15_{45}}$
& $(\overline{3},1;+\tfrac{1}{3})_{15_{45}}$
\\
\null
$(1,3;0)_{1_{45}}$
& $(1,3;0)_{1_{45}}$
& $(1,3;0)_{1_{45}}$
\\
\null
$(1,1;+1)_{1_{45}^{+}}$
& $(1,1;-1)_{1_{45}^{+}}$
& $(1,1;0)_{1_{45}^{+}}$
\\
\null
$(1,1;0)_{1_{45}^{0}}$
& $(1,1;0)_{1_{45}^{0}}$
& $(1,1;0)_{1_{45}^{0}}$
\\
\null
$(1,1;-1)_{1_{45}^{-}}$
& $(1,1;+1)_{1_{45}^{-}}$
& $(1,1;0)_{1_{45}^{-}}$
\\
\null
$(3,2;+\tfrac{1}{6})_{6_{45}^{+}}$
& $(3,2;-\tfrac{5}{6})_{6_{45}^{+}}$
& $(3,2;+\tfrac{1}{6})_{6_{45}^{+}}$
\\
\null
$(3,2;-\tfrac{5}{6})_{6_{45}^{-}}$
& $(3,2;+\tfrac{1}{6})_{6_{45}^{-}}$
& $(3,2;+\tfrac{1}{6})_{6_{45}^{-}}$
\\
\null
$(\overline{3},2;+\tfrac{5}{6})_{6_{45}^{+}}$
& $(\overline{3},2;-\tfrac{1}{6})_{6_{45}^{+}}$
& $(\overline{3},2;-\tfrac{1}{6})_{6_{45}^{+}}$
\\
\null
$(\overline{3},2;-\tfrac{1}{6})_{6_{45}^{-}}$
& $(\overline{3},2;+\tfrac{5}{6})_{6_{45}^{-}}$
& $(\overline{3},2;-\tfrac{1}{6})_{6_{45}^{-}}$
\\
\hline
$(3,2;+\tfrac{1}{6})_{4_{16}}$
& $(3,2;+\tfrac{1}{6})_{4_{16}}$
& $(3,2;-\tfrac{5}{6})_{4_{16}}$
\\
\null
$(1,2;-\tfrac{1}{2})_{4_{16}}$
& $(1,2;-\tfrac{1}{2})_{4_{16}}$
& $(1,2;-\tfrac{1}{2})_{4_{16}}$
\\
\null
$(\overline{3},1;+\tfrac{1}{3})_{\overline{4}_{16}^{+}}$
& $(\overline{3},1;-\tfrac{2}{3})_{\overline{4}_{16}^{+}}$
& $(\overline{3},1;-\tfrac{2}{3})_{\overline{4}_{16}^{+}}$
\\
\null
$(\overline{3},1;-\tfrac{2}{3})_{\overline{4}_{16}^{-}}$
& $(\overline{3},1;+\tfrac{1}{3})_{\overline{4}_{16}^{-}}$
& $(\overline{3},1;-\tfrac{2}{3})_{\overline{4}_{16}^{-}}$
\\
\null
$(1,1;+1)_{\overline{4}_{16}^{+}}$
& $(1,1;0)_{\overline{4}_{16}^{+}}$
& $(1,1;-1)_{\overline{4}_{16}^{+}}$
\\
\null
$(1,1;0)_{\overline{4}_{16}^{-}}$
& $(1,1;+1)_{\overline{4}_{16}^{-}}$
& $(1,1;-1)_{\overline{4}_{16}^{-}}$
\\
\hline
$(\overline{3},2;-\tfrac{1}{6})_{\overline{4}_{\overline{16}}}$
& $(\overline{3},2;-\tfrac{1}{6})_{\overline{4}_{\overline{16}}}$
& $(\overline{3},2;+\tfrac{5}{6})_{\overline{4}_{\overline{16}}}$
\\
\null
$(1,2;+\tfrac{1}{2})_{\overline{4}_{\overline{16}}}$
& $(1,2;+\tfrac{1}{2})_{\overline{4}_{\overline{16}}}$
& $(1,2;+\tfrac{1}{2})_{\overline{4}_{\overline{16}}}$
\\
\null
$(3,1;-\tfrac{1}{3})_{4_{\overline{16}}^{-}}$
& $(3,1;+\tfrac{2}{3})_{4_{\overline{16}}^{-}}$
& $(3,1;+\tfrac{2}{3})_{4_{\overline{16}}^{-}}$
\\
\null
$(3,1;+\tfrac{2}{3})_{4_{\overline{16}}^{+}}$
& $(3,1;-\tfrac{1}{3})_{4_{\overline{16}}^{+}}$
& $(3,1;+\tfrac{2}{3})_{4_{\overline{16}}^{+}}$
\\
\null
$(1,1;-1)_{4_{\overline{16}}^{-}}$
& $(1,1;0)_{4_{\overline{16}}^{-}}$
& $(1,1;+1)_{4_{\overline{16}}^{-}}$
\\
\null
$(1,1;0)_{4_{\overline{16}}^{+}}$
& $(1,1;-1)_{4_{\overline{16}}^{+}}$
& $(1,1;+1)_{4_{\overline{16}}^{+}}$
\\
\hline \hline
\end{tabular}
\mycaption{{The same} as in Table \ref{tab:27decompPS} for the $78$ representation.}
\label{tab:78decompPS}
\end{table}

In analogy with the flipped $SO(10)$ discussion, we shall label the SM-singlets contained in the $27$ as
$e \equiv \left(1,1;0\right)_{1_{1}}$ and $\nu \equiv \left(1,1;0\right)_{1_{16}}$.

As we are going to show, the little group of a supersymmetric
$\vev{78\oplus 27_1\oplus 27_2\oplus \overline{27}_1\oplus \overline{27}_2}$ vacuum is $SU(5)$ {in the renormalizable case}.
This is just a consequence of the larger $E_6$ algebra.
In order to obtain a SM vacuum, we need to resort to a NR scenario
that allows for a disentanglement of the $\vev{78_H}$ directions, and, consistently, for a flipped $SO(10)\otimes U(1)$ intermediate stage.
We shall make the case for an $E_6$ gauge symmetry broken near the
Planck scale, leaving an effective flipped $SO(10)$ scenario down
to the $10^{16}$ GeV.

\subsection{$Y$ and $B-L$ into $E_6$}
\label{YandBLintoE6}

Interpreting the different possible definitions of the SM hypercharge in terms of the $E_6$ maximal subalgebra $SU(3)_c \otimes SU(3)_L \otimes SU(3)_R$, one finds that the three assignments in \eqs{standardabc}{flippedSO10abc}
are each orthogonal to the three possible ways of embedding $SU(2)_I$ (with $I=R,R',E$) into $SU(3)_R$ \cite{Harada:2003sb}.
Working in the Gell-Mann basis (c.f. \app{app:SU33formalism}) the $SU(3)_R$ Cartan generators read
\bea
&& T^{(3)}_R = \tfrac{1}{2} \left( T^{1'}_{1'} - T^{2'}_{2'} \right) \, \label{T3R}
, \\
&& T^{(8)}_R = \tfrac{1}{2\sqrt{3}} \left( T^{1'}_{1'} + T^{2'}_{2'} - 2 T^{3'}_{3'} \right) \, ,
\label{T8R}
\eea
{which} defines the $SU(2)_R$ embedding.
The $SU(2)_{R'}$ and $SU(2)_E$ embeddings are obtained from \eqs{T3R}{T8R} by flipping
respectively $2' \leftrightarrow 3'$ and $3' \leftrightarrow 1'$.
Considering the standard and flipped $SO(10)$ embeddings of the hypercharge in \eq{standardabc}
and \eq{flippedSO10abc}, in the $SU(3)^3$ notation they are respectively given by
\bea
\label{Ystandard}
Y &=& \tfrac{1}{\sqrt{3}} T^{(8)}_L + T^{(3)}_R + \tfrac{1}{\sqrt{3}} T^{(8)}_R \, , \nn \\
&=& \tfrac{1}{\sqrt{3}} T^{(8)}_L - \tfrac{2}{\sqrt{3}} T^{(8)}_E \, ,
\eea
and
\bea
\label{Yso10}
Y &=& \tfrac{1}{\sqrt{3}} T^{(8)}_L - \tfrac{2}{\sqrt{3}} T^{(8)}_R \, , \nn \\
&=& \tfrac{1}{\sqrt{3}} T^{(8)}_L + T^{(3)}_E + \tfrac{1}{\sqrt{3}} T^{(8)}_E \, .
\eea
Analogously, the three SM-compatible assignments of
$B-L$ in \eqs{BLst}{BLtw2} are as well orthogonal to the three possible
ways of embedding $SU(2)_I$ into $SU(3)_R$.
However, once we fix the embedding of the hypercharge we have
only two consistent choices for $B-L$ available. They correspond to the pairs where $Y$ and $B-L$
are not orthogonal to the same $SU(2)_I$ \cite{Harada:2003sb}.

For the standard hypercharge embedding, the $B-L$ assignment in \eq{BLst} reads
\bea
\label{BLstSU3}
B-L &=& \tfrac{2}{\sqrt{3}} \left( T^{(8)}_L + T^{(8)}_R \right) \, , \nn \\
&=& \tfrac{2}{\sqrt{3}} T^{(8)}_L - T^{(3)}_E - \tfrac{1}{\sqrt{3}} T^{(8)}_E \, ,
\eea
while the $B-L$ assignment in \eq{BLtw2}, consistent
with the flipped $SO(10)$ embedding of the hypercharge, reads
\bea
\label{BLtw2SU3}
B-L &=& \tfrac{2}{\sqrt{3}} T^{(8)}_L - T^{(3)}_R - \tfrac{1}{\sqrt{3}} T^{(8)}_R \, , \nn \\
&=& \tfrac{2}{\sqrt{3}} \left( T^{(8)}_L + T^{(8)}_E \right) \, .
\eea

\subsection{The $E_6$ vacuum manifold}
\label{vacuumE6}

The most general renormalizable Higgs superpotential, made of the representations $78 \oplus 27_1 \oplus 27_2 \oplus \overline{27}_1 \oplus \overline{27}_2$, is given by
\bea
\label{WHE6}
W_H &=& \frac{\mu}{2} \Tr 78^2 + \rho_{ij} 27_i \overline{27}_j + \tau_{ij} 27_i 78 \overline{27}_j \nn \\
&+& \alpha_{ijk} 27_i 27_j 27_k + \beta_{ijk} \overline{27}_i \overline{27}_j \overline{27}_k \, ,
\eea
where $i,j = 1,2$. The couplings $\alpha_{ijk}$ and $\beta_{ijk}$ are totally symmetric in $ijk$, so that each one of them contains four complex parameters.
Without loss of generality we can take $\mu$ real by a phase redefinition of the superpotential,
while $\tau$ can be diagonalized by a bi-unitary transformation acting on the indices of the $27$ and the $\overline{27}$.
We take, $\tau_{ij} = \tau_i \delta_{ij}$, with $\tau_i$ real.
Notice that $\alpha$ and $\beta$ are not relevant for the present study, since the corresponding invariants vanish on the SM orbit.

In the standard hypercharge embedding of \eq{Ystandard}, the SM-preserving vacuum directions are parametrized by
\bea
\label{vev78}
\vev{78}&=& a_1 T^{3'}_{2'} + a_2 T^{2'}_{3'} + \frac{a_3}{\sqrt{6}} (T^{1'}_{1'} + T^{2'}_{2'} - 2T^{3'}_{3'}) \nn \\
&+& \frac{a_4}{\sqrt{2}} (T^{1'}_{1'} - T^{2'}_{2'}) + \frac{b_3}{\sqrt{6}} (T^{1}_{1} + T^{2}_{2} - 2T^{3}_{3}) \, ,
\eea
and
\bea
\label{vev2712}
\vev{27_i} &=&  (e_i) v^{3}_{3'} + (\nu_i) v^{3}_{2'} \, , \\
\label{vev27bar12}
\vev{\overline{27}_i} &=& (\overline{e}_i) u^{3'}_{3} + (\overline{\nu}_i) u^{2'}_{3} \, .
\eea
where $a_1$, $a_2$, $a_3$, $a_4$, $b_3$, $e_{1,2}$, $\overline{e}_{1,2}$, $\nu_{1,2}$ and $\overline{\nu}_{1,2}$
are 13 SM-singlet VEVs (see \app{app:SU33formalism} for notation).
Given the $B-L$ expression in \eq{BLstSU3} and the fact that we can rewrite the Cartan part of $\vev{78}$ as
\begin{multline}
\sqrt{2} a_4 T^{(3)}_R + \tfrac{1}{\sqrt{2}} (a_3 + b_3) \left( T^{(8)}_R + T^{(8)}_L \right) \\
+ \tfrac{1}{\sqrt{2}} (a_3 - b_3) \left( T^{(8)}_R - T^{(8)}_L \right) \, ,
\end{multline}
we readily identify the standard $SO(10)$ VEVs used in the previous section with the present $E_6$ notation as
$\omega_R \propto a_4$, $\omega_Y \propto a_3 + b_3$, while
$\Omega \propto a_3 - b_3$ is the $SO(10) \otimes U(1)_X$ singlet VEV in $E_6$ ($T_X\propto T^{(8)}_R - T^{(8)}_L$).

We can also write the vacuum manifold in such a way that it is manifestly invariant under the flipped $SO(10)$
hypercharge in \eq{Yso10}. This can be obtained by flipping $1' \leftrightarrow 3'$ in \eqs{vev78}{vev27bar12}, yielding
\bea
\label{vev78flip}
&& \vev{78} = a_1 T^{1'}_{2'} + a_2 T^{2'}_{1'} + \sqrt{2} a'_4 T^{(3)}_E \nn \\
&&\hspace{2em} + \tfrac{1}{\sqrt{2}} (a'_3 + b_3) \left( T^{(8)}_E + T^{(8)}_L \right) \nn \\
&&\hspace{2em} + \tfrac{1}{\sqrt{2}} (a'_3 - b_3) \left( T^{(8)}_E - T^{(8)}_L \right) \, , \\
\label{vev2712flip}
&& \vev{27_i} =  (e_i) v^{3}_{1'} + (\nu_i) v^{3}_{2'} \, , \\
\label{vev27bar12flip}
&& \vev{\overline{27}_i} = (\overline{e}_i) u^{1'}_{3} + (\overline{\nu}_i) u^{2'}_{3} \, ,
\eea
where we recognize the $B-L$ generator defined in \eq{BLtw2SU3}.
Notice that the Cartan subalgebra is actually invariant both under the standard and the flipped $SO(10)$ form of $Y$.
We have
\be
a'_3 T^{(8)}_E + a'_4 T^{(3)}_E
= a_3 T^{(8)}_R + a_4 T^{(3)}_R \, ,
\label{RvsEcartans}
\ee
with
\bea
2 a'_3 &=& - a_3 - \sqrt{3} a_4 \, , \\
2 a'_4 &=& - \sqrt{3} a_3 + a_4 \,
\label{a34prime}
\eea
thus making the use of $a_{3,4}$ or $a_{3,4}'$ directions in the flipped or standard vacuum manifold completely equivalent.
We can now complete the identification of the notation used for $E_6$ with that of the flipped $SO(10) \otimes U(1)_X$ model studied in \sect{sect:minimalflippedSO10}, by $\omega^{\pm}\propto a_{1,2}$.

From the $E_6$ stand point, the analyses of the standard and flipped vacuum manifolds given, respectively,
in \eqs{vev78}{vev27bar12} and \eqs{vev78flip}{vev27bar12flip}, lead,
as expected, to the same results with the roles of standard and flipped hypercharge interchanged (see \app{app:E6vacuum}). In order to determine the vacuum little group we may therefore proceed with the explicit discussion of the {standard} setting.

By writing the superpotential in \eq{WHE6} on the SM-preserving vacuum in \eqs{vev78}{vev27bar12}, we find
\bea
\label{vevWHE6}
\vev{W_H} &=& \mu \left(a_1 a_2+\frac{a_3^2}{2}+\frac{a_4^2}{2}+\frac{b_3^2}{2}\right) \nn \\
&+& \rho _{11} \left(e _1 \overline{e }_1+\nu_1 \overline{\nu}_1\right)+\rho _{21} \left(e _2\overline{e }_1+\nu_2 \overline{\nu}_1\right) \nn \\
&+& \rho _{12} \left(e _1 \overline{e }_2+\nu_1 \overline{\nu}_2\right)+\rho _{22} \left(e _2 \overline{e }_2+\nu_2 \overline{\nu}_2\right) \nn \\
&+& \tau _1 \left[-a_1 e _1 \overline{\nu}_1-a_2 \nu_1 \overline{e }_1 + \sqrt{\frac{2}{3}} a_3 \left(e _1 \overline{e }_1 - \frac{1}{2} \nu_1 \overline{\nu}_1\right)  \right. \nn \\
&+& \left.  \frac{a_4 \nu_1 \overline{\nu}_1}{\sqrt{2}}-\sqrt{\frac{2}{3}} b_3 \left(e _1 \overline{e }_1 + \nu_1 \overline{\nu}_1 \right)\right] \nn \\
&+& \tau _2 \left[-a_1 e _2 \overline{\nu}_2-a_2 \nu_2 \overline{e}_2 + \sqrt{\frac{2}{3}} a_3 \left( e _2 \overline{e }_2 - \frac{1}{2} \nu_2 \overline{\nu}_2 \right)\right. \nn \\
&+& \left. \frac{a_4 \nu_2 \overline{\nu}_2}{\sqrt{2}}-\sqrt{\frac{2}{3}} b_3 (e _2 \overline{e }_2 + \nu_2\overline{\nu}_2)\right] \, .
\eea
When applying the constraints coming from $D$- and $F$-term equations, a nontrivial vacuum exists {if $\rho$ and $\tau$ are hermitian, as in} the flipped $SO(10)$ case.
This is a consequence of the fact that $D$-flatness implies
$\vev{27_i} = \vev{\overline{27}_i}^*$ (see \app{DFtermE6} for details).

After imposing all the constraints due to $D$- and $F$-flatness, the $E_6$ vacuum manifold can be finally written as
\bea
\label{vacmanifoldE6}
2 \mu a_1 &=& \tau _1 r_1^2 \sin{2\alpha_1}\ e^{i(\phi_{\nu_1}-\phi_{e_1})} \nn \\
&+& \tau _2 r_2^2 \sin{2\alpha_2}\ e^{i(\phi_{\nu_2}-\phi_{e_2})} \, , \nn \\
2 \mu a_2 &=& \tau _1 r_1^2 \sin{2\alpha_1}\ e^{-i(\phi_{\nu_1}-\phi_{e_1})} \nn \\
&+& \tau _2 r_2^2 \sin{2\alpha_2}\ e^{-i(\phi_{\nu_2}-\phi_{e_2})} \, , \nn \\
2 \sqrt{6} \mu a_3 &=& - \tau _1 r_1^2 (3 \cos{2\alpha_1} + 1) - \tau _2 r_2^2 (3 \cos{2\alpha_2} + 1) \, , \nn \\
\sqrt{2} \mu a_4 &=& -\tau _1 r_1^2 \sin^2{\alpha_1} - \tau _2 r_2^2 \sin^2{\alpha_2} \, , \nn \\
\sqrt{3} \mu b_3 &=& \sqrt{2} \tau _1 r_1^2 + \sqrt{2} \tau _2 r_2^2 \, , \nn \\
e_{1,2} &=& r_{1,2} \cos \alpha_{1,2}\ e^{i \phi_{e_{1,2}}} \, , \nn \\
\nu_{1,2} &=& r_{1,2} \sin \alpha_{1,2}\ e^{i \phi_{\nu_{1,2}}} \, , \nn \\
\overline{e}_{1,2} &=& r_{1,2} \cos \alpha_{1,2}\ e^{- i \phi_{e_{1,2}}} \, , \nn \\
\overline{\nu}_{1,2} &=& r_{1,2} \sin \alpha_{1,2}\ e^{- i \phi_{\nu_{1,2}}} \, ,
\eea
where $r_{1,2}$ and $\alpha^{\pm} \equiv \alpha_1 \pm \alpha_2$ are fixed in terms of superpotential parameters, as follows
\bea
\label{r1sq}
&& r_{1}^{2} = - \frac{\mu (\rho _{22} \tau _1 -4 \rho _{11} \tau _2) }{5 \tau _1^2 \tau _2} \, , \\
\label{r2sq}
&& r_{2}^{2} = - \frac{\mu (\rho _{11} \tau _2 -4 \rho _{22} \tau _1) }{5 \tau _1 \tau _2^2} \, , \\
\label{cosalpha1m2}
&& \cos{\alpha^{-}} = \xi \ \frac{\sin \Phi_\nu - \sin \Phi_e}{\sin \left(\Phi_\nu - \Phi_e \right)} \, , \\
\label{cosalpha1p2}
&& \cos{\alpha^{+}} = \xi \ \frac{\sin \Phi_\nu + \sin \Phi_e}{\sin \left(\Phi_\nu - \Phi_e \right)} \, ,
\eea
with
\be
\label{xiE6}
\xi = \frac{5 |\rho _{12}|}{\sqrt{-\frac{4 \rho _{11}^2 \tau _2}{\tau _1}-\frac{4 \rho _{22}^2 \tau _1}{\tau _2}+17 \rho _{22} \rho _{11}}} \, .
\ee
The phase factors $\Phi_\nu$ and $\Phi_e$ are defined as
\be
\Phi_\nu \equiv \phi_{\nu_1}-\phi_{\nu_2}+\phi_{\rho _{12}} \, , \quad \Phi_e \equiv \phi_{e_1}-\phi_{e_2}+\phi_{\rho _{12}} \, .
\ee

In \app{formalproof} we show that the little group of the
the vacuum manifold in \eq{vacmanifoldE6} is $SU(5)$.

It is instructive to look at the configuration in which one pair of $27_H$, let us say $27_2 \oplus \overline{27}_2$, is decoupled.
This case can be obtained by setting
$\tau_2=\rho_{12}=\rho_{22}=0$ {in the relevant equations}.
In agreement with Ref.~\cite{Buccella:1987kc}, we find that $\alpha_{1}$ turns out to be undetermined by the $F$-term constraints,
thus parametrizing a set of isomorphic solutions.
We may therefore take in \eq{vacmanifoldE6} $\alpha_1=\alpha_2=0$
and show that the little group corresponds in this case to $SO(10)$ (see \app{formalproof}), thus recovering the result of Ref.~\cite{Buccella:1987kc}.

The same result is obtained in the case in which the vacua of the two copies of $27_{H} \oplus \overline{27}_{H}$ are
aligned, i.e. $\alpha^-=0$ and $\Phi_\nu =\Phi_e$.
Analogously to the discussion in Sect.~\ref{vacuumFSO10},
$\alpha^+$ is in this case undetermined and it can be set to zero,
that leads us again to the one $27_{H} \oplus \overline{27}_{H}$ case,
with $SO(10)$ as the preserved algebra.

These results are intuitively understood by considering that in case there is just one pair of $27_{H} \oplus \overline{27}_{H}$
(or the vacua of the two pairs of $27_i \oplus \overline{27}_i$ are aligned) the SM-singlet directions $e$ and $\nu$ are connected by
an $SU(2)_R$ transformation which can be used to rotate one of the VEVs to zero, so that the little group is locked to an $SO(10)$ configuration.
On the other hand, two misaligned $27_{H} \oplus \overline{27}_{H}$ VEVs in the $e-\nu$ plane lead (just by inspection of the VEV quantum numbers) to an $SU(5)$ little group.

In analogy with the flipped $SO(10)$ case, the Higgs adjoint plays the role of a renormalizable agent that prevents the two pairs of $\vev{27_i \oplus \overline{27}_i}$ from aligning within each other along the $SO(10)$ vacuum.
Actually, by decoupling the adjoint Higgs, $F$-flatness makes the (aligned) $27_i\oplus 27_i$ vacuum trivial, as one verifies by inspecting the $F$-terms in \eq{FtermsE6} of \app{DFtermE6} for $\vev{78_H}=0$ and $\det{\rho}\neq 0$.

In conclusion, due to the larger $E_6$ algebra, the vacuum little group remains $SU(5)$, never landing to the SM.
In this respect we guess that the authors of Ref. \cite{Kawase:2010na}, who advocate a $78{_H} \oplus 2 \times \left( 27{_H} \oplus \overline{27}{_H} \right)$ Higgs sector, implicitly refer to a NR setting.

\subsection{Breaking the residual $SU(5)$ via effective interactions}
\label{NRopsE6}

In this section we consider the
possibility of breaking the residual $SU(5)$ symmetry
of the renormalizable $E_6$ vacuum through the inclusion of effective
adjoint Higgs interactions near the Planck scale $M_P$.
We argue that an effective flipped $SO(10)\otimes U(1)_X\equiv SO(10)_f$ may survive down to the $M_f\approx 10^{16}$ GeV scale, with thresholds spread in between $M_P$ and $M_f$ in such a way not to affect proton stability and lead to realistic neutrino masses.

The relevant part of the nonrenormalizable superpotential {at the $E_6$} scale $M_E < M_P$ can be written as
\be
\label{WHE6NR}
W^{\text{NR}}_H = \frac{1}{M_P} \left[ \lambda_1 \left( \Tr 78^2 \right)^2 + \lambda_2 \Tr 78^4 + \ldots \right] \, ,
\ee
where the ellipses stand for terms which include powers of the $27$'s representations and $D \ge 5$ operators.
Projecting \eq{WHE6NR} along the SM-singlet vacuum directions in \eqs{vev78}{vev27bar12}
we obtain
\begin{multline}
\label{WHE6NRvev}
\vev{W^{\text{NR}}_H} = \frac{1}{M_P} \left\{ \lambda_1 \left( 2 a_1 a_2 + a_3^2 + a_4^2 + b_3^2 \right)^2 \right. \\
+ \lambda_2 \left[ 2 a_1 a_2 \left( a_1^2 a_2^2 + a_3^2 + a_4^2 + \tfrac{1}{\sqrt{3}} a_3 a_4 \right) \right. \\
+ \left. \left. \tfrac{1}{2} \left( a_3^2 + a_4^2 \right)^2 + \tfrac{1}{2} b_3^4 \right] + \ldots \right\} \, .
\end{multline}
One verifies that including the NR contribution in the $F$-term equations
allows for a disentanglement of the $\vev{78}$ and
$\vev{27_1 \oplus \overline{27}_1 \oplus 27_2 \oplus \overline{27}_2}$ VEVs, so that the breaking to the SM is achieved.
In particular, the SUSY vacuum allows for an intermediate
$SO(10)_f$ stage (that is prevented by the simple renormalizable
vacuum manifold in \eq{vacmanifoldE6}).
By including \eq{WHE6NRvev} in the $F$-term equations, we can consistently neglect
all VEVs but the $SO(10) \otimes U(1)$ singlet $\Omega$,
that reads
\be
\Omega^2 = - \frac{\mu M_P}{5 \lambda_1 + \frac{1}{2} \lambda_2} \, .
\ee
It is therefore possible to envisage a scenario where the $E_6$ symmetry is broken at a scale $M_E < M_P$ leaving an effective
flipped $SO(10) \otimes U(1)_X$ scenario down to the $10^{16}$ GeV,
as discussed in \sect{sect:minimalflippedSO10}.
All remaining SM singlet VEVs are contained in
$45 \oplus 16_1 \oplus \overline{16}_1 \oplus 16_2 \oplus \overline{16}_2$ that are the only Higgs multiplets required
to survive at the $M_f\ll M_E$ scale.
It is clear that this is a plausibility argument and that a detailed
study of the $E_6$ vacuum and related thresholds is needed to ascertain
the feasibility of the scenario.

The NR breaking of $E_6$ through an intermediate $SO(10)_f$ stage driven by $\Omega\gg M_f$,
while allowing (as we shall discuss next) for a consistent unification pattern, avoids the issues arising {within} a one-step breaking.
As a matter of fact, the colored triplets responsible for $D=5$ proton decay live naturally at the $\Omega^2/M_P > M_f$ scale,
while the masses of the SM-singlet neutrino states which enter the "extended" type-I seesaw formula
are governed by the $\vev{27}\sim M_f$ (see the discussion in \sect{TRflavor}).

\subsection{A unified $E_6$ scenario}
\label{UnifiedE6}

{Let us} examine the plausibility of the two-step gauge unification scenario
discussed in the previous subsection. We consider here just a simplified description that neglects thresholds effects.
As a first quantitative estimate {of the running effects} on the
$SO(10)_f$ couplings let us introduce the quantity
\begin{multline}
\Delta (M_f) \equiv \frac{\alpha_{\hat{X}}^{-1} (M_f) - \alpha_{10}^{-1} (M_f)}{\alpha_{E}^{-1}} = \\
\frac{1}{\alpha_{E}^{-1}} \frac{b_{\hat{X}} - b_{10}}{2 \pi} \log \frac{M_E}{M_f} \, ,
\end{multline}
where $M_{E}$ is the $E_6$ unification scale and $\alpha_{E}$ is the $E_6$ gauge coupling. The $U(1)_X$ charge has been properly normalized to $\hat{X}=X/\sqrt{24}$.
The one-loop beta coefficients for the superfield content
$45_H \oplus 2 \times \left( 16_H \oplus \overline{16}_H \right)
\oplus 3 \times (16_F \oplus 10_F \oplus 1_F) \oplus 45_G$
are found to be $b_{10} = 1$ and $b_{\hat{X}} = 67/24$.

Taking, for the sake of an estimate, a typical MSSM value for the GUT coupling
$\alpha_{E}^{-1} \approx 25$, for $M_E/M_f <  10^2$
one finds
$\Delta (M_f) <  5\%$.

In order to match the $SO(10)_f$ couplings with the measured SM couplings, we consider as a typical setup
the two-loop MSSM gauge running with a 1 TeV SUSY scale. The (one-loop) matching of the non abelian gauge couplings (in dimensional reduction) at the scale $M_f$ reads
\be
\alpha_{10}^{-1} (M_f) = \alpha_2^{-1} (M_f) = \alpha_3^{-1} (M_f) \, ,
\ee
while for the properly normalized hypercharge $\hat{Y}$ one obtains
\begin{multline}
\label{alphaXext}
 \alpha_{\hat{Y}}^{-1} (M_f) =
\left( \hat{\alpha}^2 + \hat{\beta}^2 \right)  \alpha_{10}^{-1} (M_f)
+ \hat{\gamma}^2 \alpha_{\hat{X}}^{-1} (M_f) \, .
\end{multline}
Here we have implemented the relation among the properly normalized U(1) generators (see \eq{flippedSO10abc})
\be
\label{runY}
\hat{Y} = \hat{\alpha} \hat{Y}' + \hat{\beta} \hat{Z} + \hat{\gamma} \hat{X} \, ,
\ee
with $\{\hat{\alpha}, \hat{\beta}, \hat{\gamma}\} = \{-\frac{1}{5}, -\frac{1}{5}\sqrt{\frac{3}{2}}, \frac{3}{\sqrt{10}}\}$.

\begin{figure}[t]
\includegraphics[width=8.5cm]{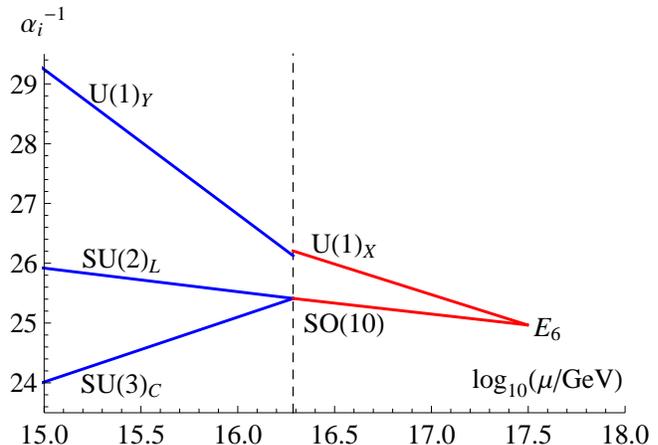}
\caption{\label{E6unification}
Sample picture of {the} gauge coupling unification in the {$E_6$-embedded}
$SO(10)\otimes U(1)_X$ model.
}
\end{figure}

The result of this simple {exercise is depicted} in \fig{E6unification}.
Barring detailed threshold effects,
it is interesting to see that the qualitative behavior
of the relevant gauge couplings is, indeed, consistent with
the basic picture of  the flipped $SO(10)\otimes U(1)_{X}$
embedded into a genuine $E_{6}$ GUT emerging below the Planck scale.

\section{Towards a realistic flavor}
\label{TRflavor}

The aim of this section is {to provide an elementary} discussion of the main features and of the possible issues arising in the Yukawa sector of {the} flipped $SO(10) \otimes U(1)_X$ model {under consideration}.
In order to keep the discussion {simple} we shall consider a {basic} Higgs contents with just one pair of $16_H \oplus \overline{16}_H$.
{As a
complement} of the tables given in Sect. \ref{sect:minimalflippedSO10}, we summarize  the SM-decomposition of the
representations relevant to the Yukawa sector {in Table \ref{tab:stanVSflipYukawa}}.

\renewcommand{\arraystretch}{1.3}
\begin{table*}[t]
\begin{tabular}{lll}
\hline \hline
\ \
& $SO(10)$
& $SO(10)_f$
\\
\hline
$16_F$ &
$\left( D^c \oplus L \right)_{\overline{5}} \oplus \left( U^c \oplus Q \oplus E^c \right)_{10} \oplus (N^c)_1$ \quad\quad &
$\left( D^c \oplus \Lambda^c \right)_{\overline{5}} \oplus \left( \Delta^c \oplus Q \oplus S \right)_{10} \oplus (N^c)_1$
\\
\null
$10_F$ &
$\left( \Delta \oplus \Lambda^c \right)_{5} \oplus \left( \Delta^c \oplus \Lambda \right)_{\overline{5}}$ &
$\left( \Delta \oplus L \right)_{5} \oplus \left( U^c \oplus \Lambda \right)_{\overline{5}}$
\\
\null
$1_F$ &
$(S)_1$ &
$(E^c)_1$
\\
$\vev{16_H}$ &
$\left( 0 \oplus \vev{H_d} \right)_{\overline{5}} \oplus \left( 0 \oplus 0 \oplus 0 \right)_{10} \oplus (\nu_H)_1$ &
$\left( 0 \oplus \vev{H_u} \right)_{\overline{5}} \oplus \left( 0 \oplus 0 \oplus s_H \right)_{10} \oplus (\nu_H)_1$
\\
\null
$\vev{\overline{16}_H}$ \quad\quad &
$\left( 0 \oplus \vev{H_u} \right)_{5} \oplus \left( 0 \oplus 0 \oplus 0 \right)_{\overline{10}} \oplus (\nu_H)_1$ &
$\left( 0 \oplus \vev{H_d} \right)_{5} \oplus \left( 0 \oplus 0 \oplus s_H \right)_{\overline{10}} \oplus (\nu_H)_1$
\\
\hline \hline
\end{tabular}
\mycaption{
SM decomposition of $SO(10)$ representations relevant for the Yukawa sector in the standard and flipped hypercharge embedding.
In the $SO(10)_f$ case $B-L$ is assigned according to \eq{BLtw2}.
A self-explanatory SM notation is used, with the outer subscripts labeling the $SU(5)$ origin.
The $SU(2)_L$ doublets decompose as $Q = (U, \ D)$, $L = (N, \ E)$, $\Lambda = (\Lambda^0, \ \Lambda^-)$ and
$\Lambda^{c} = (\Lambda^{c+}, \ \Lambda^{c0})$.
Accordingly, $\vev{H_u} = (0, \ v_u)$ and $\vev{H_d} = (v_d, \ 0)$.
The $D$-flatness constraint on the SM-singlet VEVs, $s_H$ and $\nu_H$, is taken into account.
}
\label{tab:stanVSflipYukawa}
\end{table*}

For what follows, we refer to
\cite{Nandi:1985uh,Frank:2004vg,Malinsky:2007qy,Heinze:2010du} and references therein where the basic features of models with extended matter sector are discussed in the $E_{6}$ and the standard $SO(10)$ context. For a scenario employing flipped $SO(10) \otimes U(1)$ (with an additional anomalous $U(1)$) see Ref. \cite{Maekawa:2003wm}.

\subsection{Yukawa sector of the flipped $SO(10)$ model}
\label{YukawaFSO10}

Considering for simplicity just one pair of spinor Higgs multiplets and imposing a $Z_2$ matter-parity {(negative for matter and positive for Higgs superfields) the Yukawa superpotential (up to $d= 5$ operators)} reads
\begin{multline}
\label{YukFSO10}
W_Y =  Y_{U} 16_F 10_F 16_H \\
+ \frac{1}{M_P} \left[ Y_{E} 10_F 1_F \overline{16}_H \overline{16}_H
+  Y_{D} 16_F 16_F \overline{16}_H \overline{16}_H \right] \, ,
\end{multline}
where family indexes are understood.
Notice ({c.f.} Table \ref{tab:invdec}) that {due to} the flipped
embedding the up-quarks receive mass at the renormalizable level,
while all the other fermion masses {need Planck-suppressed effective contributions in order to achieve a realistic texture}.

\renewcommand{\arraystretch}{1.3}
\begin{table*}[t]
\begin{tabular}{lll}
\hline \hline
$16_F 10_F \vev{16_{H}}$
& $10_F 1_F \vev{\overline{16}_H} \vev{\overline{16}_H}$
& $16_F 16_F \vev{\overline{16}_{H}} \vev{\overline{16}_{H}}$
\\
\hline
$(1)\ 10_F \overline{5}_F \vev{\overline{5}_H} \supset (Q U^c + S \Lambda) \vev{H_u}$
& $(2)\  \overline{5}_F 1_F \vev{5_H} \vev{\overline{1}_H} \supset \Lambda E^c \vev{H_{d}} \nu_H$
& $(1)\ 1_F 1_F \vev{\overline{1}_H} \vev{\overline{1}_H} \supset N^c N^c \nu_H^2$
\\
\null
$(1)\ 1_F 5_F \vev{\overline{5}_H} \supset N^c L \vev{H_u}$
& $(2)\ 5_F 1_F \vev{\overline{10}_H} \vev{5_H} \supset L E^c \vev{H_{d}} s_H $
& $(1)\ 10_F 10_F \vev{\overline{10}_H} \vev{\overline{10}_H} \supset S S s_H^2$
\\
\null
$(1)\ \overline{5}_F 5_F \vev{1_H} \supset (D^c \Delta + \Lambda^c L) \nu_H$
& 
& $(4)\ 10_F 1_F \vev{\overline{10}_H} \vev{\overline{1}_H} \supset S N^c s_H \nu_H$
\\
\null
$(1)\ \overline{5}_F \overline{5}_F \vev{10_H} \supset \Lambda^c \Lambda s_H$
&
& $(1)\ \overline{5}_F \overline{5}_F \vev{5_H} \vev{5_H} \supset \Lambda^c \Lambda^c \vev{H_d} \vev{H_d}$
\\
\null
$(1)\ 10_F 5_F \vev{10_H} \supset \Delta^c \Delta s_H$
&
& $(4)\ 10_F \overline{5}_F \vev{\overline{10}_H} \vev{5_H} \supset ( \Lambda^c S + Q D^c ) \vev{H_d} s_H$
\\
\null
&
& $(2)\ 10_F 10_F \vev{5_H} \vev{\overline{1}_H} \supset Q \Delta^c \vev{H_d} \nu_H$
\\
\null
&
& $(4)\ \overline{5}_F 1_F \vev{5_H} \vev{\overline{1}_H} \supset \Lambda^c N^c \vev{H_d} \nu_H$
\\
\hline \hline
\end{tabular}
\mycaption{Decomposition
of the invariants in \eq{YukFSO10} {according to flipped $SU(5)$ and SM}. The {number in the round brackets stands for the multiplicity of the invariant}. {The contractions}
$\overline{5}_{10_F} 1_{1_F} \vev{\overline{10}_{H}} \vev{\overline{10}_{H}}$ and
$\overline{5}_{16_F} 1_{16_F} \vev{\overline{10}_{H}} \vev{\overline{10}_{H}}$ {yield} no SM invariant.
}
\label{tab:invdec}
\end{table*}

\subsubsection{Mass matrices}
\label{GUTmassmatrices}

In order to avoid the recursive $1/M_P$ factors we introduce the following notation
for the relevant VEVs (see Table \ref{tab:stanVSflipYukawa}): $\hat{v}_d \equiv v_d / M_P$, $\hat{\nu}_H \equiv \nu_H / M_P$ and $\hat{s}_H \equiv s_H / M_P$.
The $M_f$-scale mass matrices for the matter fields
sharing the same unbroken $SU(3)_c \otimes U(1)_Q$ quantum numbers
can be extracted readily
by inspecting the SM decomposition of the relevant $1+10+16$ matter multiplets in the flipped SO(10) setting:
\bea
\label{MuGUT}
&& M_u = Y_U v_u \, , \nn \\[0.2ex]
\label{MdGUT}
&& M_d =
\left(
\begin{array}{cc}
 Y_D \hat{\nu}_H v_d & Y_D \hat{s}_H v_d \\
 Y_U s_H & Y_U \nu_H \\[0.2ex]
\end{array}
\right) \, , \nn \\[1ex]
\label{MeGUT}
&& M_e =
\left(
\begin{array}{cc}
 Y_E \hat{\nu}_H v_d  & Y_U s_H \\
 Y_E \hat{s}_H v_d & Y_U \nu_H
\end{array}
\right) \, ,
\eea
\begin{multline}
\label{MnuGUT}
M_\nu = \\
\left(
\begin{array}{ccccc}
 0 & 0 & Y_U s_H & 0 & Y_U v_u \\
 0 & 0 & Y_U \nu_H & Y_U v_u & 0 \\
 Y_U s_H & Y_U \nu_H & Y_D \hat{v}_d v_d & 2 Y_D \hat{v}_d \nu_H & 2 Y_D \hat{v}_d s_H \\
 0 & Y_U v_u & 2 Y_D \hat{\nu}_H v_d & Y_D \hat{\nu}_H \nu_H & 2 Y_D \hat{\nu}_H s_H \\
 Y_U v_u & 0 & 2 Y_D \hat{s}_H v_d & 2 Y_D \hat{s}_H \nu_H & Y_D \hat{s}_H s_H
\end{array}
\right) ,
\end{multline}
where, for convenience, we redefined $Y_D \rightarrow Y_D / 2$ and $Y_E \rightarrow Y_E / 2$.
The basis
$(U)(U^c)$ is used for $M_u$, $(D, \Delta)(\Delta^c, D^c)$ for $M_d$ and
$(\Lambda^-, E)(E^c, \Lambda^{c+})$ for $M_e$.
The Majorana mass matrix $M_\nu$ is {written in} the basis $(\Lambda^0, N, \Lambda^{c0}, N^c, S)$.

\subsubsection{Effective mass matrices}
\label{Effmassmatrices}

Below the $M_{f}\sim s_H \sim \nu_H $ scale,
the exotic (vector) part of the matter spectrum decouples
{and one is left with} the three standard MSSM families.
In what follows, we shall use the calligraphic symbol $\mathcal{M}$
for the $3\times 3$ effective MSSM fermion mass matrices in order to distinguish them from the mass matrices in \eqs{MuGUT}{MnuGUT}.
\vskip 0.2mm
\emph{i) Up-type quarks:}
The effective up-quark mass matrix coincides with the mass matrix in \eq{MuGUT}
\be
\label{Mueff}
\mathcal{M}_u = Y_U v_u \, .
\ee
\indent
\emph{ii) Down-type quarks and charged leptons:}
The $6 \times 6$ mass matrices in \eqs{MdGUT}{MeGUT} can be brought into a convenient
form by means of the transformations
\be
\label{Mderot}
M_d \rightarrow M_d U^{\dag}_d \equiv M'_d \, , \quad
M_e \rightarrow {U^\ast_e} M_e \equiv M'_e \, ,
\ee
where $U_{d,e}$ are $6 \times 6$ unitary matrices such that $M'_d$ and $M'_e$ are block-triangular
\be
\label{Mtriangular}
M'_d =
\mathcal{O} \left(
\begin{array}{cc}
 v & v \\
 0 & M_f
\end{array}
\right)
\, , \quad
M'_e =
\mathcal{O}
\left(
\begin{array}{cc}
 v & 0 \\
 v & M_f
\end{array}
\right)
\, .
\ee
Here $v$ denotes weak scale entries.
This corresponds to the change of basis
\be
\label{changeofbasis}
\left(
\begin{array}{c}
 d^c \\
 \tilde{\Delta}^c
\end{array}
\right)
\equiv U_d
\left(
\begin{array}{c}
 \Delta^c \\
 D^c
\end{array}
\right)
\, , \quad
\left(
\begin{array}{c}
 e \\
 \tilde{\Lambda}^-
\end{array}
\right)
\equiv {U_e}
\left(
\begin{array}{c}
 \Lambda^- \\
 E
\end{array}
\right)
\, ,
\ee
in the right-handed (RH) down quark and left-handed (LH) charged lepton sectors, respectively.
The upper components of the rotated vectors ($d^c$ and $e$) correspond
to the light MSSM degrees of freedom.
Since the residual rotations acting on the LH down quark and RH charged lepton components, that transform
the $M'_{d,e}$ matrices into fully block-diagonal forms, are extremely tiny (of $\mathcal{O}(v / M_f)$), the $3 \times 3$ upper-left blocks (ULB) in \eq{Mtriangular} can
be identified with the effective light down-type quark and charged lepton mass matrices,
i.e., $\mathcal{M}_d \equiv \left( M'_d \right)_{ULB}$ and $\mathcal{M}_e \equiv \left( M'_e \right)_{ULB}$.

{It is instructive} to work out the explicit form of the unitary matrices $U_d$ and $U_e$. For the sake of simplicity, in what follows we shall stick to the
single family case and assume the reality of all the relevant parameters. Dropping same order Yukawa factors as well, one
writes \eqs{MdGUT}{MeGUT} as
\be
M_d =
\left(
\begin{array}{cc}
 v_{\nu} & v_{s} \\
 s_H & \nu_H
\end{array}
\right) \, ,
\quad
M_e =
\left(
\begin{array}{cc}
 v_{\nu} & s_H \\
 v_{s} & \nu_H
\end{array}
\right) \, ,
\ee
and the matrices $U_d$ and $U_e$ are explicitly given by
\be
\label{Ude2by2}
U_{d,e} =
\left(
\begin{array}{cc}
 \cos{\alpha} & - \sin{\alpha} \\
 \sin{\alpha} & \cos{\alpha}
\end{array}
\right) \, .
\ee

By applying \eq{Mderot} we get that $M'_d$ and $M'_e$ have the form in
\eq{Mtriangular} provided that $\tan {\alpha} = s_H / \nu_H$.
In particular, with a specific choice of the global phase, we can write
\be
\cos{\alpha} = \frac{\nu_H}{\sqrt{s^2_H + \nu^2_H}} \, , \quad \sin{\alpha} = \frac{s_H}{\sqrt{s^2_H + \nu^2_H}} \, ,
\ee
so that the mass eigenstates (up to $\mathcal{O}(v/M_f)$ effects) are finally given by (see \eq{changeofbasis})
\be
\label{deigenstates}
\left(
\begin{array}{c}
 d^c \\
 \tilde{\Delta}^c
\end{array}
\right)
=
\frac{1}{\sqrt{s^2_H + \nu^2_H}}
\left(
\begin{array}{c}
 \nu_H \Delta^c - s_H D^c \\
 s_H \Delta^c + \nu_H D^c
\end{array}
\right)
\, ,
\ee
and
\be
\label{eeigenstates}
\left(
\begin{array}{c}
 e \\
 \tilde{\Lambda}^-
\end{array}
\right)
=
\frac{1}{\sqrt{s^2_H + \nu^2_H}}
\left(
\begin{array}{c}
 \nu_H \Lambda^-  - s_H E  \\
 s_H \Lambda^- + \nu_H E
\end{array}
\right)
\, ,
\ee
where the upper (SM) components have mass of $\mathcal{O}(v_{\nu,s})$ and the lower (exotic) ones of $\mathcal{O}(M_f)$.

\emph{iii) Neutrinos:}
{Working} again in the same approximation,
the lightest eigenvalue of $M_\nu$ in \eq{MnuGUT} is given by

\bea
\label{lighteigen}
m_{\nu}
&\sim& \frac{(\nu_H^2 + s_H^2)^2 + 2 s_H^2 \nu_H^2}{3 s_H^2 \nu_H^2  (s_H^2 + \nu_H^2)} M_P v_u^2 \, .
\eea
{For} $s_H\sim \nu_H\sim M_f\sim 10^{16}$ GeV
$M_P \sim 10^{18} \ \text{GeV}$ and $v_u \sim 10^2 \ \text{GeV}$
{one obtains}
\be
m_{\nu} \sim \frac{ v_u^2}{M_f^2/M_P} \sim 0.1 \ \text{eV} \, ,
\ee
{which is within the ballpark of the current lower bounds on the light neutrino masses set by the oscillation experiments.}

{It is also useful to examine} the composition of the lightest neutrino eigenstate $\nu$.
{At the leading order, the light neutrino eigenvector obeys the equation $M_\nu \nu =0$
which, in the components $\nu=(x_1,x_2,x_3,x_4,x_5)$, reads}
\bea
\label{eqsimplv1}
&& s_H x_3 = 0 \, , \\
\label{eqsimplv2}
&& \nu_H x_3 = 0 \, , \\
\label{eqsimplv3}
&& s_H x_1 + \nu_H x_2 = 0 \, , \\
\label{eqsimplv4}
&& \hat{\nu}_H \nu_H x_4 + 2 \hat{\nu}_H s_H x_5 = 0 \, , \\
\label{eqsimplv5}
&& 2 \hat{s}_H \nu_H x_4 + \hat{s}_H s_H x_5 = 0 \, .
\eea
By inspection, \eqs{eqsimplv4}{eqsimplv5} are compatible only if $x_4 = x_5 = 0$, while \eqs{eqsimplv1}{eqsimplv2} imply $x_3 = 0$.
Thus, the non-vanishing components of the neutrino eigenvector
are just $x_1$ and $x_2$. From
\eq{eqsimplv3}, up to a phase factor, we obtain
\be
\label{nueigenstate}
\nu = \frac{\nu_H}{\sqrt{\nu_H^2 + s_H^2}} \Lambda^{0} + \frac{- s_H}{\sqrt{\nu_H^2 + s_H^2}} N \, .
\ee
Notice that the lightest neutrino eigenstate $\nu$
and the lightest charged lepton show the same admixtures
of the {corresponding electroweak doublet components}.
Actually, this can be easily understood by taking the limit $v_u=v_d=0$ in which the preserved $SU(2)_L$
gauge symmetry imposes the same $U_e$ transformation
on the $(\Lambda^0, N)$ components.
Explicitly, given the form of $U_e$ in \eq{Ude2by2}, one obtains
in the rotated basis

\be
\label{Mnublockdiag}
M'_\nu =
\left(
\begin{array}{ccccc}
 0 & 0 & 0 & 0 & 0 \\
 0 & 0 & M_f & 0 & 0 \\
 0 & M_f & 0 & 0 & 0 \\
 0 & 0 & 0 & \frac{M_f^2}{M_P} & 2 \frac{M_f^2}{M_P} \\
 0 & 0 & 0 & 2 \frac{M_f^2}{M_P} & \frac{M_f^2}{M_P}
\end{array}
\right) \, ,
\ee
where we have taken $s_H \sim \nu_H\sim M_f$.
$M'_\nu$ is defined on the basis $(\nu, \tilde{\Lambda}^0, \Lambda^{c0}, N^c, S)$,
where
\be
\label{nueigenstates}
\left(
\begin{array}{c}
 \nu \\
 \tilde{\Lambda}^0
\end{array}
\right)
=
\frac{1}{\sqrt{2}}
\left(
\begin{array}{c}
 \Lambda^0 - N \\
 \Lambda^0 + N
\end{array}
\right)
\, .
\ee
In conclusion, we see that the "light" eigenstate $\nu$
decouples from the heavy spectrum,
\bea
\label{heavyneutrinos}
&& m_{\nu_{\text{M}_1}} \sim - M_f^2 / M_P \quad\ \, \nu_{\text{M}_1} \sim \tfrac{1}{\sqrt{2}} (N^c - S) \, , \\
&& m_{\nu_{\text{M}_2}} \sim 3 \cdot M_f^2 / M_P \quad \nu_{\text{M}_2} \sim \tfrac{1}{\sqrt{2}} (N^c + S) \, , \\
&& m_{\nu_{\text{PD}_1}} \sim - M_f \quad\quad\quad\, \nu_{\text{PD}_1} \sim \tfrac{1}{\sqrt{2}} (\tilde{\Lambda}^0 - \Lambda^{c0}) \, , \\
&& m_{\nu_{\text{PD}_2}} \sim M_f \quad\quad\quad\ \ \ \nu_{\text{PD}_2} \sim \tfrac{1}{\sqrt{2}} (\tilde{\Lambda}^0 + \Lambda^{c0}) \, ,
\eea
where $\nu_{\text{M}_1}$ and $\nu_{\text{M}_2}$ are two Majorana neutrinos of intermediate mass, $O(10^{14})$ GeV,
while the states $\nu_{\text{PD}_1}$ and $\nu_{\text{PD}_2}$ form a pseudo-Dirac neutrino of mass of $O(10^{16})$ GeV.

Notice finally that the
charged current $W_L \bar\nu_L e_L$ coupling is unaffected (c.f. \eq{nueigenstate} with \eq{eeigenstates}),
contrary to the claim in Refs. \cite{Nandi:1985uh} and \cite{Frank:2004vg},
that are based on the unjustified assumption that the physical electron $e$ is {predominantly made of} $E$.

\section{Conclusions}

In this paper we attempted to pin down the minimal Higgs setting within the framework of the supersymmetric $SO(10)$ and $E_{6}$ unifications,
consistent with a breaking of the unified gauge symmetry down to the $SU(3)_{c}\otimes SU(2)_{L}\otimes U(1)_{Y}$ of the Standard Model driven by renormalizable interactions.

The breaking of the GUT symmetries down to the SM at the renormalizable-level is a very interesting option which, simplicity apart, is supported by the success of the single-step gauge unification {inherent to} the TeV-scale minimal SUSY extension of the SM. Indeed, if any part of the GUT $\to$ SM symmetry breakdown were due to non-renormalizable (Planck induced) operators, one {has to face} a plethora of thresholds spread below the GUT scale, which {may dramatically affect the gauge running and also the proton lifetime}.

On top of that, the $B-L$ breaking scale in the vicinity of $M_{G}\sim 10^{16}$ GeV is particularly favored by the experimental lower limit on the light neutrino mass scale {($\sqrt{\Delta m^{2}_{A}}\sim 0.05$ eV)} in models in which the RH neutrinos, driving {the singlet (type-I)} variant of the seesaw mechanism, receive their masses from Planck-suppressed operators, as in the scenarios discussed {in this work}.

We argued that the simplest SUSY $SO(10)$  Higgs model that can support a full breaking of the unified symmetry down to the SM at the renormalizable level, {corresponds to the} flipped $SO(10)\otimes U(1)$ {scenario} with a $2\times (16_H\oplus \overline{16}_H)\oplus{45_H}$ Higgs sector. The enhanced breaking power of the spinorial pairs $16_H\oplus \overline{16}_H$ {and} the adjoint $45_H$ in the flipped case, each with twice as many SM singlets as the same multiplet in the standard $SO(10)$ context, does open up a room for
the desired single-step breaking of the rank~$=6$ $SO(10)\otimes U(1)$ gauge symmetry  down to the rank~$=4$ SM. These results follow {from} a detailed analysis of the relevant $F$- and $D$-flatness constraints on the gauge boson spectrum.

We {also} considered the  natural embedding of the flipped $SO(10)\otimes U(1)$ {model} into the exceptional group $E_{6}$. With an extra {copy} of the fundamental conjugated pair of $27_H\oplus \overline{27}_H$ of $E_{6}$ (comprising $16_H\oplus \overline{16}_H$ of its $SO(10)$ subgroup) on top of the simplest non-trivial renormalizable SUSY $E_{6}$ Higgs sector spanned over $27_H\oplus \overline{27}_H\oplus{78}_H$, the original symmetry is reduced to rank$=4$. However, due to the rich structure {of} $E_{6}$ as compared to its $SO(10)\otimes U(1)$ subgroup, the breaking chain stops at the $SU(5)$ level and non-renormalizable operators are still needed for a full $E_{6}\to$ SM breaking.

We made the case for a two-step breaking of an $E_6$ GUT realized in the vicinity of the Planck scale
via an intermediate flipped $SO(10)\otimes U(1)$ stage. Remarkably enough, even in the simplest picture, the few percent mismatch observed within the two-loop MSSM gauge coupling evolution at the scale of the ``one-step'' grand unification is naturally accommodated in this scheme, and it is understood as an artefact of a ``delayed'' $E_{6}$ unification superseding the flipped $SO(10)\otimes U(1)$ partial unification.
A study of GUT threshold effects and a detailed discussion of the matter spectrum will be part of future work.

\subsection*{Acknowledgments}

{We thank Michele Frigerio for helpful comments.}
S.B. is partially supported by MIUR and the EU UNILHC-grant agreement PITN-GA-2009-237920.
The work of M.M. was supported by a Marie Curie Intra European Fellowship within the 7th European Community Framework
Programme FP7-PEOPLE-2009-IEF, contract number PIEF-GA-2009-253119,  by the EU Network grant UNILHC PITN-GA-2009-237920, by the Spanish MICINN grants FPA2008-00319/FPA and MULTIDARK CAD2009-00064 and by the Generalitat Valenciana grant Prometeo/2009/091. He also warmly acknowledges the support of the Elementary particle theory group of the School of Engineering Sciences of the Royal Institute of Technology in Stockholm where a significant part of his contribution was conceived.


\appendix

\section{Flipped $SO(10)$ vacuum}
\label{app:flippedSO10vacuum}

\subsection{Flipped $SO(10)$ notation}
\label{flippedSO10notation}

We work in the basis of Ref. \cite{Rajpoot:1980xy},
where the adjoint is projected along the positive-chirality {spinorial} generators
\be
45 \equiv 45_{ij} \Sigma^{+}_{ij} \, ,
\ee
with {$i,j = 1,..,10$}. {Here}
\be
\left(
\begin{array}{c}
\Sigma^+ \\
\Sigma^-
\end{array}
\right) \equiv \frac{1}{2} \left( I_{32} \pm \Gamma_{\chi} \right) \Sigma \, ,
\ee
where $I_{32}$ is the $32$-dimensional identity matrix {and} $\Gamma_{\chi}$ {is the 10-dimensional analogue of the Dirac $\gamma_{5}$ matrix} defined as
\be
\Gamma_{\chi} \equiv - i
\Gamma_{1} \Gamma_{2} \Gamma_{3} \Gamma_{4} \Gamma_{5}
\Gamma_{6} \Gamma_{7} \Gamma_{8} \Gamma_{9} \Gamma_{10} \, .
\ee
The $\Gamma_i$ {factors} are given by the following tensor products \hyphenation{pro-ducts} of ordinary Pauli matrices {$\sigma_{i}$} and the $2$-dimensional identity {$I_{2}$}:
\bea
&& \Gamma_{1} \equiv \sigma_1 \otimes \sigma_1 \otimes I_2 \otimes I_2 \otimes \sigma_2 \, , \nn \\
&& \Gamma_{2} \equiv \sigma_1 \otimes \sigma_2 \otimes I_2 \otimes \sigma_3 \otimes \sigma_2 \, , \nn \\
&& \Gamma_{3} \equiv \sigma_1 \otimes \sigma_1 \otimes I_2 \otimes \sigma_2 \otimes \sigma_3 \, , \nn \\
&& \Gamma_{4} \equiv \sigma_1 \otimes \sigma_2 \otimes I_2 \otimes \sigma_2 \otimes I_2 \, , \nn \\
&& \Gamma_{5} \equiv \sigma_1 \otimes \sigma_1 \otimes I_2 \otimes \sigma_2 \otimes \sigma_1 \, , \nn \\
&& \Gamma_{6} \equiv \sigma_1 \otimes \sigma_2 \otimes I_2 \otimes \sigma_1 \otimes \sigma_2 \, , \nn \\
&& \Gamma_{7} \equiv \sigma_1 \otimes \sigma_3 \otimes \sigma_1 \otimes I_2 \otimes I_2 \, , \nn \\
&& \Gamma_{8} \equiv \sigma_1 \otimes \sigma_3 \otimes \sigma_2 \otimes I_2 \otimes I_2 \, , \nn \\
&& \Gamma_{9} \equiv \sigma_1 \otimes \sigma_3 \otimes \sigma_3 \otimes I_2 \otimes I_2 \, , \nn \\
&& \Gamma_{10} \equiv \sigma_2 \otimes I_2 \otimes I_2 \otimes I_2 \otimes I_2 \, ,
\eea
{which} satisfy the Clifford algebra
\be
\left\{ \Gamma_i, \Gamma_j \right\} = 2 \delta_{ij} I_{32} \, .
\ee
The {spinorial} generators, $\Sigma_{ij}$, are {then} defined as
\be
\Sigma_{ij} \equiv \frac{i}{4} \left[ \Gamma_i , \Gamma_j \right] \, .
\ee

On the flipped $SO(10)$ vacuum the adjoint representation reads
\be
\label{45vevFSO10}
\vev{45} = \left(
\begin{array}{cc}
\vev{45}_{L} & \cdot \\
\cdot & \vev{45}_{R}
\end{array}
\right) \, ,
\ee
where
\be
\vev{45}_{L} =
\text{diag} \left( \lambda_{1}, \lambda_{2}, \lambda_{3}, \lambda_{4}, \lambda_{5}, \lambda_{6}, \lambda_{7}, \lambda_{8} \right) \, ,
\ee
and
\begin{multline}
\vev{45}_{R} = \\
\left(
\begin{array}{cccccccc}
 \lambda_{9} & \cdot & \cdot & \cdot & \omega ^+ & \cdot & \cdot & \cdot \\
 \cdot & \lambda_{10} & \cdot & \cdot & \cdot & \omega ^+ & \cdot & \cdot \\
 \cdot & \cdot & \lambda_{11} & \cdot & \cdot & \cdot & \omega ^+ & \cdot \\
 \cdot & \cdot & \cdot & \lambda_{12} & \cdot & \cdot & \cdot & \omega ^+ \\
 \omega ^- & \cdot & \cdot & \cdot & \lambda_{13} & \cdot & \cdot & \cdot \\
 \cdot & \omega ^- & \cdot & \cdot & \cdot & \lambda_{14} & \cdot & \cdot \\
 \cdot & \cdot & \omega ^- & \cdot & \cdot & \cdot & \lambda_{15} & \cdot \\
 \cdot & \cdot & \cdot & \omega ^- & \cdot & \cdot & \cdot & \lambda_{16}
\end{array}
\right) \, .
\end{multline}
{In the convention defined in section \ref{vacuumFSO10} (c.f. also caption of Table \ref{tab:45decomp}),} the diagonal entries are given by
\bea
\lambda_{1} &=& \lambda_{2} = \lambda_{3} = \lambda_{5} = \lambda_{6} = \lambda_{7} = \frac{\omega _Y}{2 \sqrt{2}} \, , \\
\lambda_{4} &=& \lambda_{8} = -\frac{3 \omega _Y}{2 \sqrt{2}} \, , \nn \\
\lambda_{9} &=& \lambda_{10} = \lambda_{11} = -\frac{\omega _Y}{2 \sqrt{2}}-\frac{\omega _R}{\sqrt{2}} \, ,
\quad \lambda_{12} = \frac{3 \omega _Y}{2 \sqrt{2}}-\frac{\omega _R}{\sqrt{2}} \, , \nn \\
\lambda_{13} &=& \lambda_{14} = \lambda_{15} = -\frac{\omega _Y}{2 \sqrt{2}}+\frac{\omega _R}{\sqrt{2}} \, ,
\quad \lambda_{16} = \frac{3 \omega _Y}{2 \sqrt{2}}+\frac{\omega _R}{\sqrt{2}} \, . \nn
\eea
where $\omega _Y$ and $\omega _R$ are real, while $\omega^+ = \omega^{-*}$.

Analogously, the spinor and the anti-spinor {SM-obedient} vacuum directions are given by
\bea
\label{16vevFSO10}
&& \vev{16}^T = (\cdot \cdot \cdot \cdot \cdot \cdot \cdot \cdot \cdot \cdot \cdot \ e \ \cdot \cdot \cdot \ -\nu) \, , \\
\label{16barvevFSO10}
&& \vev{\overline{16}}^T = (\cdot \cdot \cdot \ \overline{\nu} \ \cdot \cdot \cdot \ \overline{e} \ \cdot \cdot \cdot \cdot \cdot \cdot \cdot \cdot) \, ,
\eea
where the dots stand for zeros, and the non-vanishing VEVs are generally complex.

It is worth {reminding that the shorthand} notation $16 \, \overline{16}$ and $16 \, 45 \, \overline{16}$ in \eq{WHFSO10} stands for
$16^T \mathcal{C} \, \overline{16}$ and $16^T 45^T \mathcal{C} \, \overline{16}$, where $\mathcal{C}$ {is the ``charge conjugation'' matrix obeying}
$(\Sigma^{+})^T \mathcal{C} + \mathcal{C} \, \Sigma^{-} = 0$. {In the current convention,} $\mathcal{C}$ is given by
\be
\label{CmatrixFSO10}
\mathcal{C} =
\left(
\begin{array}{cccc}
 \cdot & \cdot & \cdot & -I_{4} \\
 \cdot & \cdot & I_{4} & \cdot \\
 \cdot & I_{4} & \cdot & \cdot \\
 -I_{4} & \cdot & \cdot & \cdot
\end{array}
\right) \, ,
\ee
where $I_4$ is the four-dimensional identity matrix.

\subsection{Supersymmetric vacuum manifold}
\label{DFtermFSO10}

In order for SUSY to survive the spontaneous GUT symmetry breakdown at $M_{G}$
the vacuum manifold must be $D$- and $F$-flat at the GUT scale.
The relevant superpotential $W_H$ given in \eq{WHFSO10}, with the $\GSM$-preserving vacuum parametrized by Eq. (\ref{45vevFSO10}) and \eqs{16vevFSO10}{16barvevFSO10}, yields
the following $F$-flatness equations:
\bea
F_{\omega_R} &=&
- 4 \mu  \omega _R + \frac{\tau _1}{\sqrt{2}} ( e_1 \overline{e}_1 -\nu _1\overline{\nu }_1 ) + \frac{\tau _2}{\sqrt{2}} ( e_2 \overline{e}_2-\nu _2 \overline{\nu }_2 ) = 0 \, , \nn \\
\tfrac{2}{3}
F_{\omega_Y} \hspace{-0.3em}&=&
 4\mu  \omega _Y + \frac{\tau _1}{\sqrt{2}} ( e_1 \overline{e}_1 +\nu _1 \overline{\nu }_1 ) + \frac{\tau _2}{\sqrt{2}}  ( e_2 \overline{e}_2+\nu _2 \overline{\nu }_2 )
= 0 \, , \nn \\
F_{\omega^+} &=&
4 \mu  \omega ^- - \tau _1 \nu _1 \overline{e}_1- \tau _2\nu _2 \overline{e}_2 = 0 \, , \nn \\[1ex]
F_{\omega^-} &=&
4 \mu  \omega ^+ - \tau _1 e_1 \overline{\nu }_1 - \tau _2 e_2 \overline{\nu }_2 = 0 \, , \nn
\\[1ex]
F_{e_1} &=&
\tau _1 \left( - \omega ^-
   \overline{\nu }_1 - \frac{ \overline{e}_1 \omega _R}{\sqrt{2}}+\frac{3 \overline{e}_1 \omega _Y}{2 \sqrt{2}} \right) +\rho _{11} \overline{e}_1+\rho _{12} \overline{e}_2 = 0 \, , \nn \\
F_{e_2} &=&
\tau _2 \left( - \omega ^-
   \overline{\nu }_2 - \frac{ \overline{e}_2 \omega _R}{\sqrt{2}}+\frac{3 \overline{e}_2 \omega _Y}{2 \sqrt{2}} \right) +\rho _{21} \overline{e}_1+\rho _{22} \overline{e}_2 = 0 \, , \nn \\
F_{\nu_1} &=&
\tau _1 \left( - \omega ^+ \overline{e}_1+\frac{\overline{\nu }_1 \omega _R}{\sqrt{2}}+\frac{3 \overline{\nu }_1 \omega _Y}{2 \sqrt{2}} \right) +\rho _{11} \overline{\nu }_1+\rho
   _{12} \overline{\nu }_2 = 0 \, , \nn \\
F_{\nu_2} &=&
\tau _2 \left( - \omega ^+ \overline{e}_2+\frac{\overline{\nu }_2 \omega _R}{\sqrt{2}}+\frac{3 \overline{\nu }_2 \omega _Y}{2 \sqrt{2}} \right) +\rho _{21} \overline{\nu }_1+\rho
   _{22} \overline{\nu }_2 = 0 \, , \nn
\\
\label{FtermsFSO10}
F_{\overline{e}_1} &=&
\tau _1  \left( - \omega ^+ \nu _1  - \frac{e_1 \omega _R}{\sqrt{2}}+\frac{3 e_1 \omega _Y}{2 \sqrt{2}} \right) + \rho _{11} e_1 + \rho _{21} e_2  = 0 \, , \nn \\[1ex]
F_{\overline{e}_2} &=&
\tau _2 \left( - \omega ^+ \nu _2  - \frac{e_2 \omega _R}{\sqrt{2}}+\frac{3 e_2 \omega _Y}{2 \sqrt{2}} \right) + \rho _{12} e_1 + \rho _{22} e_2  = 0 \, , \nn \\[1ex]
F_{\overline{\nu}_1} &=&
\tau _1 \left( - \omega ^- e_1 +\frac{\nu _1 \omega _R}{\sqrt{2}}+\frac{3 \nu _1 \omega _Y}{2 \sqrt{2}} \right) + \rho _{11}\nu _1+ \rho _{21}\nu _2 = 0 \, , \nn \\[1ex]
F_{\overline{\nu}_2} &=&
\tau _2 \left( - \omega ^- e_2  +\frac{\nu _2 \omega _R}{\sqrt{2}}+\frac{3 \nu _2 \omega _Y}{2 \sqrt{2}} \right) +\rho _{12}\nu _1 + \rho _{22} \nu _2  = 0 \, . \nn \\
\eea
One can use the first four equations above to replace $\omega_R$, $\omega_Y$, $\omega^+$ and $\omega^-$ in the remaining eight {(complex)} relations
which can be rewritten in the form
\bea
16 \mu F^{\omega}_{e_1} &=& 16 \mu \left( \rho _{11} \overline{e}_1+\rho _{12} \overline{e}_2 \right) - 5 \tau _1^2 \left( \nu _1 \overline{\nu }_1 + e_1 \overline{e}_1\right) \overline{e}_1 \nn \\[1ex]
&-& \tau _1 \tau _2 \left( \nu _2 \overline{\nu }_2 \overline{e}_1 + \left( 4 \nu _2 \overline{\nu }_1 + 5 e_2 \overline{e}_1 \right) \overline{e}_2 \right)
= 0 \, , \nn \\[1ex]
16 \mu F^{\omega}_{\overline{e}_1} &=& 16 \mu \left( \rho _{11} e_1 + \rho _{21} e_2 \right)  - 5 \tau _1^2 \left( \overline{\nu }_1 \nu _1 + \overline{e}_1 e_1 \right) e_1 \nn \\
&-& \tau _1 \tau _2 \left( \overline{\nu }_2 \nu _2 e_1 + \left( 4 \overline{\nu }_2 \nu _1 + 5 \overline{e}_2 e_1 \right) e_2  \right)
= 0 \, , \nn
\eea
\bea
16 \mu F^{\omega}_{\nu_1} &=& 16 \mu \left( \rho _{11} \overline{\nu }_1+\rho _{12}
   \overline{\nu }_2 \right) - 5 \tau _1^2 \left( e_1 \overline{e}_1 + \nu _1 \overline{\nu }_1\right) \overline{\nu }_1 \nn \\
&-& \tau _1 \tau _2 \left( e_2 \overline{e}_2 \overline{\nu }_1 + \left( 4 e_2 \overline{e}_1 +5 \nu _2 \overline{\nu }_1 \right) \overline{\nu }_2 \right)
= 0 \, , \nn \\[1ex]
16 \mu F^{\omega}_{\overline{\nu}_1} &=& 16 \mu \left( \rho _{11} \nu _1 + \rho _{21} \nu _2 \right) - 5 \tau _1^2 \left( \overline{e}_1 e_1 +  \overline{\nu }_1 \nu _1 \right) \nu _1 \nn \\
&-& \tau _1 \tau _2 \left( \overline{e}_2 e_2 \nu _1 + \left( 4 \overline{e}_2 e_1 + 5 \overline{\nu }_2 \nu _1 \right) \nu _2  \right)
= 0 \, , \nn \\
\label{Fterms16}
\eea
{where the other four equations are obtained from these by exchanging
$1 \leftrightarrow 2$.}

{There are two classes of $D$-flatness conditions corresponding, respectively, to the VEVs of the $U(1)_{X}$ and the $SO(10)$ generators.
For the $X$-charge one finds}
\begin{multline}
\label{DtermsX}
D_{X} = \vev{45}^{\dag} X \vev{45} \\
+ \vev{16_{1}}^{\dag} X \vev{16_{1}} + \vev{\overline{16}_{1}}^{\dag} X \vev{\overline{16}_{1}} \\
+ \vev{16_{2}}^{\dag} X \vev{16_{2}} + \vev{\overline{16}_{2}}^{\dag} X \vev{\overline{16}_{2}} \\
= |e_1|^2 + |\nu_1|^2 - |\overline{e}_1|^2 - |\overline{\nu}_1|^2 \\
+ |e_2|^2 + |\nu_2|^2 - |\overline{e}_2|^2 - |\overline{\nu}_2|^2 = 0 \, ,
\end{multline}
while for the $SO(10)$ generators {one has}
\be
\label{DtermsSO10}
D_{ij} \equiv \ D_{ij}^{45} + D_{ij}^{16 \oplus \overline{16}} = 0 \, ,
\ee
where
\be
D_{ij}^{45} = \Tr \vev{45}^{\dag} \left[ \Sigma^{+}_{ij} , \vev{45} \right] \, ,
\ee
and
\begin{multline}
D_{ij}^{16 \oplus \overline{16}} = \vev{16_{1}}^{\dag} \Sigma^{+}_{ij} \vev{16_{1}} + \vev{\overline{16}_{1}}^{\dag} \Sigma^{-}_{ij} \vev{\overline{16}_{1}} \\
+ \vev{16_{2}}^{\dag} \Sigma^{+}_{ij} \vev{16_{2}} + \vev{\overline{16}_{2}}^{\dag} \Sigma^{-}_{ij} \vev{\overline{16}_{2}} \, .
\end{multline}
Given that
\be
\Tr \vev{45}^{\dag} \left[ \Sigma^{+}_{ij} , \vev{45} \right]
= \Tr \Sigma^{+}_{ij} \left[ \vev{45} , \vev{45}^\dag \right] \, ,
\ee
we obtain
\be
\left[ \vev{45} , \vev{45}^\dag \right] =
\left(
\begin{array}{cc}
\cdot & \cdot \\
\cdot & D_{R}
\end{array}
\right) \, ,
\ee
where
\begin{widetext}
\be
D_{R} =
\left(
\begin{array}{cccccccc}
 A & \cdot & \cdot & \cdot & \sqrt{2} B^\ast & \cdot & \cdot & \cdot \\
 \cdot & A & \cdot & \cdot & \cdot & \sqrt{2} B^\ast & \cdot & \cdot \\
 \cdot & \cdot & A & \cdot & \cdot & \cdot & \sqrt{2} B^\ast & \cdot \\
 \cdot & \cdot & \cdot & A & \cdot & \cdot & \cdot & \sqrt{2} B^\ast \\
 \sqrt{2} B & \cdot & \cdot & \cdot & -A & \cdot & \cdot & \cdot \\
 \cdot & \sqrt{2} B & \cdot & \cdot & \cdot & -A & \cdot & \cdot \\
 \cdot & \cdot & \sqrt{2} B & \cdot & \cdot & \cdot & -A & \cdot \\
 \cdot & \cdot & \cdot & \sqrt{2} B & \cdot & \cdot & \cdot & -A
\end{array}
\right) \, ,
\ee
\end{widetext}
and
\bea
A &=& \left|\omega ^+\right|^2 - \left|\omega ^-\right|^2 \, , \nn \\
B &=& \left(\omega ^+\right)^{\ast} \omega _R - \left(\omega_R\right)^{\ast} \omega ^- \, .
\eea
{Since $\omega_R$ is real and $\omega^+ = (\omega^-)^\ast$, $D_{ij}^{45} =0$ as it should be. Notice that $F_{\omega^\pm}$-flatness
implies}
\be
\tau _1 e_1 \overline{\nu }_1+\tau _2 e_2 \overline{\nu }_2 =
\tau _1 (\nu _1 \overline{e}_1)^* + \tau _2 (\nu _2 \overline{e}_2)^*
\label{Dterms45}
\ee
where the reality of $\tau_{1,2}$ {has been} taken into account.

For the {spinorial} contribution in (\ref{DtermsSO10}) we find
\begin{multline}
D_{ij}^{16 \oplus \overline{16}} =\\
(\Sigma^{+}_{ij})_{12,12}  \left( |e_1|^2 + |e_2|^2 \right) + (\Sigma^{+}_{ij})_{16,16} \left( |\nu_1|^2 + |\nu_2|^2 \right) \\
+ (\Sigma^{-}_{ij})_{4,4} \left( |\overline{\nu}_1|^2 + |\overline{\nu}_2|^2 \right) + (\Sigma^{-}_{ij})_{8,8}  \left( |\overline{e}_1|^2 + |\overline{e}_2|^2 \right) \\
- (\Sigma^{+}_{ij})_{12,16}  \left( e_1^{\ast} \nu_1 + e_2^{\ast} \nu_2 \right) - (\Sigma^{+}_{ij})_{16,12} \left( \nu_1^{\ast} e_1 + \nu_2^{\ast} e_2 \right) \\
+ (\Sigma^{-}_{ij})_{4,8} \left( \overline{\nu}_1^{\ast} \overline{e}_1 + \overline{\nu}_2^{\ast} \overline{e}_2 \right)
+ (\Sigma^{-}_{ij})_{8,4} \left( \overline{e}_1^{\ast} \overline{\nu}_1 + \overline{e}_2^{\ast} \overline{\nu}_2 \right) \, .
\end{multline}
{Given} $\Sigma^{-} = - \mathcal{C}^{-1} (\Sigma^{+})^T \mathcal{C}$ and the explicit {form} of $\mathcal{C}$ in \eq{CmatrixFSO10},
one {can verify readily} that
\bea
&& (\Sigma^{-}_{ij})_{4,4} = - (\Sigma^{+}_{ij})_{16,16} \, , \nn \\
&& (\Sigma^{-}_{ij})_{8,8} = - (\Sigma^{+}_{ij})_{12,12} \, , \nn \\
&& (\Sigma^{-}_{ij})_{4,8} = + (\Sigma^{+}_{ij})_{12,16} \, .
\eea
Thus, $D_{ij}^{16 \oplus \overline{16}}$ {can be simplified to}
\begin{multline}
\label{DtermsSO10nosigmam}
(\Sigma^{+}_{ij})_{12,12}  ( |e_1|^2 + |e_2|^2 - |\overline{e}_1|^2 - |\overline{e}_2|^2 ) \\
+ (\Sigma^{+}_{ij})_{16,16} ( |\nu_1|^2 + |\nu_2|^2 - |\overline{\nu}_1|^2 - |\overline{\nu}_2|^2 ) \\
- \left[ (\Sigma^{+}_{ij})_{12,16} ( e_1^{\ast} \nu_1 + e_2^{\ast} \nu_2 - \overline{\nu}_1^{\ast} \overline{e}_1 - \overline{\nu}_2^{\ast} \overline{e}_2 )
+ \text{c.c.} \right] = 0 \, ,
\end{multline}
or, {with \eq{DtermsX} at hand, to}
\begin{multline}
\label{DtermsSO10nosigmamred}
\left[ (\Sigma^{+}_{ij})_{16,16}  - (\Sigma^{+}_{ij})_{12,12} \right] ( |\nu_1|^2 + |\nu_2|^2 - |\overline{\nu}_1|^2 - |\overline{\nu}_2|^2 ) \\
- \left[ (\Sigma^{+}_{ij})_{12,16} ( e_1^{\ast} \nu_1 + e_2^{\ast} \nu_2 - \overline{\nu}_1^{\ast} \overline{e}_1 - \overline{\nu}_2^{\ast} \overline{e}_2 )
+ \text{c.c.} \right] = 0 \, .
\end{multline}

{Taking into account the basic features of the spinorial generators
$\Sigma^{+}_{ij}$ (e.g., the bracket [$(\Sigma^{+}_{ij})_{16,16}  - (\Sigma^{+}_{ij})_{12,12}]$ and $(\Sigma^{+}_{ij})_{12,16}$ can never act against each other because at least one of them always vanishes, or the fact that $(\Sigma^{+}_{ij})_{12,16}$ is complex) \eq{DtermsSO10nosigmamred} can be satisfied for all $ij$ if and only if}
\bea
\label{Dterms16s}
|e_1|^2 + |e_2|^2 - |\overline{e}_1|^2 - |\overline{e}_2|^2 & = & 0 \, , \nn \\
|\nu_1|^2 + |\nu_2|^2 - |\overline{\nu}_1|^2 - |\overline{\nu}_2|^2 & = & 0 \, , \nn \\
e_1^{\ast} \nu_1 + e_2^{\ast} \nu_2 - \overline{\nu}_1^{\ast} \overline{e}_1 - \overline{\nu}_2^{\ast} \overline{e}_2 & = & 0 \, ,
\eea
{Combining this with} \eq{Dterms45}, the required $D$- and $F$-flatness
{can be in general maintained only if} $e_{1,2}^{\ast} = \overline{e}_{1,2}$ and $\nu_{1,2}^{\ast} = \overline{\nu}_{1,2}$.
Hence, we can write
\bea
\label{complexDtermsFSO10}
&& e_{1,2} \equiv |e_{1,2}| e^{i \phi_{e_{1,2}}} \, , \qquad \overline{e}_{1,2} \equiv |e_{1,2}| e^{- i \phi_{e_{1,2}}} \, , \nn \\
&& \nu_{1,2} \equiv |\nu_{1,2}| e^{i \phi_{\nu_{1,2}}} \, , \qquad \overline{\nu}_{1,2} \equiv |\nu_{1,2}| e^{- i \phi_{\nu_{1,2}}} \, .
\eea
{With this at hand, one can further simplify the $F$-flatness conditions \eq{Fterms16}.
To this end, it is convenient to define the}
following linear combinations
\bea
\label{LmlincombFSO10}
&& L^{-}_{V} \equiv C_{1}^{V}\cos{\phi_{V}}  - C_{2}^{V}\sin{\phi_{V}}  \, , \\
\label{LplincombFSO10}
&& L^{+}_{V} \equiv C_{1}^{V}\sin{\phi_{V}} + C_{2}^{V}\cos{\phi_{V}}  \, ,
\eea
where
\bea
C_{1}^{V} \equiv \frac{1}{2i} \left( F^{\omega}_{\overline{V}} - F^{\omega}_{V} \right) \, , \qquad C_{2}^{V} \equiv \frac{1}{2} \left( F^{\omega}_{\overline{V}} + F^{\omega}_{V} \right) \, ,\nn
\eea
with $V$ running over {the spinorial VEVs} $e_1$, $e_2$, $\nu_1$ and $\nu_2$.
{For $\mu$, $\tau_1$ and $\tau_2$ real by definition, the requirement of $L^{\pm}_{V}=0$ for all $V$ is equivalent to}
\begin{multline}
4 \mu \Re L^{-}_{e_1} = \ \\
\left|e_2\right| \left(\tau _1 \tau _2 \left|\nu _1\right| \left|\nu _2\right| \sin \left(\phi_{e_1}-\phi_{e_2}-\phi_{\nu _1}+\phi_{\nu _2}\right) \right. \\
-2 \mu  \left(\left|\rho _{21}\right| \sin \left(\phi_{e_1}-\phi_{e_2}-\phi_{\rho _{21}}\right) \right. \\ \left. \left.
+\left|\rho _{12}\right| \sin \left(\phi_{e_1}-\phi_{e_2}+\phi_{\rho _{12}}\right)\right)\right)
= 0 \, , \nn
\end{multline}
\begin{multline}
4 \mu \Re L^{-}_{\nu_1} = \ \\
\left|\nu_2\right| \left(\tau _1 \tau _2 \left|e _1\right| \left|e _2\right| \sin \left(\phi_{\nu_1}-\phi_{\nu_2}-\phi_{e _1}+\phi_{e _2}\right) \right. \\
-2 \mu  \left(\left|\rho _{21}\right| \sin \left(\phi_{\nu_1}-\phi_{\nu_2}-\phi_{\rho _{21}}\right) \right. \\ \left. \left.
+\left|\rho _{12}\right| \sin \left(\phi_{\nu_1}-\phi_{\nu_2}+\phi_{\rho _{12}}\right)\right)\right)
= 0 \, ,
\end{multline}
\begin{multline}
- 2 \Im L^{-}_{e_1} = \ \\
\left|e_2\right| \left(\left|\rho _{21}\right| \cos \left(\phi_{e_1}-\phi_{e_2}-\phi_{\rho
   _{21}}\right) \right. \\ \left.
   -\left|\rho _{12}\right| \cos \left(\phi_{e_1}-\phi_{e_2}+\phi_{\rho
   _{12}}\right)\right)
= 0 \, , \nn
\end{multline}
\begin{multline}
- 2 \Im L^{-}_{\nu_1} = \ \\
\left|\nu_2\right| \left(\left|\rho _{21}\right| \cos \left(\phi_{\nu_1}-\phi_{\nu_2}-\phi_{\rho
   _{21}}\right) \right. \\ \left.
   -\left|\rho _{12}\right| \cos \left(\phi_{\nu_1}-\phi_{\nu_2}+\phi_{\rho
   _{12}}\right)\right)
= 0 \, ,
\end{multline}
and
\begin{widetext}
\begin{multline}
- 16 \mu \Re L^{+}_{e_1} = \ \\
- 16 \mu \left|e_1\right| \left|\rho _{11}\right| \cos \left(\phi_{\rho _{11}}\right)
- 8 \mu \left|e_2\right| \left(\left|\rho _{21}\right| \cos \left(\phi_{e_1}-\phi_{e_2}-\phi_{\rho
   _{21}}\right)+\left|\rho _{12}\right| \cos \left(\phi_{e_1}-\phi_{e_2}+\phi_{\rho
   _{12}}\right)\right) \\
   +5 \tau _1^2 \left( \left|e_1\right|{}^2 + \left|\nu_1\right|{}^2 \right) \left|e_1\right|
   +\tau _1 \tau _2 \left( \left(5  \left|e_2\right|{}^2 + \left|\nu _2\right|{}^2\right) \left|e_1\right|
   +4 \left|\nu _1\right| \left|\nu _2\right| \left|e_2\right| \cos \left(\phi_{e_1}-\phi_{e_2}-\phi_{\nu _1}+ \phi_{\nu _2}\right)\right)
    = 0 \, , \nn
\end{multline}
\begin{multline}
- 16 \mu \Re L^{+}_{\nu_1} = \ \\
- 16 \mu \left|\nu_1\right| \left|\rho _{11}\right| \cos \left(\phi_{\rho _{11}}\right)
- 8 \mu \left|\nu_2\right| \left(\left|\rho _{21}\right| \cos \left(\phi_{\nu_1}-\phi_{\nu_2}-\phi_{\rho
   _{21}}\right)+\left|\rho _{12}\right| \cos \left(\phi_{\nu_1}-\phi_{\nu_2}+\phi_{\rho
   _{12}}\right)\right) \\
   +5 \tau _1^2 \left( \left|\nu_1\right|{}^2 + \left|e_1\right|{}^2 \right) \left|\nu_1\right|
   +\tau _1 \tau _2 \left( \left(5  \left|\nu_2\right|{}^2 + \left|e _2\right|{}^2\right) \left|\nu_1\right|
   +4 \left|e _1\right| \left|e _2\right| \left|\nu_2\right| \cos \left(\phi_{\nu_1}-\phi_{\nu_2}-\phi_{e _1}+ \phi_{e _2}\right)\right)
    = 0 \, ,
\end{multline}
\end{widetext}
\begin{multline}
2 \Im L^{+}_{e_1} = \ \\
2 \left|e_1\right| \left|\rho _{11}\right| \sin \left(\phi_{\rho _{11}}\right)+\left|e_2\right| \left(\left|\rho
   _{12}\right| \sin \left(\phi_{e_1}-\phi_{e_2}+\phi_{\rho _{12}}\right) \right. \\ \left.
   -\left|\rho _{21}\right| \sin
   \left(\phi_{e_1}-\phi_{e_2}-\phi_{\rho _{21}}\right)\right)
= 0 \, , \nn
\end{multline}
\begin{multline}
2 \Im L^{+}_{\nu_1} = \ \\
2 \left|\nu_1\right| \left|\rho _{11}\right| \sin \left(\phi_{\rho _{11}}\right)+\left|\nu_2\right| \left(\left|\rho
   _{12}\right| \sin \left(\phi_{\nu_1}-\phi_{\nu_2}+\phi_{\rho _{12}}\right) \right. \\ \left.
   -\left|\rho _{21}\right| \sin
   \left(\phi_{\nu_1}-\phi_{\nu_2}-\phi_{\rho _{21}}\right)\right)
= 0 \, ,
\end{multline}
{where, as before, the remaining eight real equations for V=$e_{2},\,\nu_{2}$ are obtained  by swapping $1 \leftrightarrow 2$.}

Focusing first on $L^{-}$, one finds that
$|e_1| L^{-}_{e_1} + |e_2| L^{-}_{e_2} = 0$ and $|\nu_1| L^{-}_{\nu_1} + |\nu_2| L^{-}_{\nu_2} = 0$. Thus, we can consider just $L^{-}_{e_1}$ and $L^{-}_{\nu_1}$ as independent equations. {For instance,} from {$\Im L^{-}_{e_1}=0$} one readily gets
\bea
\label{rhoratioFSO10}
\frac{|\rho_{21}|}{|\rho_{12}|} = \frac{ \cos \left(\phi_{e _1}-\phi_{e _2}+\phi_{\rho _{12}}\right)}
{\cos \left(\phi_{e _1}-\phi_{e_2}-\phi_{\rho _{21}}\right)} \,.
\eea
{On top of that, the remaining ${\rm Re\,} L^{-}_{V}={\rm Im\,}L^{-}_{V}=0$ equations can be solved only for} $\phi_{\rho _{12}} = - \phi_{\rho _{21}}$, which, plugged into \eq{rhoratioFSO10} gives $|\rho_{12}| = |\rho_{21}|$. Thus, we end up with {the following condition for the off-diagonal entries of the $\rho$ matrix:}
\be
\label{rho21eqrho12stFSO10}
\rho_{21} = \rho_{12}^{\ast} \, .
\ee
{Inserting this into the $\Re L^{-}_{e_1}=0$ and $\Re L^{-}_{\nu_1}=0$ equations, they simplify to}
\bea
\label{rho12constFSO10}
-4 \mu |\rho_{12}| &=& \tau _1 \tau _2 \left|\nu_1\right| \left|\nu_2\right| \sin \left(\Phi_\nu - \Phi_e \right) \csc \Phi_e \, , \;\;\\
\label{rho12constbisFSO10}
4 \mu |\rho_{12}| &=& \tau _1 \tau _2 \left|e _1\right| \left|e _2\right|  \sin \left(\Phi_\nu - \Phi_e \right) \csc \Phi_\nu \, ,\;\;
\eea
{where we have denoted}
\be
\Phi_\nu \equiv \phi_{\nu_1}-\phi_{\nu_2}+\phi_{\rho _{12}} \, , \quad \Phi_e \equiv \phi_{e_1}-\phi_{e_2}+\phi_{\rho _{12}} \, .
\ee
These, taken together, yield
\bea
\label{nu1nu2e1e2constFSO10}
&& |e_1| |e_2| \sin \Phi_e  =  - |\nu_1| |\nu_2| \sin \Phi_\nu \, ,
\eea
and
\bea
\label{nu1nu2e1e2constBISFSO10}
&& |\nu_1| |\nu_2| + |e_1| |e_2| = \frac{4 \mu |\rho_{12}|}{\tau _1 \tau _2} \frac{\sin \Phi_\nu - \sin \Phi_e}
{\sin \left(\Phi_\nu - \Phi_e \right)} \, .
\eea
Notice that in the zero phases limit the constraint {(\ref{nu1nu2e1e2constFSO10})} is trivially relaxed, while $\tfrac{\sin \Phi_\nu - \sin \Phi_e}
{\sin \left(\Phi_\nu - \Phi_e \right)} \rightarrow 1$.

{Returning to the $L^{+}_{V}=0$ equations, the constraint (\ref{rho21eqrho12stFSO10}) implies, e.g.}
\bea
\Im L^{+}_{e_1} = \ &&
\left|e _1\right| \left|\rho _{11}\right| \sin \left(\phi_{\rho _{11}}\right)
= 0 \, , \nn \\
\Im L^{+}_{e_2} = \ &&
\left|e _2\right| \left|\rho _{22}\right| \sin \left(\phi_{\rho _{22}}\right)
= 0 \, , \nn \\
\Im L^{+}_{\nu_1} = \ &&
\left|\nu_1\right| \left|\rho _{11}\right| \sin \left(\phi_{\rho _{11}}\right)
= 0 \, , \nn \\
\Im L^{+}_{\nu_2} = \ &&
\left|\nu_2\right| \left|\rho _{22}\right| \sin \left(\phi_{\rho _{22}}\right)
= 0 \, .
\eea
{For generic VEVs, these relations} require $\phi_{\rho _{11}}$ and $\phi_{\rho _{22}}$ to vanish. In conclusion, a nontrivial vacuum requires $\rho$ (and hence $\tau$ for consistency) to be hermitian. This is a consequence of the fact that {$D$-flatness} for the flipped $SO(10)$ embedding implies
$\vev{16_i} = \vev{\overline{16}_i}^*$, {c.f. \eq{complexDtermsFSO10}}.
{Let us also note that} such a setting is preserved by supersymmetric wavefunction renormalization.

Taking {$\rho=\rho^\dag$} in the remaining $\Re L^{+}_{V}=0$ equations and trading $|\rho_{12}|$ for $|\nu_1||\nu_2|$
{in $\Re L^+_{e_{1,2}}=0$} by means of \eq{rho12constFSO10} {and} for $|e_1||e_2|$ {in $\Re L^+_{\nu_{1,2}}=0$} via \eq{rho12constbisFSO10}, one obtains
\begin{widetext}
\bea
- 16 \mu \Re L^{+}_{e_1} = \ &&
\left|e _1\right| \left[ -16 \mu \rho _{11} + 5 \tau _1^2 \left( \left|\nu_1\right|{}^2 + \left|e _1\right|{}^2 \right)
+ \tau _1 \tau _2 \left( \left|\nu_2\right|{}^2  +5 \left|e _2\right|{}^2 \right) \right] + 4 \tau _1 \tau _2 \left|\nu_1\right| \left|\nu_2\right| \left|e _2\right| \sin \Phi_\nu \csc \Phi_e
= 0 \, , \nn \\ \nn \\
- 16 \mu \Re L^{+}_{e_2} = \ &&
\left|e _2\right| \left[ -16 \mu \rho _{22} + 5 \tau _2^2 \left( \left|\nu_2\right|{}^2 + \left|e _2\right|{}^2 \right)
+ \tau _1 \tau _2 \left( \left|\nu_1\right|{}^2  +5 \left|e _1\right|{}^2 \right) \right] + 4 \tau _1 \tau _2 \left|\nu_1\right| \left|\nu_2\right| \left|e _1\right| \sin \Phi_\nu \csc \Phi_e
= 0 \, , \nn \\ \nn \\
- 16 \mu \Re L^{+}_{\nu_1} = \ &&
\left|\nu _1\right| \left[ -16 \mu \rho _{11} + 5 \tau _1^2 \left( \left|e_1\right|{}^2 + \left|\nu _1\right|{}^2 \right)
+ \tau _1 \tau _2 \left( \left|e_2\right|{}^2  +5 \left|\nu _2\right|{}^2 \right) \right] + 4 \tau _1 \tau _2 \left|\nu_2\right| \left|e _1\right| \left|e _2\right| \csc \Phi_\nu \sin \Phi_e
= 0 \, , \nn \\ \nn \\
- 16 \mu \Re L^{+}_{\nu_2} = \ &&
\left|\nu _2\right| \left[ -16 \mu \rho _{22} + 5 \tau _2^2 \left( \left|e_2\right|{}^2 + \left|\nu _2\right|{}^2 \right)
+ \tau _1 \tau _2 \left( \left|e_1\right|{}^2  +5 \left|\nu _1\right|{}^2 \right) \right] + 4 \tau _1 \tau _2 \left|\nu_1\right| \left|e _1\right| \left|e _2\right| \csc \Phi_\nu \sin \Phi_e
= 0 \nn \, . \\
\label{ReLprho12subFSO10}
\eea
\end{widetext}
Since only two out of these four are independent constraints, it is convenient to consider the following linear combinations

\begin{multline}
C_{3} \equiv |\nu_1|^2 \left(|e_1| \Re L^{+}_{e_1} - |e_2| \Re L^{+}_{e_2}\right) \\
- |e_1|^2 \left(|\nu_1| \Re L^{+}_{\nu_1} - |\nu_2| \Re L^{+}_{\nu_2}\right) \, ,
\end{multline}

\begin{multline}
C_{4} \equiv |\nu_2|^2 \left(|e_1| \Re L^{+}_{e_1} - |e_2| \Re L^{+}_{e_2}\right) \\
- |e_2|^2 \left(|\nu_1| \Re L^{+}_{\nu_1} - |\nu_2| \Re L^{+}_{\nu_2}\right) \, ,
\end{multline}
which admit for a simple factorized form
\begin{widetext}
\bea
\label{CK3FSO10}
16 \mu C_{3} = \ && \left(\left|\nu_2\right|{}^2 \left|e _1\right|{}^2-\left|\nu_1\right|{}^2 \left|e _2\right|{}^2\right) \left[5 \tau _2^2 \left(\left|\nu_2\right|{}^2+\left|e
   _2\right|{}^2\right)+\tau _1 \tau _2 \left(\left|\nu_1\right|{}^2+\left|e _1\right|{}^2\right)-16 \mu \rho _{22}\right] = 0 \, ,  \\
\label{CK4FSO10}
16 \mu C_{4} = \ && \left(\left|\nu_2\right|{}^2 \left|e _1\right|{}^2-\left|\nu_1\right|{}^2 \left|e _2\right|{}^2\right) \left[5 \tau _1^2 \left(\left|\nu_1\right|{}^2+\left|e
   _1\right|{}^2\right)+\tau _1 \tau _2 \left(\left|\nu_2\right|{}^2+\left|e _2\right|{}^2\right)-16 \mu \rho _{11}\right]  = 0 \, .
\eea
\end{widetext}
{These relations can be generically satisfied only if the square brackets are zero, providing}
\begin{eqnarray}
16 \mu \rho_{11} &= &
5 \tau _1^2 \left(\left|\nu_1\right|{}^2+\left|e _1\right|{}^2\right) + \tau _1 \tau _2 \left(\left|\nu_2\right|{}^2+\left|e _2\right|{}^2\right) \, ,\nn\\
16 \mu \rho_{22} &= & 5 \tau _2^2 \left(\left|\nu_2\right|{}^2+\left|e _2\right|{}^2\right) + \tau _1 \tau _2 \left(\left|\nu_1\right|{}^2+\left|e _1\right|{}^2\right) \, .\nn\\
\label{absrhoFSO10}
\end{eqnarray}
By introducing a pair of symbolic 2-dimensional vectors $\vec{r}_{1}=(|\nu_{1}|,|e_{1}|)$ and $\vec{r}_{2}=(|\nu_{2}|,|e_{2}|)$ one can write
\bea
&& r_{1}^{2} = |\nu_1|^2+|e_1|^2 \, , \nn\\
&& r_{2}^{2} = |\nu_2|^2+|e_2|^2 \, , \nn \\
\label{r1dotr2vevFSO10}
&& \vec{r}_{1}.\vec{r}_{2} = |\nu_1| |\nu_{2}|+|e_1||e_{2}| \, .
\eea
{which, in combination with eqs.
(\ref{nu1nu2e1e2constBISFSO10}) and (\ref{absrhoFSO10}) yields}
\bea
&& r_{1}^{2}  =  - \frac{2 \mu (\rho _{22} \tau _1 -5 \rho _{11} \tau _2) }{3 \tau _1^2 \tau _2} \, , \nn\\
&& r_{2}^{2}  =  - \frac{2 \mu (\rho _{11} \tau _2 -5 \rho _{22} \tau _1) }{3 \tau _1 \tau _2^2} \, ,  \nn\\
\label{r1dotr2supparFSO10}
&& \vec{r}_{1}.\vec{r}_{2} = \frac{4 \mu |\rho _{12}|}{\tau _1 \tau _2} \frac{\sin \Phi_\nu - \sin \Phi_e}{\sin \left(\Phi_\nu - \Phi_e \right)} \, .
\eea
{With this at hand, the vacuum manifold can be conveniently parametrized by means of} two angles $\alpha_1$ and $\alpha_2$
\bea
&& |\nu_1| = r_1 \sin \alpha_1 \, , \qquad |e_1| = r_1 \cos \alpha_1 \, , \nn \\
&& |\nu_2| = r_2 \sin \alpha_2 \, , \qquad |e_2| = r_2 \cos \alpha_2 \, .
\label{alpha12paramFSO10}
\eea
{which are} fixed in terms of the superpotential parameters.
By defining $\alpha^{\pm} \equiv \alpha_1 \pm \alpha_2$, \eqs{r1dotr2vevFSO10}{alpha12paramFSO10} give
\be
\label{cosalphamFSO10}
\cos{\alpha^{-}} = \frac{\vec{r}_{1}.\vec{r}_{2}}{r_{1} r_{2}} =
\xi \ \frac{\sin \Phi_\nu - \sin \Phi_e}{\sin \left(\Phi_\nu - \Phi_e \right)} \, ,
\ee
where
\be
\xi = \frac{6 |\rho _{12}|}{\sqrt{-\frac{5 \rho _{11}^2 \tau _2}{\tau _1}-\frac{5 \rho _{22}^2 \tau _1}{\tau _2}+26 \rho _{22} \rho _{11}}} \, .
\ee
Analogously, \eq{nu1nu2e1e2constFSO10} can be rewritten as
\be
\cos \alpha_1 \cos \alpha_2 \sin \Phi_e  =  - \sin \alpha_1 \sin \alpha_2 \sin \Phi_\nu \, ,
\ee
which gives
\be
\frac{\sin \Phi_e}{\sin \Phi_\nu}  =
\frac{\cos{\alpha^+} - \cos{\alpha^-}}{\cos{\alpha^-} + \cos{\alpha^+}} \, ,
\ee
and thus, using \eq{cosalphamFSO10}, we obtain
\be
\label{cosalphapFSO10}
\cos \alpha^{+} = \xi \ \frac{\sin \Phi_\nu + \sin \Phi_e}{\sin \left(\Phi_\nu - \Phi_e \right)} \, .
\ee
Notice {also} that in the real {case} (i.e., $\Phi_\nu = \Phi_e = 0$) $\alpha^{+}$ is undetermined, while $\cos \alpha^{-} = \xi$.

{This justifies the shape of the vacuum manifold given} in \eq{vacmanifoldFSO10} of \sect{vacuumFSO10}.

\subsection{Gauge boson spectrum}
\label{gaugespectrum}

In order to determine {the residual symmetry corresponding to a specific vacuum configuration} we compute explicitly the gauge spectrum.
Given the $SO(10)\otimes U(1)_X$ covariant derivatives for the
scalar components of the Higgs chiral superfields
\bea
\label{covder16}
&& D_{\mu} 16 = \partial_{\mu} 16 - i g (A_{\mu})_{(ij)} \Sigma^+_{(ij)} 16 - i g_X X_{\mu} 16 \, , \nn \\
\label{covder16bar}
&& D_{\mu} \overline{16} = \partial_{\mu} \overline{16} - i g (A_{\mu})_{(ij)} \Sigma^-_{(ij)} \overline{16} + i g_X X_{\mu} \overline{16} \, , \nn \\
\label{covder45}
&& D_{\mu} 45 = \partial_{\mu} 45 - i g (A_{\mu})_{(ij)} \left[ \Sigma^+_{(ij)},  45 \right]  \, ,
\eea
where the indices in brackets $(ij)$ stand for ordered pairs,
and the properly normalized kinetic terms
\be
D_{\mu} 16^{\dag} D_{\mu} 16 \, , \ \
D_{\mu} \overline{16}^{\dag} D_{\mu} \overline{16} \, , \ \
\tfrac{1}{4} \Tr D_{\mu} 45^{\dag} D_{\mu} 45 \, , \\
\ee
one can write the 46-dimensional gauge boson mass matrix {governing the mass bilinear of the form}
\be
\label{masstermgauge}
\frac{1}{2} \left( (A_{\mu})_{(ij)}\ ,  X_{\mu} \right) \mathcal{M}^2 (A,X) \left( (A^{\mu})_{(kl)}\ ,  X^{\mu} \right)^{T} \,
\ee
as
\bea
&& \mathcal{M}^2(A,X) =
\left(
\begin{array}{ll}
 \mathcal{M}_{(ij)(kl)}^2 & \mathcal{M}_{(ij)X}^2  \\
 \mathcal{M}_{X(kl)}^2 & \mathcal{M}_{XX}^2
\end{array}
\right) \, .
\eea
The relevant matrix elements are given by
\bea
\mathcal{M}_{(ij)(kl)}^2 &=&
g^2\left(\vev{16}^{\dag} \{\Sigma^+_{(ij)},\Sigma^+_{(kl)}\} \vev{16} \right. \nn \\
&+&\vev{\overline{16}}^{\dag} \{\Sigma^-_{(ij)},\Sigma^-_{(kl)}\} \vev{\overline{16}} \nn \\
&+& \left. \frac{1}{2} \Tr  \left[ \Sigma^+_{(ij)},  \vev{45} \right]^\dag \left[ \Sigma^+_{(kl)}, \vev{45} \right]  \right) \, , \nn
\\
\mathcal{M}_{(ij) X}^2 &=&
2 g g_X \left(\vev{16}^{\dag} \Sigma^+_{(ij)} \vev{16} - \vev{\overline{16}}^{\dag} \Sigma^-_{(ij)} \vev{\overline{16}}\right) \, , \nn \\
\mathcal{M}_{X (kl)}^2 &=&
2 g g_X \left(\vev{16}^{\dag} \Sigma^+_{(kl)} \vev{16} - \vev{\overline{16}}^{\dag} \Sigma^-_{(kl)} \vev{\overline{16}}\right) \, , \nn \\
\mathcal{M}_{XX}^2 &=&
2 g_X^{2} \left(\vev{16}^{\dag} \vev{16} + \vev{\overline{16}}^{\dag} \vev{\overline{16}}\right) \, .
\eea

\subsubsection{Spinorial contribution}
\label{gaugespectrum16}

Considering first the contribution of the reducible representation $\vev{16_1 \oplus 16_2 \oplus \overline{16}_1 \oplus \overline{16}_2}$ to the gauge boson mass matrix, we find
\bea
\label{Mgauge130_FSO10}
&& \mathcal{M}_{{16}}^2 (1,3,0)_{1_{45}} = 0 \, , \\
\label{Mgauge810_FSO10}
&& \mathcal{M}_{{16}}^2 (8,1,0)_{15_{45}} = 0 \, ,
\eea
\begin{multline}
\label{Mgauge13m13_FSO10}
\mathcal{M}_{{16}}^2 (3,1,-\tfrac{1}{3})_{15_{45}} = \\
g^2 \left( |e_1|^2 + |\nu_1|^2 + |e_2|^2 + |\nu_2|^2 \right. \\
\left. + |\overline{e}_1|^2 + |\overline{\nu}_1|^2 + |\overline{e}_2|^2 + |\overline{\nu}_2|^2  \right) \, ,
\end{multline}

{In} the $(6^-_{45},6^+_{45})$ basis (see Table \ref{tab:45decomp} for the labelling of the states) we obtain
\begin{widetext}
\be
\label{Mgauge32p16_FSO10}
\mathcal{M}_{{16}}^2 (3,2,+\tfrac{1}{6}) =
\left(
\begin{array}{cc}
 g^2 \left( |\nu_1|^2 + |\nu_2|^2 + |\overline{\nu}_1|^2 + |\overline{\nu}_2|^2 \right) &
 -i g^2 \left( e_1^\ast \nu_1 + e_2^\ast \nu_2 + \overline{\nu}_1^\ast \overline{e}_1 + \overline{\nu}_2^\ast \overline{e}_2 \right) \\
 i g^2 \left( e_1 \nu_1^\ast + e_2 \nu_2^\ast + \overline{\nu}_1 \overline{e}_1^\ast + \overline{\nu}_2 \overline{e}_2^\ast \right) &
 g^2 \left( |e_1|^2 + |e_2|^2 + |\overline{e}_1|^2 + |\overline{e}_2|^2 \right)
\end{array}
\right)
\, , \\
\ee
\end{widetext}
The five dimensional SM singlet mass matrix {in} the $\left( 15_{45}, 1^-_{45}, 1^0_{45}, 1^+_{45}, 1_{1}  \right)$ basis reads
{
\begin{widetext}
\begin{equation}
\label{Mgauge110_two16}
\mathcal{M}_{{16}}^2 (1,1,0) = \\
\left(
\begin{array}{ccccc}
 \frac{3}{2} g^2 S_{1} & i \sqrt{3} g^2 S_{3} & -\sqrt{\frac{3}{2}} g^2 S_{2} & -i \sqrt{3} g^2 S_{3}^{\ast}  & -\sqrt{3} g g_X
   S_{1} \\
 -i \sqrt{3} g^2 S_{3}^{\ast}  & g^2 S_{1} & 0 & 0 & 2 i g g_X S_{3} \\
 -\sqrt{\frac{3}{2}} g^2 S_{2} & 0 & g^2 S_{1}  & 0 & \sqrt{2} g g_X S_{2} \\
 i \sqrt{3} g^2 S_{3} & 0 & 0 & g^2 S_{1} & -2 i g g_X S_{3}^{\ast}  \\
 -\sqrt{3} g g_X S_{1} & - 2 i g g_X S_{3}^{\ast} & \sqrt{2} g g_X S_{2} &2 i g g_X S_{3} &
2 g_X^2 S_{1}
\end{array}
\right)
\end{equation}
\end{widetext}
where $S_{1}\equiv |e_1|^2+|e_2|^2+|\nu _1|^2+|\nu _2|^2+|\overline{e}_1|^2+|\overline{e}_2|^2+|\overline{\nu }_1|^2+|\overline{\nu }_2|^2$, $S_{2}\equiv |e_1|^2+|e_2|^2-|\nu _1|^2-|\nu _2|^2+|\overline{e}_1|^2+|\overline{e}_2|^2-|\overline{\nu }_1|^2-|\overline{\nu }_2|^2$ and $S_{3}\equiv e_1 \nu _1^\ast+e_2 \nu
   _2^\ast+\overline{e}_1^\ast \overline{\nu }_1+\overline{e}_2^\ast \overline{\nu }_2$.}

For generic VEVs $\text{Rank}\ \mathcal{M}_{{16}}^2 (1,1,0) = 4$, and we recover 12 massless gauge bosons
with the quantum numbers of the standard model algebra.

We verified that this result is maintained when implementing the
constraints of the flipped vacuum manifold in \eq{vacmanifoldFSO10}.
Since it is, by construction, the smallest algebra that can be preserved by the whole vacuum manifold, it must be maintained when adding the $\vev{45_H}$ contribution. We can therefore claim that the invariant algebra on the {generic} vacuum is the SM.
On the other hand, the $45_H$ plays already an active role in this result since it allows for a misalignment of the VEV directions in the two $16_H\oplus \overline{16}_H$ spinors
such that the spinor vacuum preserves SM and not $SU(5)\otimes U(1)$. {More details shall be given in the next section.}

\subsubsection{Adjoint contribution}
\label{gaugespectrum45}

Considering the contribution of $\vev{45_H}$ to the
gauge spectrum, we find
\bea
\label{Mgauge130_45}
&& \mathcal{M}_{{45}}^2 (1,3,0)_{1_{45}} = 0 \, , \\
\label{Mgauge810_45}
&& \mathcal{M}_{{45}}^2 (8,1,0)_{15_{45}} = 0 \, , \\
\label{Mgauge13m13_45}
&& \mathcal{M}_{{45}}^2 (3,1,-\tfrac{1}{3})_{15_{45}} =
4 g^2 \omega _Y^2 \, .
\eea
Analogously, {in} the $(6^-_{45},6^+_{45})$ basis, we have
\begin{widetext}
\be
\label{Mgauge32p16_45}
\mathcal{M}_{{45}}^2 (3,2,+\tfrac{1}{6}) =
\left(
\begin{array}{cc}
  g^2 \left(\left(\omega _R+\omega _Y\right){}^2+2 \omega ^- \omega ^+\right) &  i 2 \sqrt{2} g^2 \omega _Y \omega ^- \\
 - i 2 \sqrt{2} g^2 \omega _Y \omega ^+ &  g^2 \left(\left(\omega _R-\omega _Y\right){}^2+2 \omega ^- \omega ^+\right)
\end{array}
\right)
\, . \\
\ee
\end{widetext}
The SM singlet mass matrix {in} the $\left( 15_{45}, 1^-_{45}, 1^0_{45}, 1^+_{45}, 1_{1}  \right)$ basis {reads}
\begin{widetext}
\be
\label{Mgauge110_45}
\mathcal{M}_{{45}}^2 (1,1,0) =
\left(
\begin{array}{ccccc}
 0 & 0 & 0 & 0 & 0 \\
 0 & 4 g^2 \left(\omega _R^2+\omega ^- \omega ^+\right) & - i 4 g^2 \omega _R \omega ^- & 4 g^2 \left(\omega ^-\right)^2 & 0 \\
 0 & i 4 g^2 \omega _R \omega ^+ & 8 g^2 \omega ^- \omega ^+ & - i 4 g^2 \omega _R \omega ^- & 0 \\
 0 & 4 g^2 \left(\omega ^+\right)^2 & i 4 g^2 \omega _R \omega ^+ & 4 g^2 \left(\omega _R^2+\omega ^- \omega ^+\right) & 0 \\
 0 & 0 & 0 & 0 & 0
\end{array}
\right) \, .
\ee
\end{widetext}
For generic VEVs we find $\text{Rank}\ \mathcal{M}_{{45}}^2 (1,1,0) = 2$
leading globally to the 14 massless gauge bosons
of the $SU(3)_c \otimes SU(2)_L \otimes U(1)^3$ algebra.

As a consistency check, by switching on just the $\omega_R$ and $\omega_Y$ VEVs , we recover the results of \cite{Bertolini:2009es} for standard $SO(10)$.

\subsubsection{Vacuum little group}
\label{FormalproofFSO10}

{With the results of sections \ref{gaugespectrum16} and \ref{gaugespectrum45} at hand the residual gauge symmetry can be readily identified from the properties of the} complete gauge boson mass matrix. For the sake of simplicity here we {shall present the results in the real VEV approximation}.

{Trading} the VEVs for the superpotential parameters,
one can {immediately identify the strong and weak gauge bosons of the SM that, as expected, remain massless:}
\bea
&& \mathcal{M}^2 (8,1,0)_{15_{45}} = 0 \, , \nn\\
&& \mathcal{M}^2 (1,3,0)_{1_{45}} = 0 \, .
\eea
{Similarly, it is straightforward to obtain}
\begin{multline}
\mathcal{M}^2 (3,1,-\tfrac{1}{3})_{15_{45}} = \\
\frac{4 g^2}{9 \tau _1^2 \tau _2^2}
\left(3 \mu  \left(\rho _{22} \tau _1 \left(5 \tau _1-\tau _2\right)  +\rho _{11} \tau _2 \left(5 \tau _2-\tau
   _1\right)\right) \right. \\ \left. +2 \left(\rho _{22} \tau _1+\rho _{11} \tau _2\right){}^2\right) \, .
\end{multline}
On the other hand, the complete matrices $\mathcal{M}^2 (3,2,+\tfrac{1}{6})$ and $\mathcal{M}^2 (1,1,0)$ turn out to be
quite involved once the vacuum constraints are imposed, and we do not show them here explicitly. {Nevertheless, it is sufficient to consider}
\begin{multline}
\Tr \mathcal{M}^2 (3,2,+\tfrac{1}{6}) =
\frac{g^2}{8 \mu ^2} \left[ 16 \mu^2 \left( r_1^2 + r_2^2 \right) + \tau _1^2 r_1^4 + \tau _2^2 r_2^4  \right. \\
\left. + \tau _1 \tau _2 r_1^2 r_2^2
\left( 1 + \cos 2 \alpha^-\right)\right]
\end{multline}
and
\begin{multline}
\label{det3216r2alpham}
\det \mathcal{M}^2 (3,2,+\tfrac{1}{6}) =
\frac{g^4 r_1^2 r_2^2}{128 \mu ^4}
\left[512 \mu ^4 + 32 \mu ^2 \left( \tau _1^2 r_1^2 + \tau _2^2 r_2^2 \right)
\right. \\ \left.
+ \tau _1^2 \tau _2^2 r_1^2 r_2^2 \left( 1 - \cos 2 \alpha^-\right) \right] \sin ^2 \alpha^- \,
\end{multline}
to see that for a generic {non-zero value of $\sin\alpha^-$} one gets Rank $\mathcal{M}^2 (3,2,+\tfrac{1}{6})=2$.
On the other hand, when $\alpha^- = 0$ (i.e., $\vev{16_{1}}\propto\vev{16_{2}}$) or $r_2 = 0$ (i.e., $\vev{16_{2}}=0$),
$\text{Rank}\ \mathcal{M}^2 (3,2,+\tfrac{1}{6}) = 1$ and one is left with an additional massless $(3,2,+\tfrac{1}{6})\oplus (\overline{3},2,-\tfrac{1}{6})$ gauge boson, corresponding to an enhanced residual symmetry.

In the case of the $5$-dimensional {matrix} $\mathcal{M}^2 (1,1,0)$ it is sufficient to {notice that for a generic non-zero $\sin\alpha^{-}$}
\be
\label{rank110r2alpham}
\text{Rank}\ \mathcal{M}^2 (1,1,0) = 4 \,,
\ee
on the vacuum manifold, which leaves a massless $U(1)_Y$ gauge boson, thus completing the SM algebra.
As before, for $\alpha^- = 0$ or for $r_2 = 0$,
we find $\text{Rank}\ \mathcal{M}^2 (1,1,0) = 3$.
Taking into account the massless states in the $(3,2,+\tfrac{1}{6})\oplus (\overline{3},2,-\tfrac{1}{6})$ sector, we recover, as expected, the flipped $SU(5) \otimes U(1)$ algebra.

\section{$E_6$ vacuum}
\label{app:E6vacuum}

\subsection{The $SU(3)^3$ formalism}
\label{app:SU33formalism}

Following closely the notation of Ref. \cite{Buccella:1987kc}, we decompose the adjoint and fundamental representations of $E_6$ under its $SU(3)_C \otimes SU(3)_L \otimes SU(3)_R$ maximal subalgebra as
\bea
\label{adjointE6}
78 &\equiv& (8,1,1) \oplus (1,8,1) \oplus (1,1,8) \oplus (\overline{3},3,3) \oplus (3,\overline{3},\overline{3}) \nn \\
&\subset& T^{\alpha}_{\beta} \oplus T^{i}_{j} \oplus T^{i'}_{j'} \oplus Q^{\alpha}_{i j'} \oplus Q^{i j'}_{\alpha} \, , \\
\label{fundE6}
27 &\equiv& (3,3,1) \oplus (1,\overline{3},3) \oplus (\overline{3},1,\overline{3}) \nn \\
&\equiv& v_{\alpha i} \oplus v^{i}_{j'} \oplus v^{\alpha j'} \, , \\
\label{antifundE6}
\overline{27} &\equiv& (\overline{3},\overline{3},1) \oplus (1,3,\overline{3}) \oplus (3,1,3) \nn \\
&\equiv& u^{\alpha i} \oplus u^{j'}_{i} \oplus u_{\alpha j'} \, ,
\eea
where the greek, latin and primed-latin {indices, corresponding to $SU(3)_c$, $SU(3)_L$ and $SU(3)_R$, respectively, run from $1$ to $3$}.
As far as the $SU(3)$ algebras in \eq{adjointE6} are concerned,
the generators {follow the standard Gell-Mann convention}
\bea
\label{gellmannbasis}
&& T^{(1)} = \tfrac{1}{2} (T^{1}_{2} + T^{2}_{1}) \, , \quad
T^{(2)} = \tfrac{i}{2} (T^{1}_{2} - T^{2}_{1})  \, , \nn \\
&& T^{(3)} = \tfrac{1}{2} (T^{1}_{1} - T^{2}_{2})  \, , \quad
T^{(4)} = \tfrac{1}{2} (T^{1}_{3} + T^{3}_{1})  \, ,  \\
&& T^{(5)} = \tfrac{i}{2} (T^{1}_{3} - T^{3}_{1}) \, , \quad
T^{(6)} = \tfrac{1}{2} (T^{2}_{3} + T^{3}_{2}) \, , \nn \\
&& T^{(7)} = \tfrac{i}{2} (T^{2}_{3} - T^{3}_{2}) \, , \quad
T^{(8)} = \tfrac{1}{2\sqrt{3}} (T^{1}_{1} + T^{2}_{2} - 2 T^{3}_{3}) \, ,\nn
\eea
{with  $(T^{a}_{b})^{k}_{l} = \delta^{k}_{b} \delta^{a}_{l}$, so they are all normalized so that
$\Tr\ T^{(a)} T^{(b)} = \frac{1}{2} \delta^{ab}$.}

Taking into account \eqs{adjointE6}{gellmannbasis},
the $E_6$ algebra can be written as
\bea
&& [ T^{\alpha}_{\beta}, T^{\gamma}_{\eta} ] = \delta^{\alpha}_{\eta} T^{\gamma}_{\beta} - \delta^{\gamma}_{\beta} T^{\alpha}_{\eta} \nn \\
&& [ T^{i}_{j}, T^{k}_{l} ] = \delta^{i}_{l} T^{k}_{j} - \delta^{k}_{j} T^{i}_{l} \nn \\
&& [ T^{i'}_{j'}, T^{k'}_{l'} ] = \delta^{i'}_{l'} T^{k'}_{j'} - \delta^{k'}_{j'} T^{i'}_{l'} \nn \\
&& [ T^{\alpha}_{\beta}, T^{i}_{j} ] = [ T^{\alpha}_{\beta}, T^{i'}_{j'} ] = [ T^{i}_{j}, T^{i'}_{j'} ] = 0 \, ,
\label{E6action78_1}
\eea
\bea
&& [ Q^{\gamma}_{i j'}, T^{\alpha}_{\beta} ] = \delta^{\gamma}_{\beta} Q^{\alpha}_{i j'} \nn \\
&& [ Q^{i j'}_{\gamma}, T^{\alpha}_{\beta} ] = - \delta^{\alpha}_{\gamma} Q^{i j'}_{\beta} \nn \\
&& [ Q^{\gamma}_{i j'}, T^{k}_{l} ] = - \delta^{k}_{i} Q^{\gamma}_{l j'} \nn \\
&& [ Q^{i j'}_{\gamma}, T^{k}_{l} ] = \delta^{i}_{l} Q^{k j'}_{\gamma} \nn \\
&& [ Q^{\gamma}_{i j'}, T^{k'}_{l'} ] = - \delta^{k'}_{j'} Q^{\gamma}_{i l'} \nn \\
&& [ Q^{i j'}_{\gamma}, T^{k'}_{l'} ] = \delta^{j'}_{l'} Q^{i k'}_{\gamma} \, ,
\label{E6action78_2}
\eea
\bea
&& [ Q^{\alpha}_{i j'}, Q^{k l'}_{\beta} ] = - \delta^{\alpha}_{\beta} \delta^{k}_{i} T^{l'}_{j'} - \delta^{\alpha}_{\beta} \delta^{l'}_{j'} T^{k}_{i} + \delta^{k}_{i} \delta^{l'}_{j'} T^{\alpha}_{\beta} \nn \\
&& [ Q^{\alpha}_{i j'}, Q^{\beta}_{k l'} ] = \epsilon^{\alpha \beta \gamma} \epsilon_{i k p} \epsilon_{j' l' q'} Q^{p q'}_{\gamma} \nn \\
&& [ Q^{i j'}_{\alpha}, Q^{k l'}_{\beta} ] = - \epsilon_{\alpha \beta \gamma} \epsilon^{i k p} \epsilon^{j' l' q'} Q^{\gamma}_{p q'} \, ,
\label{E6action78_3}
\eea
The action of the algebra on the fundamental $27$ representation reads
\bea
&& T^{\beta}_{\gamma} v_{\alpha i} = \delta^{\beta}_{\alpha} v_{\gamma i} \nn \\
&& T^{k}_{l} v_{\alpha i} = \delta^{k}_{i} v_{\alpha l} \nn \\
&& T^{k'}_{l'} v_{\alpha i} = 0 \nn \\
&& Q^{\beta}_{p q'} v_{\alpha i} = \delta^{\beta}_{\alpha} \epsilon_{p i k} v^{k}_{q'} \nn \\
&& Q^{p q'}_{\beta} v_{\alpha i} = \delta^{p}_{i} \epsilon_{\beta \alpha \gamma} v^{\gamma q'} \, ,
\label{E6action27a}
\eea
\bea
&& T^{\beta}_{\gamma} v^{i}_{j'} = 0 \nn \\
&& T^{k}_{l} v^{i}_{j'} = - \delta^{i}_{l} v^{k}_{j'} \nn \\
&& T^{k'}_{l'} v^{i}_{j'} = \delta^{k'}_{j'} v^{i}_{l'} \nn \\
&& Q^{\beta}_{p q'} v^{i}_{j'} = - \delta^{i}_{p} \epsilon_{q' j' k'} v^{\beta k'} \nn \\
&& Q^{p q'}_{\beta} v^{i}_{j'} = \delta^{q'}_{j'} \epsilon^{p i k} v_{\beta k} \, ,
\label{E6action27b}
\eea
\bea
&& T^{\beta}_{\gamma} v^{\alpha j'} = - \delta^{\alpha}_{\gamma} v^{\beta j'} \nn \\
&& T^{k}_{l} v^{\alpha j'} = 0 \nn \\
&& T^{k'}_{l'} v^{\alpha j'} = - \delta^{j'}_{l'} v^{\alpha k'} \nn \\
&& Q^{\beta}_{p q'} v^{\alpha j'} = - \delta^{j'}_{q'} \epsilon^{\beta \alpha \gamma} v_{\gamma p} \nn \\
&& Q^{p q'}_{\beta} v^{\alpha j'} = - \delta^{\alpha}_{\beta} \epsilon^{q' j' k'} v^{p}_{k'} \, ,
\label{E6action27c}
\eea
and accordingly on $\overline{27}$
\bea
&& T^{\beta}_{\gamma} u^{\alpha i} = - \delta^{\alpha}_{\gamma} u^{\beta i} \nn \\
&& T^{k}_{l} u^{\alpha i} = - \delta^{i}_{l} u^{\alpha k} \nn \\
&& T^{k'}_{l'} u^{\alpha i} = 0 \nn \\
&& Q^{\beta}_{p q'} u^{\alpha i} = - \delta^{i}_{p} \epsilon^{\beta \alpha \gamma} u_{\gamma q'} \nn \\
&& Q^{p q'}_{\beta} u^{\alpha i} = - \delta^{\alpha}_{\beta} \epsilon^{p i k} u^{q'}_{k} \, ,
\label{E6action27bara}
\eea
\bea
&& T^{\beta}_{\gamma} u^{j'}_{i} = 0 \nn \\
&& T^{k}_{l} u^{j'}_{i} = \delta^{k}_{i} u^{j'}_{l} \nn \\
&& T^{k'}_{l'} u^{j'}_{i} = - \delta^{j'}_{l'} u^{k'}_{i} \nn \\
&& Q^{\beta}_{p q'} u^{j'}_{i} = - \delta^{j'}_{q'} \epsilon_{p i k} u^{\beta k} \nn \\
&& Q^{p q'}_{\beta} u^{j'}_{i} = \delta^{p}_{i} \epsilon^{q' j' k'} u_{\beta k'} \, ,
\label{E6action27barb}
\eea
\bea
&& T^{\beta}_{\gamma} u_{\alpha j'} = \delta^{\beta}_{\alpha} u_{\gamma j'} \nn \\
&& T^{k}_{l} u_{\alpha j'} = 0 \nn \\
&& T^{k'}_{l'} u_{\alpha j'} = \delta^{k'}_{j'} u_{\alpha l'} \nn \\
&& Q^{\beta}_{p q'} u_{\alpha j'} = \delta^{\beta}_{\alpha} \epsilon_{q' j' k'} u^{k'}_{p} \nn \\
&& Q^{p q'}_{\beta} u_{\alpha j'} = \delta^{q'}_{j'} \epsilon_{\beta \alpha \gamma} u^{\gamma p} \, .
\label{E6action27barc}
\eea
Given the SM hypercharge {definition}
\be
\label{SMhyperchargeE6}
Y = \frac{1}{\sqrt{3}} T^{(8)}_L + T^{(3)}_R + \frac{1}{\sqrt{3}} T^{(8)}_R \, ,
\ee
the SM-preserving vacuum direction {corresponds to} \cite{Buccella:1987kc}
\begin{multline}
\label{78vacE6}
\vev{78} = a_1 T^{3'}_{2'} + a_2 T^{2'}_{3'} + \frac{a_3}{\sqrt{6}} (T^{1'}_{1'} + T^{2'}_{2'} - 2T^{3'}_{3'}) \\
+ \frac{a_4}{\sqrt{2}} (T^{1'}_{1'} - T^{2'}_{2'}) + \frac{b_3}{\sqrt{6}} (T^{1}_{1} + T^{2}_{2} - 2T^{3}_{3}) \, ,
\end{multline}
\be
\label{27vacE6}
\vev{27} =  e v^{3}_{3'} + \nu v^{3}_{2'} \, , \quad \vev{\overline{27}} = \overline{e} u^{3'}_{3} + \overline{\nu} u^{2'}_{3} \, ,
\ee
where $a_1$, $a_2$, $a_3$, $a_4$, $b_3$, $e$, $\overline{e}$, $\nu$ and $\overline{\nu}$ are SM-singlet VEVs.
This can be checked {by means} of \eqs{E6action78_1}{E6action27barc}. Notice that the adjoint VEVs $a_3,\ a_4$ and $b_3$ are real, while $a_1=a_2^*$. The {VEVs of $27\oplus \overline{27}$} are generally complex.

\subsection{$E_6$ vacuum manifold}
\label{DFtermE6}

Working out the {$D$-flatness} equations, one finds that the nontrivial constraints are given by
\bea
D_{E_\alpha} &=& \left( \frac{3 a_3}{\sqrt{6}} -\frac{a_4}{\sqrt{2}} \right) a_2^\ast - a_1 \left( \frac{3 a_3^\ast}{\sqrt{6}} - \frac{a_4^\ast}{\sqrt{2}} \right) \nn \\
&& + e_1^\ast \nu_1 - \overline{e}_1 \overline{\nu}_1^\ast + e_2^\ast \nu_2 - \overline{e}_2 \overline{\nu}_2^\ast = 0 \, , \nn \\[1ex]
D_{T^{(8)}_R} &=&\! 3 \left( |a_1|^2 - |a_2|^2 \right)\! +\! 2 \left( |\overline{e}_1|^2 - |e_1|^2 \right)\! +\! 2 \left( |\overline{e}_2|^2 - |e_2|^2 \right) \nn \\
&& + |\nu_1|^2 - |\overline{\nu}_1|^2 + |\nu_2|^2 - |\overline{\nu}_2|^2 = 0 \, , \nn \\[1ex]
D_{T^{(3)}_R} &=& |a_2|^2 - |a_1|^2 + |\overline{\nu}_1|^2 - |\nu_1|^2 + |\overline{\nu}_2|^2 - |\nu_2|^2 = 0 \, , \nn \\
D_{T^{(8)}_L} &=& |e_1|^2 + |\nu_1|^2 + |e_2|^2 + |\nu_2|^2 \nn \\
&& - |\overline{e}_1|^2 - |\overline{\nu}_1|^2 - |\overline{e}_2|^2 - |\overline{\nu}_2|^2 = 0 \, ,
\label{E6Dterms}
\eea
where $D_{E_\alpha}$ is the {ladder operator from the $(1,1,8)$ sub-multiplet} of $78$.
Notice that {the relations corresponding to} $D_{T^{(8)}_R}$, $D_{T^{(3)}_R}$ and $D_{T^{(8)}_L}$ are linearly dependent, since the linear combination associated to the SM hypercharge in Eq. (\ref{SMhyperchargeE6}) vanishes.

The superpotential $W_H$ in \eq{WHE6} {evaluated on the vacuum manifold (\ref{78vacE6})-(\ref{27vacE6})} yields \eq{vevWHE6}.
Accordingly, one finds the following {$F$-flatness} equations
\bea
&& F_{a_1} = \ \mu a_2 - \tau _1 e _1 \overline{\nu}_1- \tau _2 e _2 \overline{\nu}_2
= 0 \, , \nn \\
&& F_{a_2} = \ \mu a_1 - \tau _1 \nu_1 \overline{e }_1- \tau _2 \nu_2 \overline{e }_2
= 0 \, , \nn \\
&& F_{a_3} = \ \mu a_3 - \frac{1}{\sqrt{6}} \left( \tau _1 ( \nu_1 \overline{\nu}_1 - 2 e _1 \overline{e }_1 ) + \tau _2 ( \nu_2 \overline{\nu}_2 - 2 e _2 \overline{e }_2 ) \right)
= 0 \, , \nn \\
&& F_{a_4} = \ \mu a_4 + \frac{1}{\sqrt{2}} \left( \tau _1 \nu_1 \overline{\nu}_1+ \tau _2 \nu_2 \overline{\nu}_2 \right)
= 0 \, , \nn \\
&& F_{b_3} = \ \mu b_3 -\sqrt{\frac{2}{3}} \left( \tau _1 (\nu_1 \overline{\nu}_1+e _1 \overline{e }_1) + \tau _2 (\nu_2 \overline{\nu}_2+e _2 \overline{e }_2) \right)
= 0 \, , \nn
\eea
\bea
3 F_{e_1} = \ && 3 ( \rho _{11} \overline{e }_1+\rho _{12} \overline{e }_2 ) \nn \\
&& -  \tau _1 \left(\sqrt{6}  \left(b_3-a_3\right) \overline{e }_1 +3 a_1 \overline{\nu}_1\right)
= 0 \, , \nn \\
3 F_{e_2} = \ && 3 ( \rho _{21} \overline{e }_1+\rho _{22} \overline{e }_2 ) \nn \\
&& -  \tau _2 \left(\sqrt{6}  \left(b_3-a_3\right) \overline{e }_2 +3 a_1 \overline{\nu}_2\right)
= 0 \, , \nn \\
6 F_{\nu_1} = \ && 6 ( \rho _{11} \overline{\nu}_1+\rho _{12} \overline{\nu}_2 ) \nn \\
&& -  \tau _1 \left( \sqrt{2}( \sqrt{3} a_3 -3 a_4 +2 \sqrt{3} b_3 ) \overline{\nu}_1+6 a_2 \overline{e }_1\right)
= 0 \, , \nn \\
6 F_{\nu_2} = \ && 6 ( \rho _{21} \overline{\nu}_1+\rho _{22} \overline{\nu}_2 ) \nn \\
&& - \tau _2 \left( \sqrt{2}( \sqrt{3} a_3 -3 a_4 +2 \sqrt{3} b_3 ) \overline{\nu}_2+6 a_2 \overline{e }_2\right)
= 0 \, , \nn
\eea
\bea
\label{FtermsE6}
3 F_{\overline{e}_1} = \ && 3 ( \rho _{11} e _1+\rho _{21} e _2 ) \nn \\
&& - \tau _1 \left(\sqrt{6} \left(b_3-a_3\right) e _1 +3 a_2 \nu_1\right)
= 0 \, , \nn \\
3 F_{\overline{e}_2} = \ && 3 ( \rho _{12} e _1+ \rho _{22} e _2 ) \nn \\
&& - \tau _2 \left(\sqrt{6} \left(b_3-a_3\right) e _2 +3 a_2 \nu_2\right)
= 0 \, , \nn \\
6 F_{\overline{\nu}_1} = \ && 6 ( \rho _{11} \nu_1+ \rho _{21}\nu_2 ) \nn \\
&& - \tau _1 \left( \sqrt{2} ( \sqrt{3} a_3 -3 a_4 +2 \sqrt{3} b_3 ) \nu_1 +6 a_1 e _1\right)
= 0 \, , \nn \\
6 F_{\overline{\nu}_2} = \ && 6 ( \rho _{12} \nu_1+ \rho _{22} \nu_2 ) \nn \\
&& - \tau _2 \left( \sqrt{2} ( \sqrt{3} a_3 -3 a_4 +2 \sqrt{3} b_3 ) \nu_2+6 a_1 e _2\right)
= 0 \, . \nn \label{E6Fterms}\\
\eea
{Following the strategy of Appendix \ref{DFtermFSO10} one can solve the first five equations above for   $a_1$, $a_2$, $a_3$, $a_4$ and $b_3$}:
\bea
\label{a1tob3asfunof27}
&& \mu a_1 = \tau _1 \nu_1 \overline{e }_1+ \tau _2\nu_2 \overline{e }_2 \, , \\
&& \mu a_2 = \tau _1 e _1 \overline{\nu}_1+ \tau _2e _2 \overline{\nu}_2 \, , \nn \\
&& \sqrt{6} \mu a_3 = \tau _1 \left(\nu_1 \overline{\nu}_1-2 e _1 \overline{e }_1\right)+\tau _2 \left(\nu_2 \overline{\nu}_2-2 e _2 \overline{e }_2\right) \, ,  \nn \\
&& \sqrt{2} \mu a_4 = - \tau _1 \nu_1 \overline{\nu}_1- \tau _2 \nu_2 \overline{\nu}_2 \, , \nn \\
&& \sqrt{3} \mu b_3 = \sqrt{2} \left( \tau _1 \left(\nu_1 \overline{\nu}_1+ e _1 \overline{e }_1\right)+ \tau _2 \left(\nu_2 \overline{\nu}_2+e _2 \overline{e
   }_2\right) \right) \, . \nn
\eea
{Since $a_{1}=a_{2}^{*}$ and $\tau_1$ and $\tau_2$ can be taken real
without loss of generality (see \sect{vacuumE6}), the first two equations above} imply
\be
\tau _1 \nu_1 \overline{e }_1+ \tau _2\nu_2 \overline{e }_2 =
\tau _1 (e _1 \overline{\nu}_1)^* + \tau _2 (e _2 \overline{\nu}_2)^*
\, ,
\label{Dterma1a2}
\ee
{Using (\ref{a1tob3asfunof27}) the remaining $F$-flatness conditions
in \eq{E6Fterms} can be rewritten in the form}
\bea
3 \mu F^{a}_{e_1} &=& 3 \mu ( \rho _{11} \overline{e }_1+\rho _{12} \overline{e }_2 ) - 4 \tau _1^2 \left( \nu_1 \overline{\nu}_1 + e _1 \overline{e }_1\right)\overline{e }_1 \nn \\
&-& \tau _1 \tau _2 \left(3\nu_2 \overline{\nu}_1 \overline{e }_2+ ( \nu_2 \overline{\nu}_2 +4 e _2 \overline{e }_2 ) \overline{e }_1\right)
= 0 \, , \nn \\[1ex]
3 \mu F^{a}_{\overline{e}_1} &=& 3 \mu ( \rho _{11} e _1 + \rho _{21} e _2 ) - 4 \tau _1^2 \left( \overline{\nu}_1 \nu_1  + \overline{e }_1 e _1 \right)e _1 \nn \\
&-& \tau _1 \tau _2 \left(3 \overline{\nu}_2 \nu_1 e _2 + ( \overline{\nu}_2 \nu_2 +4  \overline{e }_2 e _2  ) e _1 \right)
= 0 \, , \nn \\[1ex]
3 \mu F^{a}_{\nu_1} &=& 3 \mu ( \rho _{11} \overline{\nu}_1+\rho _{12} \overline{\nu}_2 ) - 4 \tau _1^2 \left( e _1  \overline{e }_1+ \nu_1 \overline{\nu}_1\right)\overline{\nu}_1 \nn \\
&-& \tau _1 \tau _2 \left(3 e _2 \overline{e}_1 \overline{\nu}_2 + ( e _2  \overline{e }_2+4 \nu_2  \overline{\nu}_2 ) \overline{\nu}_1\right)
= 0 \, , \nn \\[1ex]
3 \mu F^{a}_{\overline{\nu}_1} &=& 3 \mu ( \rho _{11} \nu_1+ \rho _{21} \nu_2  ) - 4 \tau _1^2 \left( \overline{e }_1 e _1 + \overline{\nu}_1 \nu_1 \right) \nu_1 \nn \\
&-& \tau _1 \tau _2 \left(3 \overline{e}_2 e _1 \nu_2 + ( \overline{e }_2 e _2 +4  \overline{\nu}_2 \nu_2  ) \nu_1\right)
= 0 \, , \nn \\
\label{Faterms}
\eea
{and the additional four relations can be again obtained} by exchanging
$1 \leftrightarrow 2$.
{Similarly, the triplet of linearly independent {$D$-flatness conditions in \eq{E6Dterms}} can be brought to the form}
\bea
D_{E_\alpha} = \ && e_1^\ast \nu_1 - \overline{e}_1 \overline{\nu}_1^\ast + e_2^\ast \nu_2 - \overline{e}_2 \overline{\nu}_2^\ast = 0 \, , \nn \\
D_{T^{(3)}_R} = \ && |\overline{\nu}_1|^2 - |\nu_1|^2 + |\overline{\nu}_2|^2 - |\nu_2|^2 = 0 \, , \nn \\
D_{T^{(8)}_L} = \ && |e_1|^2 + |\nu_1|^2 + |e_2|^2 + |\nu_2|^2 \nn \\
&& - |\overline{e}_1|^2 - |\overline{\nu}_1|^2 - |\overline{e}_2|^2 - |\overline{\nu}_2|^2 = 0 \, .
\eea
{Combining these with \eq{Dterma1a2}, the $D$-flatness is ensured if and only if} $e_{1,2}^{\ast} = \overline{e}_{1,2}$ and $\nu_{1,2}^{\ast} = \overline{\nu}_{1,2}$.
Hence, {in complete analogy with the flipped $SO(10)$ case \eq{complexDtermsFSO10},} one can write
\bea
\label{complexDterms}
&& e_{1,2} \equiv |e_{1,2}| e^{i \phi_{e_{1,2}}} \, , \qquad \overline{e}_{1,2} \equiv |e_{1,2}| e^{- i \phi_{e_{1,2}}} \, , \nn \\
&& \nu_{1,2} \equiv |\nu_{1,2}| e^{i \phi_{\nu_{1,2}}} \, , \qquad \overline{\nu}_{1,2} \equiv |\nu_{1,2}| e^{- i \phi_{\nu_{1,2}}} \, .
\eea

From now on, the discussion of the vacuum manifold follows very closely that for {the} flipped $SO(10)$ in \sect{DFtermFSO10} and we shall not repeat it here. In particular the existence of a nontrivial vacuum requires the hermiticity  of the $\rho$ and $\tau$ couplings. This is related to the fact that $D$- and $F$-flatness require $\vev{27_i}=\vev{\overline{27}_i}^*$.
The detailed shape of the {resulting} vacuum manifold so obtained is
{given} in \eq{vacmanifoldE6} of \sect{vacuumE6}.

\subsection{Vacuum little group}
\label{formalproof}

In order to find the algebra left invariant by the vacuum {configurations} in \eq{vacmanifoldE6}, we need to
compute the action of the $E_6$ generators
on {the} $\vev{78 \oplus 27_1 \oplus 27_2 \oplus \overline{27}_1 \oplus \overline{27}_2}$ VEV.
From \eqs{E6action78_1}{E6action78_2} one obtains
\begin{widetext}
\bea
\label{E6actionvev78}
T^{\alpha}_{\beta} \vev{78} &=& 0 \nn \\
T^{i}_{j} \vev{78} &=& \frac{b_3}{\sqrt{6}} ( \delta^{i}_{1} T^{1}_{j} - \delta^{1}_{j} T^{i}_{1} + \delta^{i}_{2} T^{2}_{j} - \delta^{2}_{j} T^{i}_{2}
- 2 \delta^{i}_{3} T^{3}_{j} + 2 \delta^{3}_{j} T^{i}_{3} ) \nn \\
T^{i'}_{j'} \vev{78} &=& a_1 ( \delta^{i'} _{2'} T^{3'}_{j'} - \delta^{3'} _{j'} T^{i'}_{2'} ) + a_2 ( \delta^{i'} _{3'} T^{2'}_{j'} - \delta^{2'} _{j'} T^{i'}_{3'} )
+ \frac{a_3}{\sqrt{6}} ( \delta^{i'}_{1'} T^{1'}_{j'} - \delta^{1'}_{j'} T^{i'}_{1'} + \delta^{i'}_{2'} T^{2'}_{j'} - \delta^{2'}_{j'} T^{i'}_{2'}
- 2 \delta^{i'}_{3'} T^{3'}_{j'} + 2 \delta^{3'}_{j'} T^{i'}_{3'} ) \nn \\
&+& \frac{a_4}{\sqrt{2}} ( \delta^{i'}_{1'} T^{1'}_{j'} - \delta^{1'}_{j'} T^{i'}_{1'} - \delta^{i'}_{2'} T^{2'}_{j'} + \delta^{2'}_{j'} T^{i'}_{2'} ) \nn \\
Q^{\alpha}_{i j'} \vev{78} &=& - a_1 ( \delta^{3'}_{j'} Q^{\alpha}_{i 2'} ) - a_2 ( \delta^{2'}_{j'} Q^{\alpha}_{i 3'} )
- \frac{a_3}{\sqrt{6}} (\delta^{1'}_{j'} Q^{\alpha}_{i 1'} + \delta^{2'}_{j'} Q^{\alpha}_{i 2'} - 2 \delta^{3'}_{j'} Q^{\alpha}_{i 3'})
- \frac{a_4}{\sqrt{2}} (\delta^{1'}_{j'} Q^{\alpha}_{i 1'} - \delta^{2'}_{j'} Q^{\alpha}_{i 2'}) \nn \\
&-& \frac{b_3}{\sqrt{6}} (\delta^{1}_{i} Q^{\alpha}_{1 j'} + \delta^{2}_{i} Q^{\alpha}_{2 j'} - 2 \delta^{3}_{i} Q^{\alpha}_{3 j'}) \nn \\
Q^{i j'}_{\alpha} \vev{78} &=& a_1 ( \delta^{j'}_{2'} Q^{i 3'}_{\alpha} ) + a_2 ( \delta^{j'}_{3'} Q^{i 2'}_{\alpha} )
+ \frac{a_3}{\sqrt{6}} (\delta^{j'}_{1'} Q^{i 1'}_{\alpha} + \delta^{j'}_{2'} Q^{i 2'}_{\alpha} - 2 \delta^{j'}_{3'} Q^{i 3'}_{\alpha})
+ \frac{a_4}{\sqrt{2}} (\delta^{j'}_{1'} Q^{i 1'}_{\alpha} - \delta^{j'}_{2'} Q^{i 2'}_{\alpha}) \nn \\
&+& \frac{b_3}{\sqrt{6}} (\delta^{i}_{1} Q^{1 j'}_{\alpha} + \delta^{i}_{2} Q^{2 j'}_{\alpha} - 2 \delta^{i}_{3} Q^{3 j'}_{\alpha}) \, ,
\eea
on the adjoint vacuum.
For $\vev{27_1 \oplus 27_2}$ one finds
\bea
\label{E6actionvev271}
T^{\alpha}_{\beta} \vev{27_1 \oplus 27_2} &=& 0 \nn \\
T^{i}_{j} \vev{27_1 \oplus 27_2} &=&
- (e_1+e_2) [ \delta^{3}_{j} v^{i}_{3'} ] - (\nu_1+\nu_2) [ \delta^{3}_{j} v^{i}_{2'} ] \nn \\
T^{i'}_{j'} \vev{27_1 \oplus 27_2} &=&
(e_1+e_2) [ \delta^{i'}_{3'} v^{3}_{j'}  ] + (\nu_1+\nu_2) [ \delta^{i'}_{2'} v^{3}_{j'}  ] \nn \\
Q^{\alpha}_{i j'} \vev{27_1 \oplus 27_2} &=&
- (e_1+e_2) [ \delta^{3}_{i} \epsilon_{j' 3' k'} v^{\alpha k'} ] - (\nu_1+\nu_2) [ \delta^{3}_{i} \epsilon_{j' 2' k'} v^{\alpha k'}  ] \nn \\
Q^{i j'}_{\alpha} \vev{27_1 \oplus 27_2} &=&
(e_1+e_2) [ \delta^{j'}_{3'} \epsilon^{i 3 k} v_{\alpha k}  ] + (\nu_1+\nu_2) [ \delta^{j'}_{2'} \epsilon^{i 3 k} v_{\alpha k}  ] \, ,
\eea
and, accordingly, for $\vev{\overline{27}_1 \oplus \overline{27}_2}$
\bea
\label{E6actionvev272}
T^{\alpha}_{\beta} \vev{\overline{27}_1 \oplus \overline{27}_2} &=& 0 \nn \\
T^{i}_{j} \vev{\overline{27}_1 \oplus \overline{27}_2} &=&
(\overline{e}_1+\overline{e}_2) [ \delta^{i}_{3} u^{3'}_{j} ] + (\overline{\nu}_1+\overline{\nu}_2) [ \delta^{i}_{3} u^{2'}_{j} ] \nn \\
T^{i'}_{j'} \vev{\overline{27}_1 \oplus \overline{27}_2} &=&
- (\overline{e}_1+\overline{e}_2) [ \delta^{3'}_{j'} u^{i'}_{3} ] - (\overline{\nu}_1+\overline{\nu}_2) [ \delta^{2'}_{j'} u^{i'}_{3} ] \nn \\
Q^{\alpha}_{i j'} \vev{\overline{27}_1 \oplus \overline{27}_2} &=&
- (\overline{e}_1+\overline{e}_2) [ \delta^{3'}_{j'} \epsilon_{i 3 k} u^{\alpha k} ] - (\overline{\nu}_1+\overline{\nu}_2) [ \delta^{2'}_{j'} \epsilon_{i 3 k} u^{\alpha k} ] \nn \\
Q^{i j'}_{\alpha} \vev{\overline{27}_1 \oplus \overline{27}_2} &=&
(\overline{e}_1+\overline{e}_2) [ \delta^{i}_{3} \epsilon^{j' 3' k'} u_{\alpha k'} ] + (\overline{\nu}_1+\overline{\nu}_2) [ \delta^{i}_{3} \epsilon^{j' 2' k'} u_{\alpha k'} ] \, .
\eea
\end{widetext}

{On the vacuum manifold in Eq. (\ref{vacmanifoldE6})} one finds that the generators generally preserved by the VEVs of $78 \oplus 27_1 \oplus 27_2 \oplus \overline{27}_1 \oplus \overline{27}_2$ are
\bea
&& T_{c}^{(1)}\ T_{c}^{(2)}\ T_{c}^{(3)}\ T_{c}^{(4)}\ T_{c}^{(5)}\ T_{c}^{(6)}\ T_{c}^{(7)}\ T_{c}^{(8)}: \ (8,1,0) \, , \nn \\
&& T_{L}^{(1)}\ T_{L}^{(2)}\ T_{L}^{(3)}: \ (1,3,0) \, , \nn \\
&& Y : \ (1,1,0) \, , \nn \\
&& Q^{\alpha}_{1 1'} \ Q^{\alpha}_{2 1'} \ Q^{1 1'}_{\alpha} \ Q^{2 1'}_{\alpha}: \ (\overline{3},2,+\tfrac{5}{6}) \oplus (3,2,-\tfrac{5}{6}) \, ,
\label{SU5generators}
\eea
which generate an $SU(5)$ algebra. As an example showing the nontrivial constraints enforced by the vacuum manifold in Eq. (\ref{vacmanifoldE6}),
let us inspect the action {of} one of the lepto-quark generators, say $Q^{\alpha}_{1 1'}$:
\bea
&& Q^{\alpha}_{1 1'} \vev{78} = - \tfrac{1}{\sqrt{6}} \left( a_3 + \sqrt{3} a_4 + b_3 \right) Q^{\alpha}_{1 1'} \, , \label{Q11action}\\
&& Q^{\alpha}_{1 1'} \vev{27_1 \oplus 27_2} = 0 \, , \nn\\
&& Q^{\alpha}_{1 1'} \vev{\overline{27}_1 \oplus \overline{27}_2} = 0 \, . \nn
\eea
It is easy to check that $a_3 + \sqrt{3} a_4 + b_3$ vanishes on the whole vacuum manifold in Eq. (\ref{vacmanifoldE6}) and, thus, $Q^\alpha_{11'}$ is preserved. Let us also remark that the $U(1)_Y$ charges above correspond to the standard $SO(10)$ embedding (see the discussion in sect.~\ref{vacuumE6}). In the flipped $SO(10)$ embedding, the $(\overline{3},2) \oplus (3,2)$ generators in \eq{SU5generators} carry  hypercharges $\mp \tfrac{1}{6}$, respectively.

Considering instead the vacuum manifold invariant with respect to the flipped $SO(10)$ hypercharge (see \eqs{vev78flip}{vev27bar12flip}),
the preserved generators, in addition to those of the SM, are
$Q^{\alpha}_{1 3'} \ Q^{\alpha}_{2 3'} \ Q^{1 3'}_{\alpha} \ Q^{2 3'}_{\alpha}$.
These, for the standard hypercharge embedding of \eq{Ystandard}, transform as
$(\overline{3},2,-\tfrac{1}{6}) \oplus (3,2,+\tfrac{1}{6})$, whereas with the
flipped hypercharge assignment in \eq{Yso10}, the same transform as $(\overline{3},2,+\tfrac{5}{6}) \oplus (3,2,-\tfrac{5}{6})$.
{Needless to say, one finds again $SU(5)$ as the vacuum little group.}

It is interesting to consider the configuration $\alpha_1=\alpha_2=0$, which can be chosen without loss of generality once a pair, let us say $27_2 \oplus \overline{27}_2$, is decoupled {or when the two copies of $27_H \oplus \overline{27}_H$ are aligned}. According to \eq{vacmanifoldE6} this implies all VEVs equal to zero but $a_3 = -b_3$ and $e_1$ ($e_2$).
Then, from \eqs{E6actionvev78}{E6actionvev272}, one verifies that the preserved generators are (see \eq{gellmannbasis} for notation)
\bea
&& T_{c}^{(1)}\ T_{c}^{(2)}\ T_{c}^{(3)}\ T_{c}^{(4)}\ T_{c}^{(5)}\ T_{c}^{(6)}\ T_{c}^{(7)}\ T_{c}^{(8)}: \ (8,1,0) \, , \nn \\
&& T_{L}^{(1)}\ T_{L}^{(2)}\ T_{L}^{(3)}: \ (1,3,0) \, , \nn \\
&& T_{R}^{(1)}\ T_{R}^{(2)}\ T_{R}^{(3)}: \ (1,1,-1) \oplus (1,1,0) \oplus (1,1,+1) \, , \nn \\
&& T_{L}^{(8)} + T_{R}^{(8)}: \ (1,1,0) \, , \nn \\
\eea
\bea
&& Q^{\alpha}_{1 1'} \ Q^{\alpha}_{2 1'} \ Q^{1 1'}_{\alpha} \ Q^{2 1'}_{\alpha}: \ (\overline{3},2,+\tfrac{5}{6}) \oplus (3,2,-\tfrac{5}{6}) \, , \nn \\
&& Q^{\alpha}_{1 2'} \ Q^{\alpha}_{2 2'} \ Q^{1 2'}_{\alpha} \ Q^{2 2'}_{\alpha}: \ (\overline{3},2,-\tfrac{1}{6}) \oplus (3,2,+\tfrac{1}{6}) \, , \nn \\
&& Q^{\alpha}_{3 3'} \ Q^{3 3'}_{\alpha}: \ (\overline{3},1,-\tfrac{2}{3}) \oplus (3,1,+\tfrac{2}{3}) \, ,
\eea
which support an $SO(10)$ algebra.
In particular, $a_3 = -b_3$ preserves $SO(10) \otimes U(1)$, where the extra $U(1)$ generator,
which commutes with all $SO(10)$ generators, is proportional to $T_{L}^{(8)} - T_{R}^{(8)}$.
On the other hand, the VEV $e_1$ breaks $T_{L}^{(8)} - T_{R}^{(8)}$ (while preserving the sum).
We therefore recover the result of Ref. \cite{Buccella:1987kc} for
the $E_6$ setting with {$78_H \oplus 27_H \oplus \overline{27}_H$}.


\end{document}